\documentclass[apj]{emulateapj}
\usepackage{natbib,apjfonts,graphicx,threeparttable,rotating}
\pdfoutput=1

\def\ts   {\thinspace}
\def\kms  {\ifmmode{{\rm \ts km\ts s}^{-1}}\else{\ts km\ts s$^{-1}$}\fi}
\def\kmspc  {\ifmmode{{\rm \ts km\ts s}^{-1}\ts{\rm pc}^{-1}}\else{\ts km\ts s$^{-1}$\ts pc$^{-1}$}\fi}
\def\kkms  {\ifmmode{{\rm \ts K\ts km\ts s}^{-1}}\else{\ts K\ts km\ts s$^{-1}$}\fi}
\def\lcou  {\ifmmode{{\rm \ts K\ts km\ts s}^{-1}\ts {\rm pc}^{2}}\else{\ts K\ts km\ts s$^{-1}$\ts pc$^{2}$}\fi}
\def\xcou  {\ifmmode{{\rm \ts cm^{-2}\ts (K\ts km\ts s^{-1})}^{-1}}\else{\ts cm$^{-2}$\ts (K\ts km\ts s$^{-1}$)$^{-1}$}\fi}
\def\phu  {\ifmmode{{\rm \ts 10^{4}\ts k_{B}\ts cm^{-3}\ts K}}\else{\ts 10$^{4}$\ts k$_{B}$\ts cm$^{-3}$\ts K}\fi}
\def\pcmsq  {\ifmmode{{\rm \ts cm}^{-2}}\else{\ts cm$^{-2}$}\fi}
\def\cc  {\ifmmode{{\rm \ts cm}^{-2}}\else{\ts cm$^{-2}$}\fi}
\def\ccc  {\ifmmode{{\rm \ts cm}^{-3}}\else{\ts cm$^{-3}$}\fi}
\def\mo   {\ifmmode{{\rm M}_{\odot}}\else{M$_{\odot}$}\fi}
\def\lsol   {\ifmmode{{\rm L}_{\odot}}\else{L$_{\odot}$}\fi}
\def\msol   {\ifmmode{{\rm M}_{\odot}}\else{M$_{\odot}$}\fi}
\def\mpcsq   {\ifmmode{{\rm M}_{\odot}\ts {\rm pc}^{-2}}\else{M$_{\odot}$}\ts pc$^{-2}$\fi}
\def\mpcc   {\ifmmode{{\rm M}_{\odot}\ts {\rm pc}^{-3}}\else{M$_{\odot}$}\ts pc$^{-3}$\fi}
\def\myrkpcsq   {\ifmmode{{\rm M}_{\odot}\ts {\rm yr}^{-1}\ts {\rm kpc}^{-2}}\else{M$_{\odot}$}\ts \ts yr$^{-1}$\ts kpc$^{-2}$\fi}
\def\myr   {\ifmmode{{\rm M}_{\odot}\ts {\rm yr}^{-1}}\else{M$_{\odot}$}\ts yr$^{-1}$\fi}
\def\msolyr   {\ifmmode{{\rm M}_{\odot}\ts {\rm yr}^{-1}}\else{M$_{\odot}$}\ts yr$^{-1}$\fi}

\def\aco {\ifmmode{^{12}{\rm CO}(J=1\to0)}\else{$^{12}{\rm CO}(J=1\to0)$}\fi}
\def\bco {\ifmmode{^{12}{\rm CO}(J=2\to1)}\else{$^{12}{\rm CO}(J=2\to1)$}\fi}
\def\m  {\ifmmode{\mu {\rm m}}\else{$\mu$m}\fi}
\def\mg  {\ifmmode{\mu {\rm G}}\else{$\mu$G}\fi}
\def\j {\ifmmode{J=1\to0}\else{$J=1\to0$}\fi}
\def\cco {\ifmmode{^{13}{\rm CO}(J=1\to0)}\else{$^{13}{\rm CO}(J=1\to0)$}\fi}
\def\dco {\ifmmode{^{13}{\rm CO}(J=2\to1)}\else{$^{13}{\rm CO}(J=2\to1)$}\fi}
\def\eco {\ifmmode{{\rm C}^{18}{\rm O}(J=1\to0)}\else{{\rm C}$^{18}{\rm O}(J=1\to0)$}\fi}
\def\hi  {\ifmmode{{\rm H}{\rm \scriptsize I}}\else{H\ts {\scriptsize I}}\fi}
\def\nii  {\ifmmode{{\rm N}{\rm \scriptsize II}}\else{N\ts {\scriptsize II}}\fi}
\def\cii  {\ifmmode{{\rm C}{\rm \scriptsize II}}\else{C\ts {\scriptsize II}}\fi}
\def\civ  {\ifmmode{{\rm C}{\rm \scriptsize IV}}\else{C\ts {\scriptsize IV}}\fi}
\def\ovi  {\ifmmode{{\rm O}{\rm \scriptsize VI}}\else{O\ts {\scriptsize VI}}\fi}
\def\oii  {\ifmmode{{\rm O}{\rm \scriptsize II}}\else{O\ts {\scriptsize II}}\fi}
\def\Ha  {\ifmmode{{\rm H}{\alpha}}\else{H\ts {$\alpha$}}\fi}
\def\nh  {\ifmmode{N(\rm H)}\else{$N$(H)}\fi}
\def\nhi  {\ifmmode{N(\hi)}\else{$N$(\hi)}\fi}
\def\nhh  {\ifmmode{N(\rm H_{\rm 2})}\else{$N$(H$_{\rm 2}$)}\fi}
\def\nhm  {\ifmmode{N({\rm H})_{\rm mol}}\else{$N$(H)$_{\rm mol}$}\fi}
\def\hun  {\ifmmode{I_{100}}\else{$I_{100}$}\fi}
\def\sex  {\ifmmode{I_{60}}\else{$I_{60}$}\fi}
\def\hh   {\ifmmode{{\rm H}_2}\else{H$_2$}\fi}
\def\zwco  {\ifmmode{^{12}{\rm CO}}\else{$^{12}{\rm CO}$}\fi}
\def\nzwco  {\ifmmode{N(^{12}{\rm CO})}\else{$N(^{12}{\rm CO})$}\fi}
\def\wzwco  {\ifmmode{W(^{12}{\rm CO})}\else{$W(^{12}{\rm CO})$}\fi}
\def\drco  {\ifmmode{^{13}{\rm CO}}\else{$^{13}{\rm CO}$}\fi}
\def\ndrco  {\ifmmode{N(^{13}{\rm CO})}\else{$N(^{13}{\rm CO})$}\fi}
\def\wdrco  {\ifmmode{W(^{13}{\rm CO})}\else{$W(^{13}{\rm CO})$}\fi}
\def\tex  {\ifmmode{T_{ex}({\rm CO})}\else{$T_{ex}({\rm CO})$}\fi}
\def\xco   {\ifmmode{X_{\rm CO}}\else{$X_{\rm CO}$}\fi}

\def\mh     {H$_{2}$}
\def\ha   {\ifmmode{{\rm H}{\alpha}}\else{H\ts {$\alpha$}}\fi}  
\def\hii  {\ifmmode{{\rm H}{\rm \small II}}\else{H\ts {\scriptsize II}}\fi}
\def\ihi {\ifmmode{I(\hi)}\else{$I(\hi)$}\fi}
\def\ico {\ifmmode{I(\rm CO)}\else{$I(\rm CO)$}\fi}
\def\lco {\ifmmode{L_{\rm CO}}\else{$L_{\rm CO}$}\fi}
\def\tpk {\ifmmode{T_{\rm pk}}\else{$T_{\rm pk}$}\fi}
\def\tpkco {\ifmmode{T_{\rm pk}(\rm CO)}\else{$T_{\rm pk}(\rm CO)$}\fi}
\def\avtpk {\ifmmode{\langle T_{\rm pk} \rangle}\else{$\langle T_{\rm pk} \rangle$}\fi}
\def\covd {\ifmmode{\sigma_{\rm v}({\rm CO})}\else{$\sigma_{\rm v}$(CO)}\fi}
\def\tmax {\ifmmode{T_{\rm max}}\else{$T_{\rm max}$}\fi}
\def\avtpkhi {\ifmmode{\langle T_{\rm pk}(\hi) \rangle}\else{$\langle T_{\rm pk}(\hi) \rangle$}\fi}
\def\tpkhi {\ifmmode{T_{\rm pk}(\hi)}\else{$T_{\rm pk}(\hi)$}\fi}
\def\hivd {\ifmmode{\sigma_{\rm v}(\hi)}\else{$\sigma_{\rm v}(\hi)$}\fi}
\def\avhivd {\ifmmode{\langle \sigma_{\rm v}(\hi) \rangle}\else{$\langle \sigma_{\rm v}(\hi) \rangle$}\fi}
\def\avsp {\ifmmode{\langle r \rangle}\else{$\langle r \rangle$}\fi}
\def\avspp {\ifmmode{\langle p \rangle}\else{$\langle p \rangle$}\fi}
\def\D {\ifmmode{^{\circ}}\else{$^\circ$}\fi}
\def\nodata {\ifmmode{...}\else{$...$}\fi}

\shorttitle{CO PDFs in M51 by PAWS}
\slugcomment{\today\ draft}

\begin{document}

\title{Probability Distribution Functions of \aco\ Brightness and
  Integrated Intensity in M51: the PAWS View}

\author{
Annie Hughes\altaffilmark{1},
Sharon E. Meidt\altaffilmark{1},
Eva Schinnerer\altaffilmark{1},
Dario Colombo\altaffilmark{1},
Jer\^ome Pety\altaffilmark{2,3},
Adam K. Leroy\altaffilmark{4},
Clare L. Dobbs\altaffilmark{5},
Santiago Garc\'ia-Burillo\altaffilmark{6},
Todd A. Thompson\altaffilmark{7,8},
Ga\"elle Dumas\altaffilmark{2},
Karl F. Schuster\altaffilmark{2},
Carsten Kramer\altaffilmark{9}
\altaffiltext{1}{Max-Planck-Institut f\"ur Astronomie, K\"onigstuhl 17, D-69117, Heidelberg, Germany}
\altaffiltext{2}{Institut de Radioastronomie Millim\'etrique, 300 Rue de la Piscine, F-38406 Saint Martin d'H\`eres, France}
\altaffiltext{3}{Observatoire de Paris, 61 Avenue de l'Observatoire, F-75014 Paris, France}
\altaffiltext{4}{National Radio Astronomy Observatory, 520 Edgemont Road, Charlottesville, VA 22903, USA}
\altaffiltext{5}{School of Physics and Astronomy, University of Exeter, Stocker Road, Exeter EX4 4QL, UK}
\altaffiltext{6}{Observatorio Astron\'omico Nacional, Observatorio de Madrid, Alfonso XII, 3, 28014 Madrid, Spain}
\altaffiltext{7}{Department of Astronomy, The Ohio State University, 140 W. 18th Ave., Columbus, OH 43210, USA} 
\altaffiltext{8}{Center for Cosmology and AstroParticle Physics, The Ohio State University, 191 W. Woodruff Ave., Columbus, OH 43210, USA}
\altaffiltext{9}{Instituto Radioastronom\'ia Milim\'etrica, Av. Divina Pastora 7, Nucleo Central, 18012 Granada, Spain}
}

\begin{abstract}
%%%%%%%%%%%%%%%%%%%%%%%%%%%%%%%%
%%%%%%%%%%%%%%%%%%%%%%%%%%%%%%%%
%\section{Abstract}
%%%%%%%%%%%%%%%%%%%%%%%%%%%%%%%%
%%%%%%%%%%%%%%%%%%%%%%%%%%%%%%%%
\label{sect:abstract}

\noindent We analyse the distribution of CO brightness temperature and
integrated intensity in M51 at $\sim40$\,pc resolution using new
\aco\ data from the Plateau de Bure Arcsecond Whirlpool Survey
(PAWS). We present probability distribution functions (PDFs) of the CO
emission within the PAWS field-of-view, which covers the inner $\sim11
\times 7$\,kpc of M51. We find clear variations in the shape of CO
PDFs both within different M51 environments, defined according to
dynamical criteria, and between M51 and two nearby low-mass galaxies,
M33 and the Large Magellanic Cloud.  Globally, the PDFs for the inner
disk of M51 can be represented by narrow lognormal functions that
cover $\sim1$ to 2 orders of magnitude in CO brightness and integrated
intensity. The PDFs for M33 and the LMC are narrower and peak at lower
CO intensities, consistent with their lower gas surface
densities. However, the CO PDFs for different dynamical environments
within the PAWS field depart significantly from the shape of the
global distribution. The PDFs for the interarm region are
approximately lognormal, but in the spiral arms and central region of
M51, they exhibit diverse shapes with a significant excess of bright
CO emission. The observed environmental dependence on the shape of the
CO PDFs is qualitatively consistent with changes that would be
expected if molecular gas in the spiral arms is characterised by a
larger range of average densities, gas temperatures and velocity
fluctuations, though further work is required to disentangle the
relative importance of large-scale dynamical effects versus star
formation feedback in regulating these properties. We show that the
shape of the CO PDFs for different M51 environments is only weakly
related to global properties of the CO emission, e.g. the total CO
luminosity, but is strongly correlated with properties of the local
giant molecular cloud (GMC) and young stellar cluster populations,
including the shape of their mass distributions. For galaxies with
strong spiral structure such as M51, our results indicate that
galactic-scale dynamical processes play a significant role in the
formation and evolution of GMCs and stellar clusters.
\end{abstract}


%%%%%%%%%%%%%%%%%%%%%%%%%%%%%%%%%%%
%%%%%%%%%%%%%%%%%%%%%%%%%%%%%%%%%%%
\section{Introduction}
%%%%%%%%%%%%%%%%%%%%%%%%%%%%%%%%%%%
%%%%%%%%%%%%%%%%%%%%%%%%%%%%%%%%%%%
\label{sect:intro}

\noindent Although the interstellar medium (ISM) represents a minor
fraction of the baryonic matter in galaxies, it plays an important
role in their evolution, providing the raw fuel for star formation,
receiving and then redistributing heavy elements created in stellar
interiors, and mediating the exchange of matter and energy between
galaxies and the intergalactic medium (IGM). Conditions in the ISM are
influenced by multiple physical processes occurring across a range of
temporal and spatial scales, including accretion of primordial IGM
material, protostellar jets and outflows, spiral shocks, supernovae,
and thermal instability in the diffuse atomic gas. As a consequence of
this diverse physics, interstellar gas occurs in multiple phases, with
temperatures, densities and spatial structures that span five orders
of magnitude or more.\\

\noindent Due to its complex hierarchical structure, quantitative
analysis of ISM properties and dynamics is challenging. The intensity
of emission from ISM structures can be characterised in terms of a
fractal index \citep[e.g.][]{elmegreenfalgarone96} or power spectrum
\citep[e.g.][]{blocketal10}, while properties of the velocity field
can be summarised using a power spectrum, or by the use of structure
functions \citep[e.g.][]{bruntetal03}. The density (and column
density) structure of the ISM is most commonly represented using a
probability density function (PDF) \citep[e.g.][]{kainulainenetal09}.
Density PDFs of the ISM are supposed to follow a lognormal (LN)
distribution. One of the most widespread explanations for a LN
distribution is that the expected density PDF resulting from a
turbulent velocity field is lognormal
\citep[e.g.][]{padoanetal97}. Simulations of supersonically turbulent
isothermal gas find that the width of the density PDF increases with
the root-mean-square (rms) Mach number
\citep[e.g.][]{padoannordlund02}, and that the precise form of the
relationship depends on the relative importance of compressible and
solenoidal modes in the turbulence forcing
\citep{federrathetal08,federrathetal10}. Comparison with observations
of the Taurus and IC5146 molecular clouds indicates that turbulent
driving in the ISM must contain a significant compressive component,
or that the width of the density PDF depends on additional physics
that is not included in the simulations \citep{priceetal11}. However
these models apply only on the scale of individual molecular clouds -
where the assumption of isothermality is reasonable - not to whole
galaxies. The properties of ISM turbulence should depend on the energy
injection scale and the height of the galactic disk, both of which
will become relevant for galactic-scale systems. As noted by previous
authors \citep[e.g.][]{wadanorman07}, the PDFs of galaxies and
individual molecular clouds cannot be directly compared since a
galaxy's total molecular gas content cannot be characterised by a
single temperature or a spatially uniform, time-independent Mach
number.\\

\noindent In spite of these considerations, LN PDFs are still apparent
in galaxy-scale simulations
\citep[e.g.][]{wadanorman07,dobbsetal08,taskerbryan06}. The lognormal
shape of the density PDFs emerges quickly in the simulations
(i.e. within a local dynamical time), and is suprisingly robust to a
diverse range of additional input physics \citep[e.g. magnetic fields
  and energy feedback from stellar winds and
  supernovae,][]{wadanorman01,dobbsetal11}. The characteristic density
of high- and low-mass models is roughly invariant ($\langle n \rangle
\sim 1$\,\ccc) for disks with initial mass densities spanning an order
of magnitude \citep[e.g.][]{wadanorman07}. The proposed explanation
for why LN density PDFs appear to be generic is that galactic disks
are globally stable, with a hierarchical density structure that
results from the action of a large number of stochastic, independent
and nonlinear processes; in this case, the density PDF should evolve
towards a LN shape by the central limit theorem
\citep[e.g.][]{vazquezsemadeni94}. Notably, the only simulation in
\citet{dobbsetal11} that shows a strong departure from a LN PDF is the
one with very low star formation efficiency and hence a low level of
thermal and kinetic energy feedback into the ISM. In this case, the
model galaxy disk does not achieve an equilibrium state, since most of
the interstellar gas becomes quickly confined to dense,
gravitationally bound clumps (C. Dobbs, priv. comm). \\

\noindent More recently, LN density PDFs have gained more
significance, as they have been used as the basis for explaining the
Kennicutt-Schmidt (KS) relation \citep{schmidt59,kennicutt98}. The KS
relation relates the surface density of gas to the star formation rate
surface density according to $\Sigma_{SFR} = A\Sigma_{\rm
  gas}^{n}$. On global scales (i.e. averaged over entire star-forming
disks), \citet{kennicutt98} obtained $A = (2.5\pm0.7) \times 10^{-4}$
and $n=1.4\pm0.15$ for a composite sample of $\sim100$ normal and
starburst galaxies. The KS relation can also be written using the
molecular gas surface density only, in which case $n$ is nearly linear
\citep[e.g.][]{wongblitz02,bigieletal08,bigieletal11}. That a LN
density PDF for the interstellar gas might naturally yield the KS
relation was first suggested by \citet{elmegreen02}, and later
developed in more detail by \citet{kravtsov03},
\citet{krumholzmckee05} and \citet{wadanorman07}. The PDF is used to
find the fraction of gas above a given density threshold, which is
assumed to collapse and form stars. Summing up the mass of gas at each
density divided by the free fall time provides the star formation
rate. This approach also forms the basis of the universal star
formation law recently proposed by \citet{krumholzetal12}.\\

\noindent In spite of the potential importance of the ISM's
hierarchical structure to the interpretation of extragalactic star
formation laws, observational evidence to support a universal LN
density PDF for galactic disks remains
scarce. \citet{gaustadvanburen93} used the {\it Infrared Sky Survey
  Atlas} to measure gas densities near $\sim1800$ OB stars within
$\sim400$\,pc of the Sun, obtaining a density PDF that resembles a LN
function for densities in the range $0.1 < n <
10$\,\ccc. \citet{wadaetal00} showed that the PDF of \hi\ column
density in the Large Magellanic Cloud is approximately lognormal (at
$\sim15$\,pc resolution) over $\sim2$ orders of magnitude. More
recently, \citet{berkhuijsenfletcher08} derived average volume
densities in the diffuse (i.e. $n < 1$\,\ccc) ionized and atomic gas
for sightlines towards $\sim200$ pulsars and $\sim400$ stars within a
few kiloparsecs of the Sun. The resulting density PDFs were consistent
with a LN function over $\sim2$ dex, but the precise shape of the PDF
(i.e. the dispersion and density corresponding to the distribution
peak) varied with Galactic latitude. Over larger scales, neither the
density nor column density distribution of molecular hydrogen has been
widely investigated. In part, this is because observations of
\aco\ emission -- the most widely-used tracer of extragalactic
molecular gas -- have been limited to low resolution surveys of whole
galaxies, or high resolution (often interferometric) imaging over a
small fraction of a galactic disk. On much smaller scales, the PDF of
\hh\ column density has been examined for individual molecular clouds
in the solar neighbourhood, mostly through extinction of background
stars \citep[e.g.][]{kainulainenetal09, froebrichrowles10}. These
studies have shown that the column density PDFs of non-star-forming
clouds are approximately lognormal, but that star-forming clouds
exhibit power-law tails at high column densities, presumably due to
the formation of high density regions undergoing localised collapse.\\

\noindent Our aim in this paper is to provide a quantitative
description of the properties of the CO emission in our high angular
resolution survey of M51's inner disk (PAWS, Schinnerer et al., in
preparation), in order to provide simple empirical benchmarks for
models of the \hh\ content of galactic disks
\citep[e.g.][]{taskertan09,dobbsetal11}. For this purpose, we present
PDFs of CO integrated intensity and brightness temperature in M51, M33
and the Large Magellanic Cloud (LMC), as well as CO PDFs for different
dynamical environments within the PAWS field. The extent to which the
CO emission in M51 can be attributed to discrete, self-gravitating
structures akin to giant molecular clouds (GMCs) in the Milky Way is
discussed elsewhere (Colombo et al., submitted). The rest of this
paper is organized as follows. We summarize the origin and
characteristics of the CO datasets that we have used in
Section~\ref{sect:data}. Our method for constructing the PDFs is
outlined in Section~\ref{sect:analysis}, and our results are presented
in Section~\ref{sect:results}. Our discussion in
Section~\ref{sect:discussion} focusses on the relationship between
PDFs constructed from our CO observations on $\sim40$\,pc scales and
the PDFs of \hh\ density and column density predicted by numerical
models, and on the connection between GMC properties, star formation
and the shape of the PDF for different M51 environments. We conclude
with a summary of our key results in
Section~\ref{sect:conclusions}. We include three appendices, where we
describe how we tested the robustness of the PDFs to non-physical
effects such as resolution, sensitivity and signal identification, and
assessed the uncertainty associated with our estimates for the slopes
of the GMC and young stellar cluster mass distributions. \\

%%%%%%%%%%%%%%%%%%%%%%%%%%%%%%
%%%%%%%%%%%%%%%%%%%%%%%%%%%%%%
\section{Data}
%%%%%%%%%%%%%%%%%%%%%%%%%%%%%%
%%%%%%%%%%%%%%%%%%%%%%%%%%%%%%
\label{sect:data}

\subsection{M51}

\noindent The CO data for M51 were obtained by the Plateau de Bure
Arcsecond Whirlpool Survey (PAWS, Schinnerer et al., in
preparation). PAWS observations mapped a total field of view of
approximately 270\arcsec\ $\times$ 170\arcsec\ in the inner disk of
M51 in the ABCD configurations of the PdBI between August 2009 and
March 2010. Since an interferometer filters out low spatial
frequencies, the PdBI data were combined with observations of CO
emission in M51 obtained using the IRAM 30\,m single-dish telescope in
May 2010. In this section, we summarize the most important aspects of
the observations and data reduction; the PAWS observing strategy, data
reduction and combination procedures, and flux calibration are
described in detail by Pety et al. (submitted).\\

\subsubsection{PdBI Observations}

\noindent The PdBI observations consisted of two 30-field mosaics,
centered such that their combination covers the inner part of M51.
The mosaic pointings follow a hexagonal pattern, with each pointing
being separated from its nearest neighbors by the primary beam full
width at half maximum (FWHM). Each pointing was observed for
$3\times15$ seconds in turn, allowing us to cycle completely through
one mosaic pattern between calibrations, which were obtained every
22.5 minutes. The hexagonal pattern ensures Nyquist sampling along the
declination axis but slightly undersamples the beam along the right
ascension axis. The total telescope time in all four array
configurations was 169 hours. The total on-source integration time
during which useful data were obtained was 126.5 hours.\\

\noindent For the PAWS observations, the two polarizations of the
PdBI's single-sideband receiver were tuned to 115.090\,GHz, i.e. the
\aco\ rest frequency redshifted to the LSR velocity (471.26\,\kms) of
M51. Four correlator bands of 160\,MHz per polarization were
concatenated to cover a total bandwidth of 550\,MHz, corresponding to
a velocity bandwidth of 1430\,\kms. The intrinsic frequency (velocity)
channel spacing was 1.25\,MHz (3.25\,\kms). We later smoothed the data
to a velocity resolution of 5\,\kms\ to reduce the effect of
correlation between adjacent frequency channels and increase
signal-to-noise. Inspection of the data showed CO emission between
$\pm110$\,\kms\ of M51's systemic velocity. We therefore imaged and
deconvolved $120 \times 5$\,\kms\ channels covering the LSR velocity
range [174,769]\,\kms.\\

\noindent Calibration of the PdBI data was carried out using standard
methods implemented in {\tt GILDAS/CLIC}\footnote{See {\tt
    http://www.iram.fr/IRAMFR/GILDAS} for more information about the
  GILDAS software \citep{pety05}.} The bright ($\sim10$\,Jy) quasars
0851+202 and 3C279 were used as bandpass calibrators. The temporal
phase and amplitude gains were obtained from spline fits through
regular measurements of the quasars 1418+546, 1308+326 and
J1332+473. The flux scale was determined against the PdBI's primary
flux calibrator, MWC349, and were found to be accurate within
$\sim10$\%.

\subsubsection{IRAM 30\,m Observations}

\noindent CO emission in M51 was observed with the IRAM-30m single
dish telescope in order to recover the low spatial frequency
information filtered out by the PdBI. A $\sim60$ square arcminute
field, covering the entire M51 system, was mapped in position-switch
on-the-fly observing mode. For this, we divided the survey
field-of-view into seven regions. Four regions covered the central
400\arcsec $\times 400$\arcsec\ of M51; the remaining three regions
extended the coverage to include the ends of the spiral arms and M51's
companion, NGC5195. To suppress scan artifacts, each region was
scanned in orthogonal directions, i.e. along the right ascension and
declination axes, with each position in the central 400\arcsec $\times
400$\arcsec\ observed 34 times on average. The slew speed was
8\arcsec/sec and the dump time was 0.5s, yielding $\sim5.5$
integrations per beam in the scanning direction. The scanning rows
were separated by 8\arcsec, slightly oversampling the beam. We checked
the pointing every hour, and estimate the positional accuracy to be
$\sim2$\arcsec. Hot and cold loads plus the sky contribution were
observed every 12 minutes to establish the temperature scale.\\

\noindent M51 observations were conducted with the EMIR receivers and
WILMA autocorrelator backend to simultaneously record data for the
\aco\ and \cco\ emission lines. We used the upper sideband for \aco,
with a total bandwidth of 8\,GHz. The channel spacing was 2\,MHz,
corresponding to a velocity channel spacing of 5.4 and 5.2\,\kms\ at 110
and 115\,GHz respectively. \\

\noindent The {\tt GILDAS/MIRA} software was used to calibrate
 the temperature scale of the 30\,m data. ``OFF'' spectra were constructed
using {\tt GILDAS/MIRA}'s default scheme, i.e. averaging the closest
(in time) observations together. These``OFF'' spectra were then
subtracted from the corresponding on-source spectra. Visual inspection
indicated the presence of signal between $-200$ and $+300$\,\kms\ of
M51's systemic velocity. A third-order polynomial was fit and
subtracted from each spectrum. For the baseline fitting, we used an
outlier-resistant approach and excluded regions of the spectrum that
were known to contain bright emission, based on our inspection of
trial data reductions or other observations. We experimented with
higher and lower order baselines and found a third-degree fit to yield
the best results. After fitting, we compared the RMS noise about the
baseline fit in signal-free regions of each spectrum to the expected
theoretical noise. Based on this comparison, we rejected a small
number of spectra where the observed noise was much greater than
expected. \\

\noindent We gridded the calibrated, off-subtracted,
baseline-subtracted spectra into a data cube with a pixel size of
4\arcsec, weighting each spectrum by the inverse of the RMS noise. For
the gridding, we employed a gaussian convolution kernel with a FWHM of
8\arcsec. This gridding strategy increases the effective FWHM of the
beamwidth to $\sim23\farcs5$ at 115\,GHz. After gridding, we fit and
subtracted a second set of third-order polynomial baselines from each
spectrum; this procedure was a minor refinement to the initial
(pre-gridding) baseline fit.\\

\subsubsection{Combination of Single-dish and Interferometric Data}

\noindent The final PAWS data cube is a joint deconvolution of the
PdBI and IRAM 30\,m data sets. The short-spacing visibilities not
sampled by the PdBI were recreated from the single-dish map using the
{\tt GILDAS/MAPPING} software. For this, the map was deconvolved from
the IRAM 30\,m beam in the Fourier plane before multiplication by the
PdBI primary beam in the image plane \citep[as described
  by][]{rodriguezfernandezetal08}. After a further Fourier transform,
pseudo-visibilities were sampled between 0 and 15\,m (the diameter of
a PdBI antenna), and these visibilities were then merged with the
interferometric observations. For the joint deconvolution, we used an
adaption of the H\"ogbom CLEAN algorithm, as implemented in {\tt
  GILDAS/MAPPING}. Supports defining the region to search for CLEAN
components were defined for each velocity channel, based on where
significant emission was detected in the 30\,m cube. The convergence
of the deconvolution was checked in three different ways. First, the
cumulative flux as a function of the number of CLEAN components
converged in each channel. Second, the residual channel images look
like noise. Both criteria indicate a satisfying convergence of the
deconvolution. Finally, we deconvolved the data a second time using
exactly the same method except that we doubled the maximum number of
CLEAN components from 320,000 to 640,000. The subtraction of both
cubes again looks like noise.\\

\begin{figure*}
\begin{center}
\includegraphics[width=90mm,angle=270]{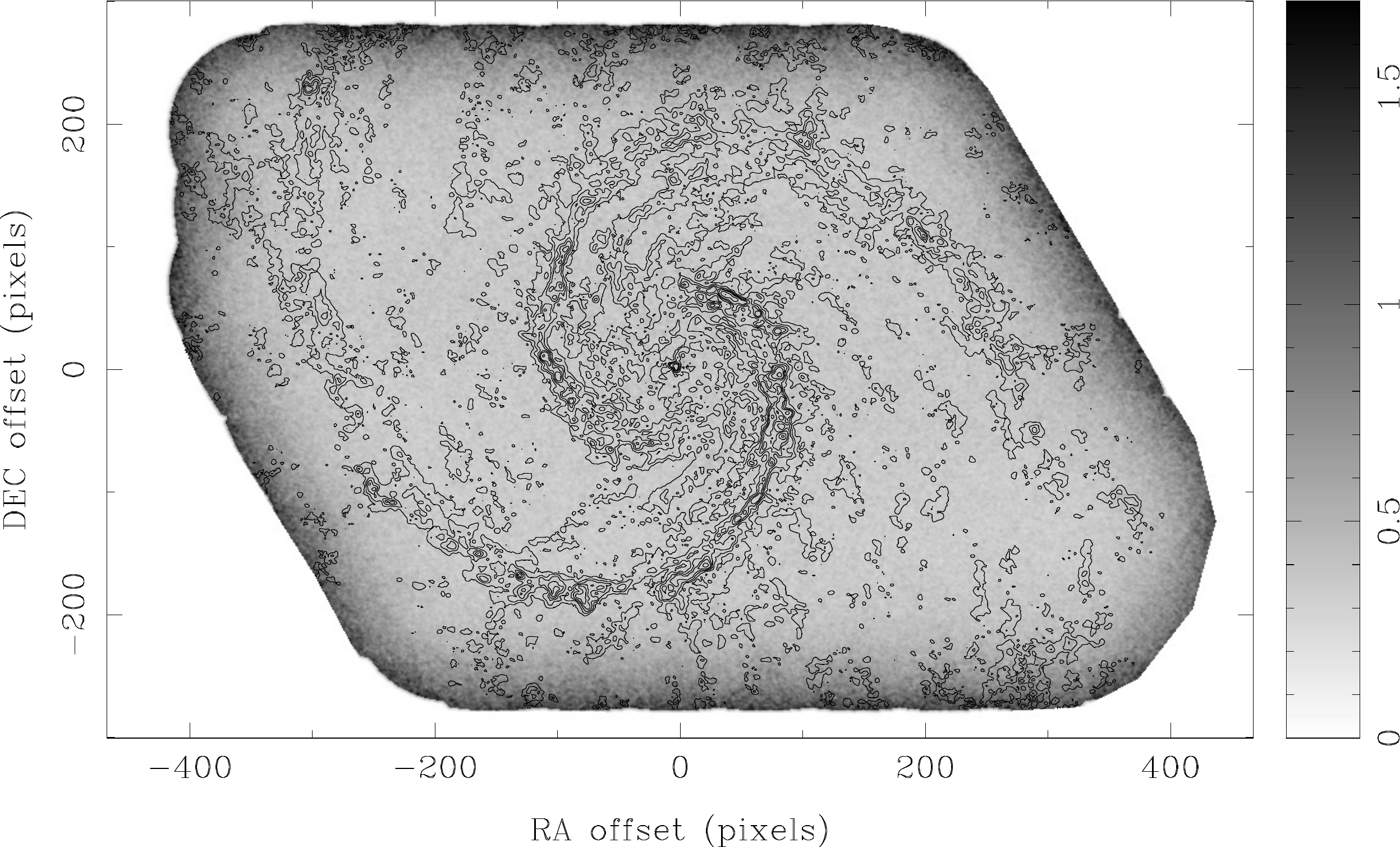}
\caption{\small The RMS of the noise fluctuations across the PAWS
  survey region. The greyscale image is shown in units of $T_{\rm mb}$
  K. The black contours indicate $\ico = 20, 70, 120, 170, 220, 320,
  320$\kkms, as measured by PAWS.}
\label{fig:icomap}
\end{center}
\end{figure*}

\noindent The effective angular resolution of the final combined PAWS
data cube is 1\farcs16 $\times$ 0\farcs97, corresponding to a spatial
resolution of $\sim40$\,pc at our assumed distance to M51
\citep[7.6\,Mpc,][]{ciardulloetal02}. The data cube covers the LSR
velocity range 173 to 769\,\kms\ and the width of each velocity
channel is 5\,\kms.  The mean RMS of the noise fluctuations across the
survey is $\sigma_{RMS} \sim0.4$\,K in a 5.0\,\kms\ channel.  For a
typical CO linewidth of 15\,\kms, this corresponds to an average
sensitivity of 3.5\,\kkms\ for the map of CO integrated intensity. A
map of the noise fluctuations across the PAWS field, overlaid with
contours of CO integrated intensity, is shown in
Figure~\ref{fig:icomap}. \\

\subsection{M33}

\noindent For M33, we use the CO data presented by
\citet{rosolowskyetal07}, which combines observations by BIMA
\citep{engargiolaetal03} and FCRAO \citep{heyeretal04}. The common
field-of-view of the single-dish and interferometer surveys is 0.25
square degrees, covering most of M33's optical disk. The angular
resolution of the combined cube is $13\farcs2 \times 12\farcs9$,
corresponding to a spatial resolution of 53\,pc for our assumed
distance to M33 of 840\,kpc \citep[e.g.][]{galletietal04}. The data
covers the LSR velocity range $[-400,40]$\,\kms, and the velocity
channel width is 2.0\,\kms. The average RMS noise per channel is
0.24\,K.

\subsection{The Large Magellanic Cloud}

\noindent The CO data for the LMC were obtained by the Magellanic
Mopra Assessement (MAGMA). The MAGMA survey design, data acquisition,
reduction procedures and calibration are described in detail by
\citet{wongetal11}. MAGMA mapped CO cloud complexes that had been
identified at lower resolution by NANTEN \citep{fukuietal08},
targeting 114 NANTEN GMCs with CO luminosities greater than
$7000$\,\lcou, and peak integrated intensities greater than
$1$\,\kkms. The combined field-of-view of the MAGMA survey is
$\sim3.6$ square degrees. Although the clouds targeted for mapping
represent only $\sim50$\% of the clouds in the NANTEN catalog, the
region surveyed by MAGMA contributes $\sim80$\% of the total CO flux
measured by NANTEN. The MAGMA LMC data cube has an effective
resolution of 45\arcsec, corresponding to a linear resolution of
$\sim11$\,pc at the distance of the LMC
\citep[$50.1$\,kpc,][]{alves04}. The velocity channel width is
0.53\,\kms, and the total LSR velocity range of the cube is 200 to
305\,\kms. The average RMS noise per channel across the MAGMA field is
$0.3$\,K.

%%%%%%%%%%%%%%%%%%%%%%%%%%%%%%
%%%%%%%%%%%%%%%%%%%%%%%%%%%%%%
\section{Methods}
%%%%%%%%%%%%%%%%%%%%%%%%%%%%%%
%%%%%%%%%%%%%%%%%%%%%%%%%%%%%%
\label{sect:analysis}

%%%%%%%%%%%%%%%%%%%%%%%%%%%%%%
%%%%%%%%%%%%%%%%%%%%%%%%%%%%%%
\subsection{CO Probability Distribution Functions}
%%%%%%%%%%%%%%%%%%%%%%%%%%%%%%
%%%%%%%%%%%%%%%%%%%%%%%%%%%%%%
\label{sect:analysis_pdfs}

\noindent We construct PDFs of CO integrated intensity \ico\ and CO
brightness $T_{\rm mb}$. The \ico\ PDF is simply a histogram of the
$(x,y)$ pixel values within a \ico\ map, while the $T_{\rm mb}$ PDF is
a histogram of the $(x,y,v)$ pixel values within a spectral line
cube. The PDFs are constructed after applying a blanking mask that
identifies genuine emission within the data cubes. For our analysis of
the PAWS data cube, we identify significant emission using the masking
method described by Pety et al. (submitted). For our comparative
analysis of M51, M33 and the LMC in Section~\ref{sect:pdfs_allgals},
we construct an initial mask that contains pixels above a
$5\sigma_{RMS}$ threshold over two or more contiguous velocity
channels. This mask defines a high significance core, which is then
expanded to include all connected pixels above $1.2\sigma_{RMS}$ over
at least two velocity channels. We discuss the rationale for using
these blanking masks, and the influence of different masking
techniques on the shape of the PDFs in
Appendix~\ref{app:pdfs_methods}. The total number of independent data
points within the \ico\ PDFs varies between $\sim15000$ and
$\sim250000$ (for the $T_{\rm mb}$ PDFs, this increases by factor of
$\sim7$). For the \ico\ PDFs, we normalise the histogram by the number
of pixels within the survey field-of-view, not by the number of pixels
where significant emission is detected. Likewise, we normalise the
$T_{\rm mb}$ histogram by the number of independent $(x,y,v)$ elements
in the data cube. \\

\noindent As we discuss in Section~\ref{sect:pdfs_enviros}, the PDFs
of CO integrated intensity and CO brightness $T_{\rm mb}$ within M51
exhibit diverse shapes that are often inconsistent with a simple
functional form such as a lognormal or power-law distribution. We
therefore parameterise the shape of each PDF using the {\it brightness
  distribution index} ($BDI$), a metric recently devised by
\citet{sawadaetal12} to characterize the ratio between faint and
bright \aco\ emission within an $0\fdg8 \times 0\fdg8$ field in the
Galactic plane. More precisely, we specify the $BDI$ of CO brightness
as:
\begin{equation}
\label{eqn:bdi}
BDI = \log  \left( \frac{\sum_{T_{2} < T_{i} < T_{3}} T_{i}}{\sum_{T_{0} <T_{i} < T_{1}} T_{i}} \right)
\end{equation}
\noindent where $T_{i}$ is the brightness of the $i^{\rm th}$ pixel,
and $(T_{0},T_{1},T_{2},T_{3}) = (1.2,2.5,5,\infty)$\,K are the
thresholds that we use to define faint and bright emission. These are
not the same thresholds adopted by \citet{sawadaetal12} for their
analysis at $\lesssim1$\,pc resolution, but are chosen such that
variations in the shape of the PDF are captured (i.e. $BDI$ is
defined) for all the M51 environments that we analyse and that all the
pixels included in the calculation contain significant CO emission
(i.e. $T_{i} > 3\sigma_{RMS}$). We define an equivalent parameter for
the \ico\ PDFs, which we refer to as the {\it integrated intensity
  distribution index} ($IDI$). Analogous to Equation~\ref{eqn:bdi}, we
specify this as:
\begin{equation}
\label{eqn:idi}
IDI = \log  \left( \frac{\sum_{I_{2} < I_{i} < I_{3}} I_{i}}{\sum_{I_{0} <I_{i} < I_{1}} I_{i}} \right)
\end{equation}
\noindent adopting $(I_{0},I_{1},I_{2},I_{3}) = (10.5,25,60,\infty)$\,\kkms. \\

\noindent When appropriate, we derive the best-fitting LN function to
a PDF using a Levenberg-Marquardt fit to the function:
\begin{equation}
\label{eqn:lmfit}
P(s) = c_{0} \times \exp \left [\frac{-(\log s - \log
    s_{0})^{2}}{2x^{2}} \right ]\,.
\end{equation}
\noindent We define the probability $P(s)$ as the number of pixels in
the bin divided by the total number of pixels in the map (or
cube). Depending on context, $s$ represents \ico\ or $T_{\rm
  mb}$. Only bins to the right of the peak of the PDF, brighter than
$4\sigma_{RMS}$ and containing ten or more pixels are used to derive
the fit. Several of the PDFs resemble power-laws more than LN
functions. In this case, we estimate the best-fitting slope of the
power-law using ordinary least squares linear regression. 

%%%%%%%%%%%%%%%%%%%%%%%%%%%%%%
%%%%%%%%%%%%%%%%%%%%%%%%%%%%%%
\subsection{Identifying and Parameterising GMC Properties}
%%%%%%%%%%%%%%%%%%%%%%%%%%%%%%
%%%%%%%%%%%%%%%%%%%%%%%%%%%%%%
\label{sect:analysis_gmcs}

\noindent In Section~\ref{sect:pdfs_vs_gmcs}, we investigate whether
there is a connection between the shape of the CO PDFs and the
properties of giant molecular clouds (GMCs) identified within the PAWS
field. For this, we use the GMC catalog presented by Colombo et
al. (submitted). The catalog was constructed using the
\textsc{CPROPS} package \citep[][henceforth
  RL06]{rosolowskyleroy06}. \textsc{CPROPS} uses a dilated mask
technique to isolate regions of significant emission within spectral
line cubes, and a modified watershed algorithm to assign the emission
into individual clouds. Moments of the emission along the spatial and
spectral axes are used to determine the size, linewidth and flux of
the clouds.\\

\noindent To generate the PAWS GMC catalog, \textsc{CPROPS} first
identifies significant emission by finding pixels with CO brightness
$T_{\rm mb}$ above a $4\sigma_{RMS}$ threshold across two adjacent
velocity channels, where the RMS noise $\sigma_{RMS}$ is estimated
from the median absolute deviation (MAD) of each spectrum. This mask
is then expanded to include all connected pixels with $T_{\rm mb} >
1.5\sigma_{RMS}$. Emission regions are then decomposed into GMCs by
identifying emission that can be uniquely associated with local
maxima. Full details of the decomposition procedure are presented in
Colombo et al. (submitted).\\

\noindent We adopt the default \textsc{CPROPS} definitions of GMC
properties. The cloud radius is defined as $R = 1.91 \sigma_{\rm R}$\,pc,
where $\sigma_{\rm R}$ is the geometric mean of the second moments of the
emission along the cloud's major and minor axes. The velocity
dispersion $\sigma_{\rm v}$ is the second moment of the emission
distribution along the velocity axis, which for a Gaussian line
profile is related to the FWHM linewidth, $\Delta v$, by $\Delta v =
\sqrt{8 \ln 2}\sigma_{\rm v}$. The CO luminosity of the cloud $L_{\rm
  CO}$ is the emission inside the cloud integrated over position and
velocity, i.e.
\begin{equation} 
L_{\rm CO} \; [\lcou] = D^{2} \left( \frac{\pi}{180 \times 3600} \right)^{2} \Sigma T \delta v \delta x \delta y\; ,
\label{eqn:lcodef}
\end{equation}
\noindent where $D$ is the distance to the galaxy in parsecs, $\delta x$
and $\delta y$ are the spatial dimensions of a pixel in arcseconds,
and $\delta v$ is the width of one channel in \kms. The mass of
molecular gas estimated from the GMC's CO luminosity $M_{\rm CO}$ is
calculated as
\begin{equation}
M_{\rm CO} \; [\msol] \equiv 4.4 \frac{\xco}{2 \times 10^{20} [\xcou]} L_{\rm CO}\; ,
\label{eqn:mcodef}
\end{equation}
\noindent where \xco\ is the assumed CO-to-\mh\ conversion factor, and
a factor of 1.36 is applied to account for the mass contribution of
helium. The fiducial value of \xco\ used by \textsc{CPROPS} is $\xco =
2.0 \times 10^{20}$\,\xcou. The virial mass is estimated as $M_{\rm
  vir} \; [\msol] = 1040 \sigma_{\rm v}^{2}R$, which assumes that
molecular clouds are spherical with truncated $\rho \propto r^{-1}$
density profiles \citep{maclarenetal88}. \textsc{CPROPS} estimates the
error associated with a cloud property measurement using a
bootstrapping method, which is described in section~2.5 of RL06.  \\

\noindent The final PAWS GMC catalog contains 1507 objects. The GMCs
have peak brightness temperatures between $\sim2$ and 16\,K, radii
between 5 and 150\,pc, and velocity dispersions between 1 and
30\,\kms. The catalog and the properties of GMCs in different
environments within the PAWS field are the subject of a companion
paper (Colombo et al., submitted). In general, the physical properties
of the cataloged GMCs are similar to the GMCs identified by CO surveys
of the inner Milky Way and other nearby galaxies, although GMCs in M51
tend to be larger, brighter and have higher velocity dispersions and
mass surface densities relative to their size than the GMCs in nearby
low-mass systems such as the Magellanic Clouds and M33 (Hughes et al.,
in preparation). The spatial resolution and sensitivity of PAWS is
sufficient to resolve structures with size and mass comparable to a
typical Galactic GMC \citep[50\,pc, 10$^{5}$\,\msol][]{blitz93}. We
note, however, that CO emission is almost ubiquitous across the PAWS
field, and that much of the emission resides in large
($\sim$kpc-sized) regions of high brightness that bear little
resemblance to Galactic GMCs. Overall, the cataloged GMCs account for
approximately half of the total CO flux within the PAWS datacube, a
fraction that varies from $\sim40$\% in the interarm region to
$\sim60$\% in the spiral arms and central zone.

%%%%%%%%%%%%%%%%%%%%%%%%%%%%%%
%%%%%%%%%%%%%%%%%%%%%%%%%%%%%%
\section{Results}
%%%%%%%%%%%%%%%%%%%%%%%%%%%%%%
%%%%%%%%%%%%%%%%%%%%%%%%%%%%%%
\label{sect:results}

%%%%%%%%%%%%%%%%%%%%%%%%%%%%%%
\subsection{\ico\ PDF}
%%%%%%%%%%%%%%%%%%%%%%%%%%%%%%
\label{sect:pdfs_ico}

\begin{figure*}
\centering
\hspace{-0.5cm}
\includegraphics[width=75mm,angle=0]{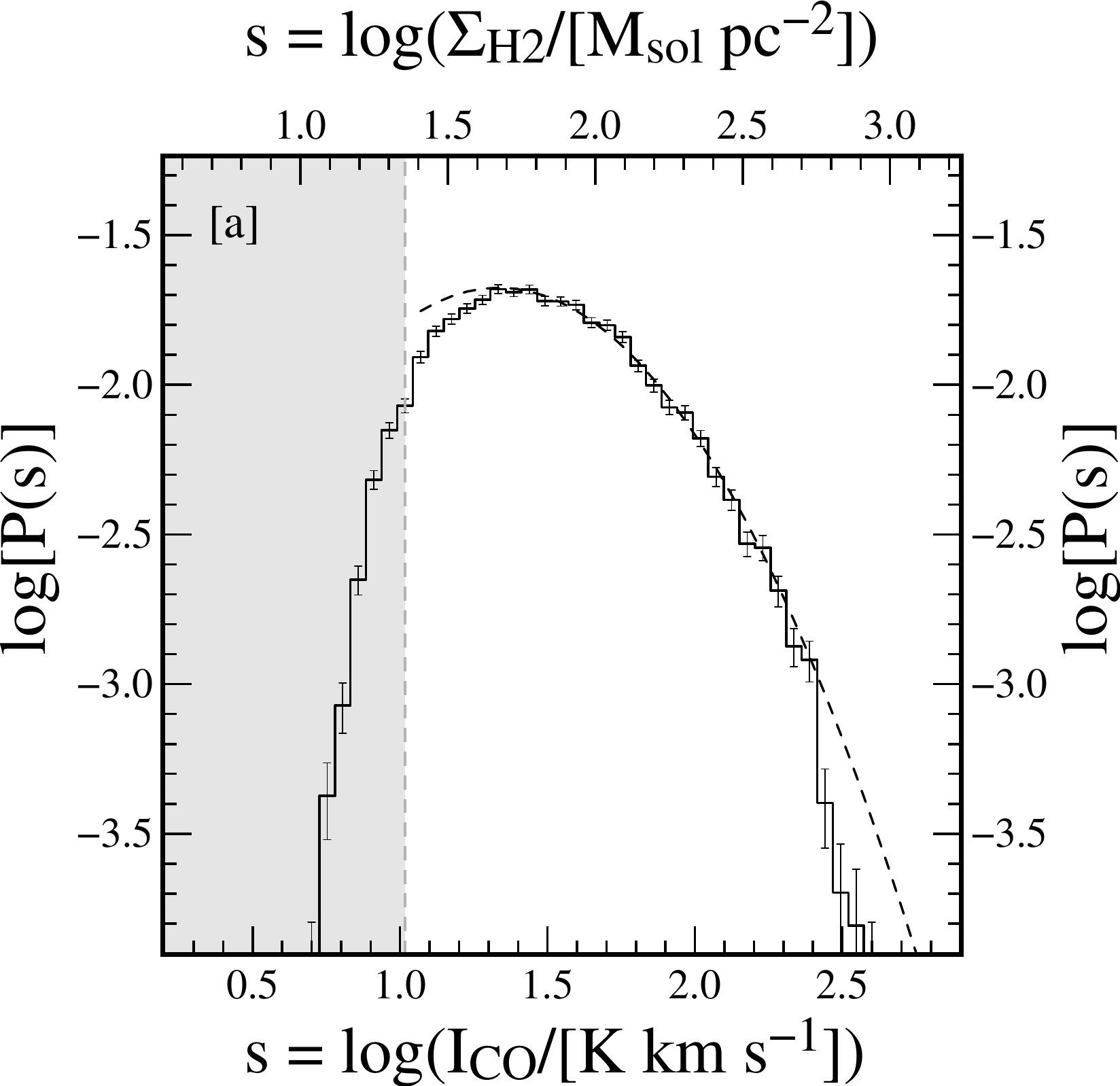}
\hspace{0.6cm}
\includegraphics[width=75mm,angle=0]{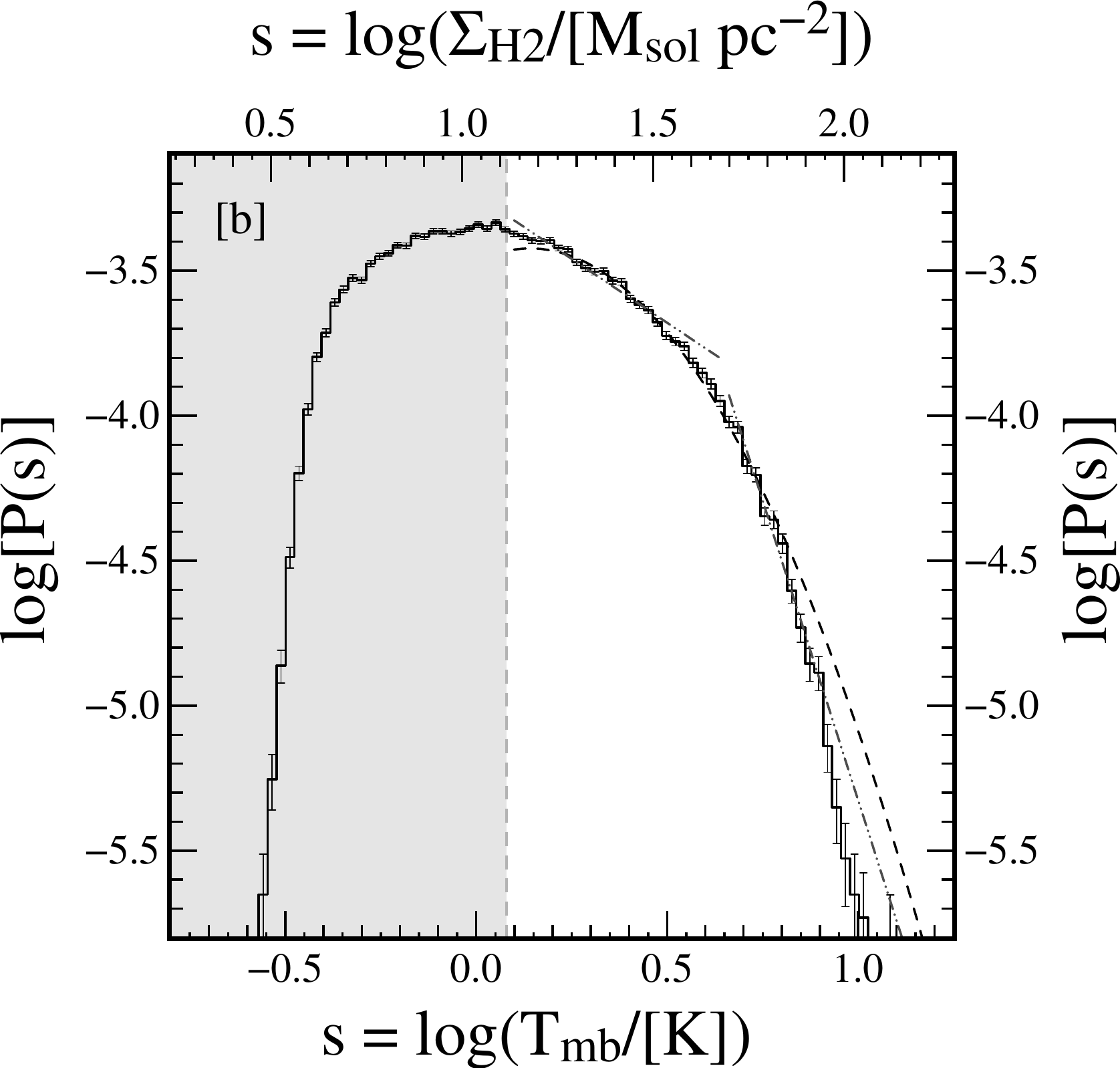}
\caption{\small The [a] \ico\ and [b] $T_{\rm mb}$ PDFs for emission
  within the PAWS field. In both panels, a dashed parabola indicates
  the LN function that provides the best-fit to the PDF. In panel [b],
  the dot-dot-dashed lines represent two segments of a broken
  power-law, which also provides a reasonable fit to the
  distribution. The grey shaded region indicates values beneath our nominal
  $3\sigma_{RMS}$ sensitivity limits of 10.5\,\kkms (panel [a]) and
  1.2\,K (panel [b]). The top horizontal axis shows the equivalent
  \mh\ mass surface density for the \ico\ or $T_{\rm mb}$ value on the
  lower axis, assuming $\xco = 2 \times 10^{20}$\,\xcou\ and a helium
  contribution of 1.36 by mass. The vertical error bars represent the
  uncertainty associated with simple counting ($\sqrt{N}$) errors. }
\label{fig:pdf_paws}
\end{figure*}

\noindent The PDF of CO integrated intensity for the entire PAWS field
is presented in Figure~\ref{fig:pdf_paws}[a]. The distribution is
adequately described by a LN function, with a mean of $\langle \ico
\rangle = 17.4$\,\kkms\ and a logarithmic width of 0.44. The
logarithmic dispersion in the fit residuals $\epsilon$ for bins above
the $3\sigma$ sensitivity limit that contain more than five counts is
0.08. Relative to this LN function, there is some evidence for a
truncation at high \ico\ values ($\gtrsim 200$\,\kkms). In principle,
this could be due to opacity of the \aco\ emission line, an effect
that is often observed on parsec scales in regions of high extinction
\citep[$A_{\rm V} \gtrsim 5 - 10$\,mag,
  e.g.][]{lombardietal06,pinedaetal08}, but has rarely been considered
for the scales probed by extragalactic observations
\citep[cf.][]{dickmanetal86}. Against this intepretation, the 99$^{\rm
  th}$ percentile of the CO peak brightness within the PAWS field is
only $\sim7$\,K, suggesting that there are negligible sightlines where
the CO emission completely fills the telescope beam. Furthermore,
recent numerical simulations that examine the ability of \ico\ to
trace the \hh\ column density on 20 to 60\,pc scales show that
saturation tends to produce a secondary peak in the \ico\ PDF rather
than a smooth truncation \citep[][see also
  Section~\ref{sect:models}]{shettyetal11,feldmannetal12}. The finite
resolution of observational data can also produce a truncation at high
intensities (we explore this effect in
Appendix~\ref{app:pdfs_resn_sens}). In this case, however, we would
expect the \ico\ PDFs for subregions within the PAWS field to exhibit
similar thresholds, whereas several of them are consistent with pure
LN functions (see Section~\ref{sect:pdfs_enviros}). \\

\noindent The truncation of the PDF is Figure~\ref{fig:pdf_paws}[a]
may therefore be physical. \citet{elmegreen11} show that the density
PDF should fall beneath a pure LN function at high gas densities if
the Mach number decreases with increasing average gas density (as
would be expected for a cloud that obeys the \citet{larson81}
size-linewidth relation). Alternatively, the truncation may reflect
the efficacy of feedback processes that prevent the molecular gas from
reaching very high mass surface densities ($\Sigma_{\rm H_{2}} \gtrsim
400\,\mpcsq$, assuming $\xco = 2.0 \times 10^{20}$\,\xcou). We discuss
physical processes that could be influencing the shape of the
\ico\ PDFs in M51 in more detail in Section~\ref{sect:models}. \\

\begin{table*}
\centering
\caption{\small Fit Parameters for CO PDFs in M51, M33, the LMC and Environments within M51}
\label{tbl:pdffits}
\par \addvspace{0.2cm}
\begin{threeparttable}
{\footnotesize
\begin{tabular}{@{}cllcccc}
\multicolumn{7}{c}{Lognormal Fits} \\
\hline 
\hline 
Figure & CO Property    &  Galaxy/Region         & Mean          & \multicolumn{2}{c}{Logarithmic Width} &  Goodness-of-fit     \\
       &                &                        & $s_{0}$        &  \multicolumn{2}{c}{$x$}              & $\epsilon$      \\
\hline
\ref{fig:pdf_paws}[a]   & \ico\           &  PAWS field      & 21.6\,\kkms      & \multicolumn{2}{c}{0.44}  & 0.08 \\
\ref{fig:pdf_paws}[b]   & $T_{\rm mb}$      &  PAWS field     & 1.4\,K           & \multicolumn{2}{c}{0.31}  & 0.18 \\

\ref{fig:pdf_ico_enviros}[a]  & \ico\          &  bar           & 50.5\,\kkms        & \multicolumn{2}{c}{0.21} & 0.41\\
\ref{fig:pdf_ico_enviros}[b]  &                &  ring          & 40.8\,\kkms        & \multicolumn{2}{c}{0.47} & 0.18\\
\ref{fig:pdf_ico_enviros}[c]  &                &  A1I           & 23.8\,\kkms        & \multicolumn{2}{c}{0.47} & 0.20\\
\ref{fig:pdf_ico_enviros}[d]  &                &  A1O           & 36.6\,\kkms        & \multicolumn{2}{c}{0.35} & 0.08\\
\ref{fig:pdf_ico_enviros}[e]  &                &  A1            & 27.4\,\kkms        & \multicolumn{2}{c}{0.44} & 0.28\\
\ref{fig:pdf_ico_enviros}[f]  &                &  A2            & 25.4\,\kkms        & \multicolumn{2}{c}{0.35} & 0.10\\
\ref{fig:pdf_ico_enviros}[g]  &                &  up            & 18.8\,\kkms        & \multicolumn{2}{c}{0.25} & 0.09\\
\ref{fig:pdf_ico_enviros}[h]  &                &  down          & 24.0\,\kkms        & \multicolumn{2}{c}{0.22} & 0.13\\

\ref{fig:pdf_cube_enviros}[a] & $T_{\rm mb}$    &  bar           & 1.5\,K              & \multicolumn{2}{c}{0.30} & 0.11\\
\ref{fig:pdf_cube_enviros}[b] &                &  ring          & 1.6\,K              & \multicolumn{2}{c}{0.37} & 0.19 \\
\ref{fig:pdf_cube_enviros}[c] &                &  A1I           & 1.0\,K              & \multicolumn{2}{c}{0.38} & 0.17\\
\ref{fig:pdf_cube_enviros}[d] &                &  A1O           & 1.5\,K              & \multicolumn{2}{c}{0.29} & 0.17 \\
\ref{fig:pdf_cube_enviros}[e] &                &  A1            & 1.3\,K              & \multicolumn{2}{c}{0.32} & 0.20 \\
\ref{fig:pdf_cube_enviros}[f] &                &  A2            & 1.2\,K              & \multicolumn{2}{c}{0.31} & 0.10 \\
\ref{fig:pdf_cube_enviros}[g] &                &  up            & 1.1\,K              & \multicolumn{2}{c}{0.24} & 0.05\\
\ref{fig:pdf_cube_enviros}[h] &                &  down          & 1.0\,K              & \multicolumn{2}{c}{0.29} & 0.06 \\

\ref{fig:pdf_3gals}[a]  & \ico\           &  M51            & 12.3\,\kkms       & \multicolumn{2}{c}{0.53} & 0.10\\
\ref{fig:pdf_3gals}[c]  &                 &  LMC            & 1.8\,\kkms        & \multicolumn{2}{c}{0.26} & 0.13\\
\ref{fig:pdf_3gals}[d]  & $T_{\rm mb}$     &  M51            & 0.8\,K            & \multicolumn{2}{c}{0.36} & 0.32 \\
\ref{fig:pdf_3gals}[f]  &                 &  LMC            & 0.1\,K            & \multicolumn{2}{c}{0.28} & 0.14 \\
\hline
\hline
%\vspace{1cm}
\\[0.5cm]
\multicolumn{7}{c}{Power-Law Fits} \\
\hline
\hline
Figure &  CO Property    &  Galaxy/Region         & Slope           & Slope 2        & Domain   & Goodness-of-fit      \\
       &                 &                        &  $\gamma_{1}$   & $\gamma_{2}$     &          & $\epsilon$           \\
\hline
\ref{fig:pdf_paws}[b]   & $T_{\rm mb}$     &  PAWS field      & $-0.88$          & $-4.13$  & $\gamma_{1}: [1 < T_{\rm mb} < 3]$\,K & 0.10 \\
                        &                 &                                     &          & $\gamma_{2}: [5 < T_{\rm mb} < 8]$\,K &      \\

\ref{fig:pdf_ico_enviros}[c]  & \ico\           &  A1I              & $-0.85$          & \nodata  & $\gamma_{1}: \ico > 10$\,\kkms       &   0.24 \\

\ref{fig:pdf_cube_enviros}[b] & $T_{\rm mb}$     &  ring           & $-0.53$          & $-6.26$  &  $\gamma_{1}: [1 < T_{\rm mb} < 5]$\,K  &  0.09 \\
                              &                 &                 &                  &          &  $\gamma_{2}: [8 < T_{\rm mb} < 12.5]$\,K &      \\
\ref{fig:pdf_cube_enviros}[c] &                 &  A1I            & $-0.97$          & $-4.68$  &  $\gamma_{1}: [1 < T_{\rm mb} < 4]$\,K  &  0.07 \\
                              &                 &                 &                  &          &  $\gamma_{2}: [5 < T_{\rm mb} < 8]$\,K &      \\
\ref{fig:pdf_cube_enviros}[d] &                 &  A1O            & $-0.96$          & $-4.97$  &  $\gamma_{1}: [1 < T_{\rm mb} < 4]$\,K  &  0.07 \\
                              &                 &                 &                  &          &  $\gamma_{2}: [5 < T_{\rm mb} < 8]$\,K &      \\
\ref{fig:pdf_cube_enviros}[e] &                 &  A1             & $-0.97$          & $-4.82$  &  $\gamma_{1}: [1 < T_{\rm mb} < 4]$\,K  &  0.08 \\
                              &                 &                 &                  &          &  $\gamma_{2}: [5 < T_{\rm mb} < 8]$\,K &      \\

\ref{fig:pdf_3gals}[d]  & $T_{\rm mb}$     &  M51            & $-1.30$           & $-4.96$  & $\gamma_{1}: [1 < T_{\rm mb} < 3]$\,K  & 0.15 \\
                        &                &                 &                   &          & $\gamma_{2}: [5 < T_{\rm mb} < 8]$\,K  &      \\
\ref{fig:pdf_3gals}[e]  &                &  M33            & $-3.63$           & \nodata  & $\gamma_{1}: [0.8 < T_{\rm mb} < 1.3]$\,K  & 0.13 \\ 
\ref{fig:pdf_3gals}[f]  &                &  LMC            & $-2.15$           & \nodata  & $\gamma_{1}: [0.2 < T_{\rm mb} < 1]$\,K  & 0.07 \\
\hline			
\hline
\end{tabular}}
{\footnotesize 
\begin{tablenotes}
\item[]{Parameters of best-fitting functions to PDFs in
  Figures~\ref{fig:pdf_paws}, \ref{fig:pdf_ico_enviros},
  \ref{fig:pdf_cube_enviros} and~\ref{fig:pdf_3gals}. The parameters
  of the LN functions are determined from a Levenberg-Marquardt fit to
  Equation~\ref{eqn:lmfit}; the power-law and broken power-law fits
  are estimated using ordinary least squares regression. We use the
  logarithmic dispersion of the fit residuals to estimate the
  goodness-of-fit.}
\end{tablenotes}}
\end{threeparttable}
\end{table*}

%%%%%%%%%%%%%%%%%%%%%%%%%%%%%%
\subsection{$T_{\rm mb}$ PDF}
%%%%%%%%%%%%%%%%%%%%%%%%%%%%%%

\noindent The PDF of CO brightness for the PAWS cube is shown in
Figure~\ref{fig:pdf_paws}[b]. The $T_{\rm mb}$ PDF is less like a LN
function than the $\ico$ PDF, with two roughly flat segments across $1
< T_{\rm mb} < 3$\,K and $T_{\rm mb} > 5$\,K. Our LN fit to the PDF
yields a mean $\langle T_{\rm mb} \rangle \sim 0.9$\,K and logarithmic
width $x \sim 0.35$. Assuming that the true distribution of $T_{\rm
  mb}$ values is LN, there are fewer high brightness pixels than would
be expected from this LN function. The truncation begins to occur at a
CO brightness temperature of $\sim5$\,K, which would seem too low to
be due to opacity effects. Instead of an LN function, a broken
power-law with a slope of $\sim-1.4$ for $1 < T_{\rm mb} < 5$\,K and a
much steeper slope of $\sim-4.9$ for $T_{\rm mb} > 5$\,K -- or,
alternatively, a pure power-law with a truncation at $\sim5$\,K -- may
provide a better description of the PDF. The fit parameters and
goodness-of-fit for the best-fitting LN and power-law functions to the
$T_{\rm mb}$ PDF in Figure~\ref{fig:pdf_paws}[b] are listed in
Table~\ref{tbl:pdffits}. \\

%%%%%%%%%%%%%%%%%%%%%%%%%%%%%%
\subsection{M51 Environments}
%%%%%%%%%%%%%%%%%%%%%%%%%%%%%%
\label{sect:pdfs_enviros}

\noindent An important question that we would like to address with the
PAWS data is whether the organization and physical properties of
molecular gas depend on galactic environment. Variations in the shape
of the PDF could reflect differences in the relative importance of
self-gravity, star formation feedback or gas flow in different parts
of the galactic disk, which in turn might influence the ability of the
molecular gas to form stars. Within the PAWS field, there are three
main regions where the gas is likely to experience distinct physical
conditions: within the strong spiral arms, the inter-arm region
situated upstream and downstream of the spiral arms, and the central
region, where the gas is influenced by the presence of a nuclear
stellar bar \citep{zaritskyetal93}. These regions can be further
classified according to their level of star formation activity (as
traced by e.g. \ha) and/or gas flows, which we determine using the
present-day torque profile (Meidt et al., submitted).\\

\begin{figure*}
\begin{center}
\includegraphics[width=50mm,angle=270]{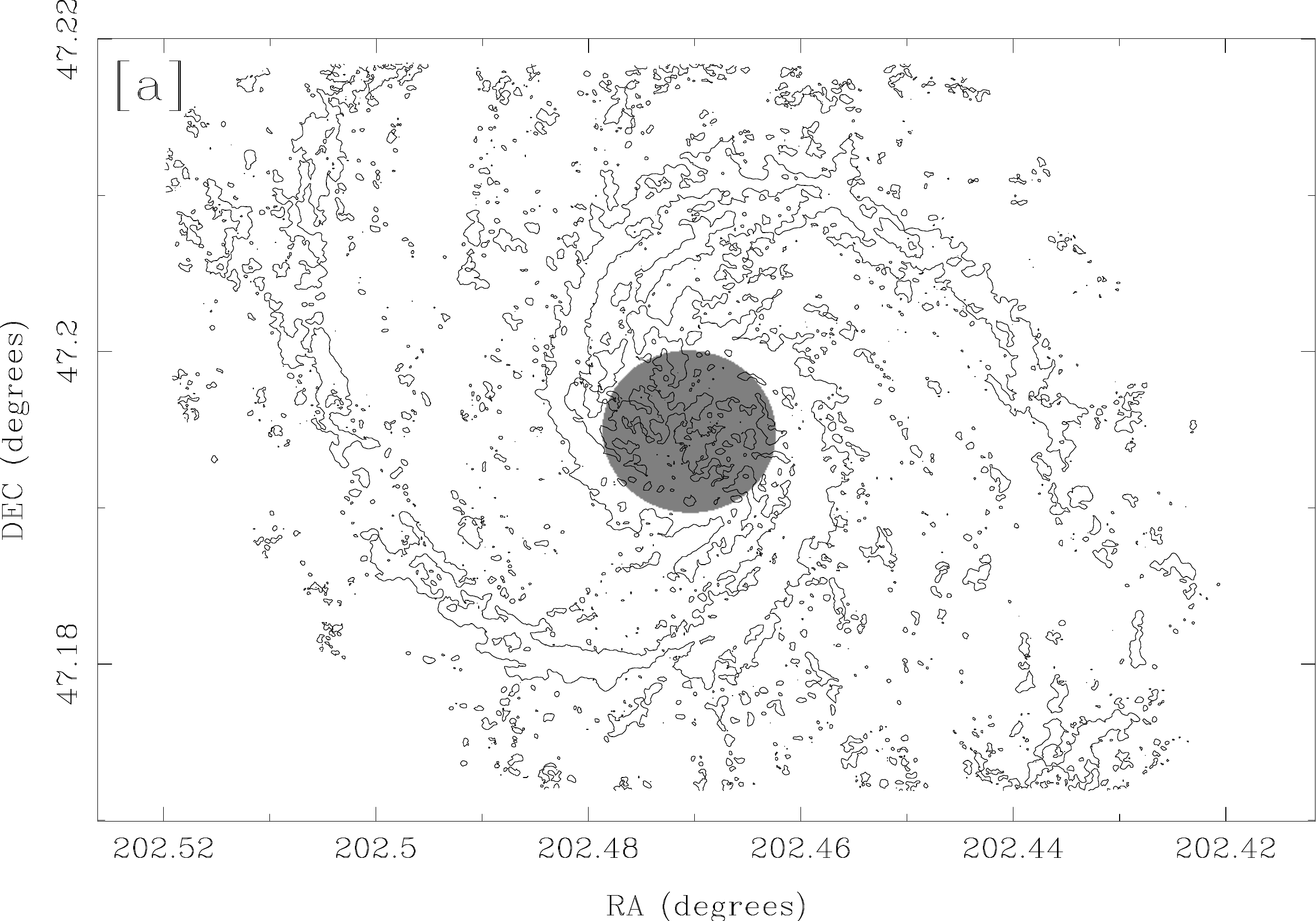}
\includegraphics[width=50mm,angle=270]{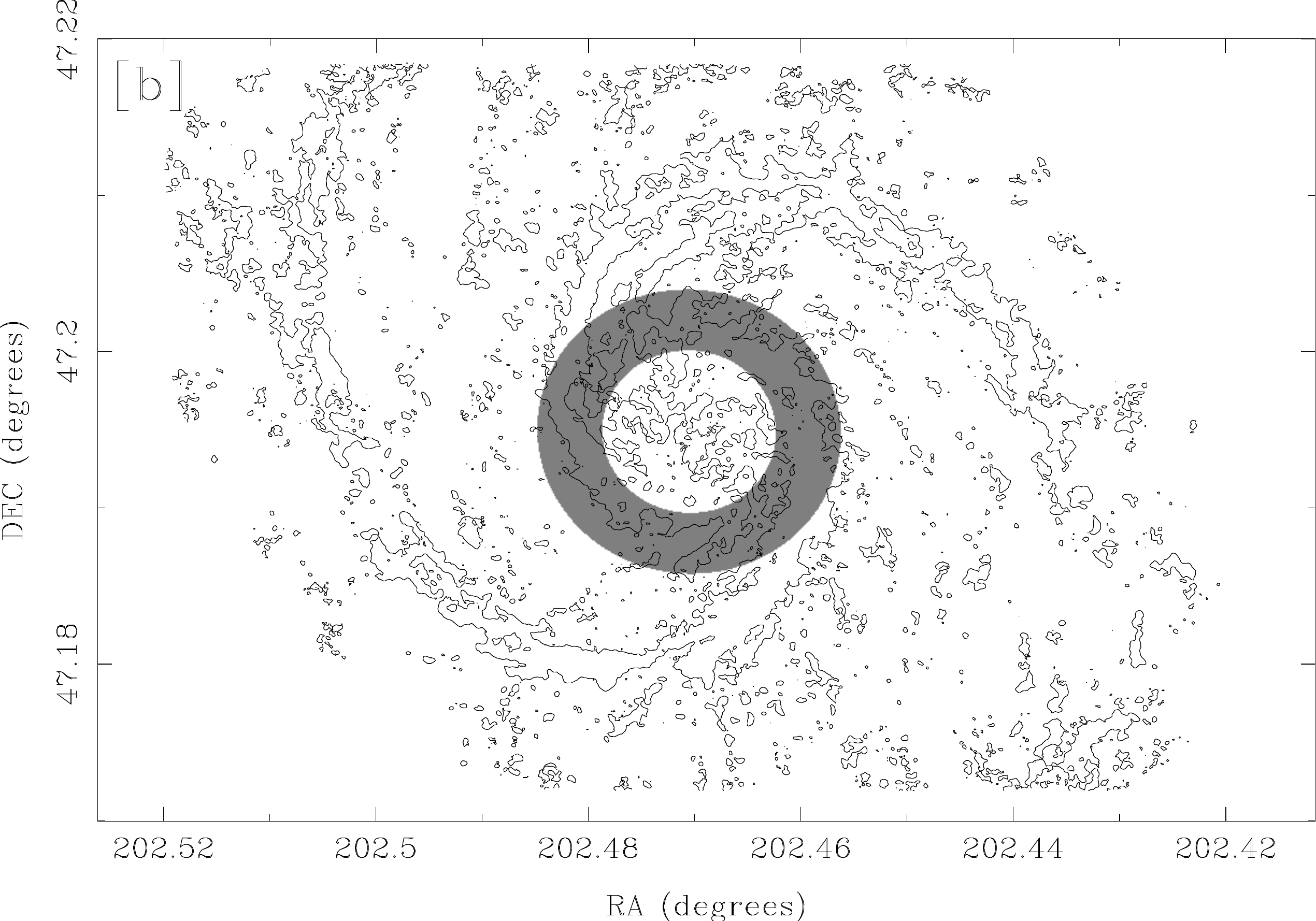}
\includegraphics[width=50mm,angle=270]{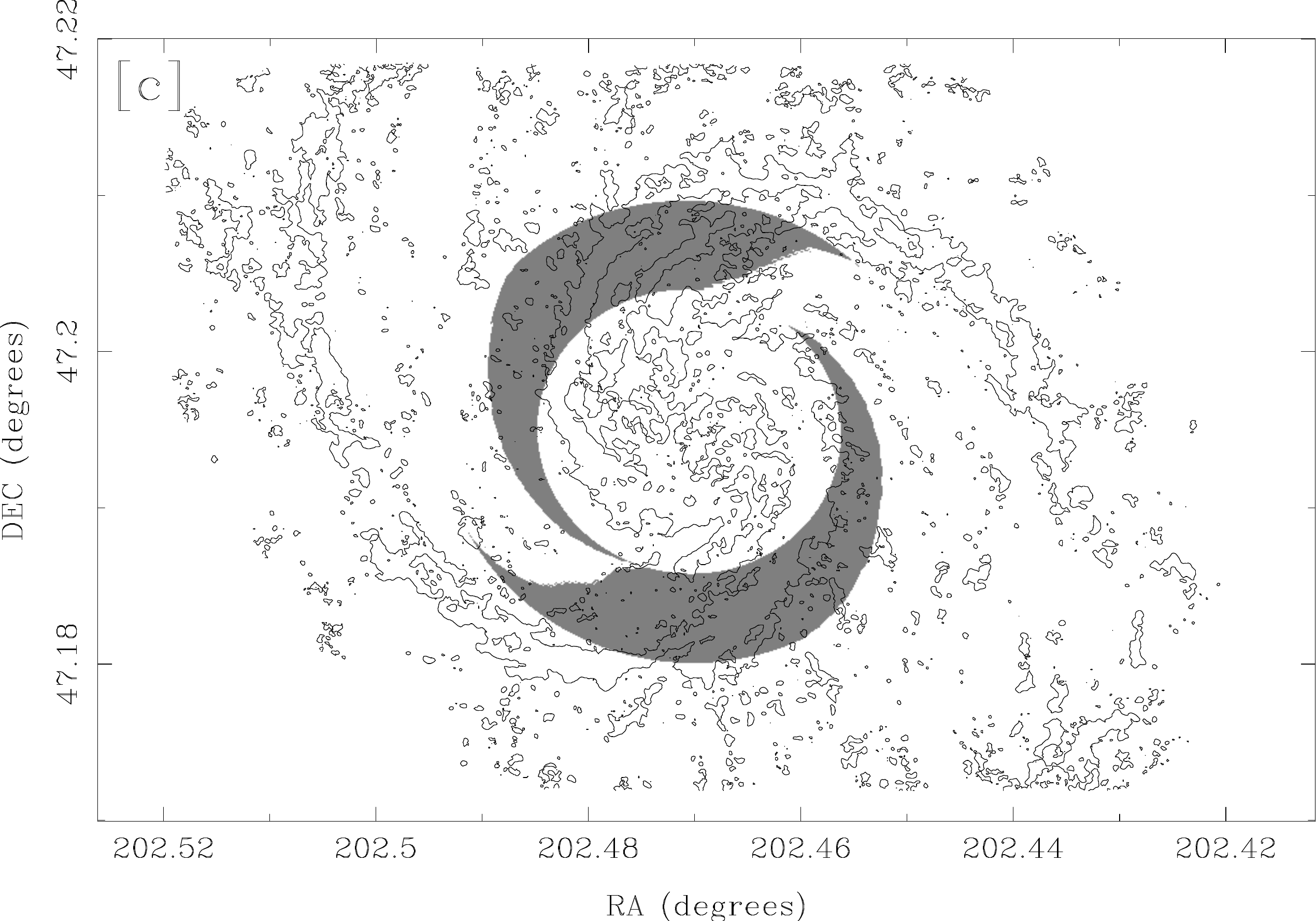}
\includegraphics[width=50mm,angle=270]{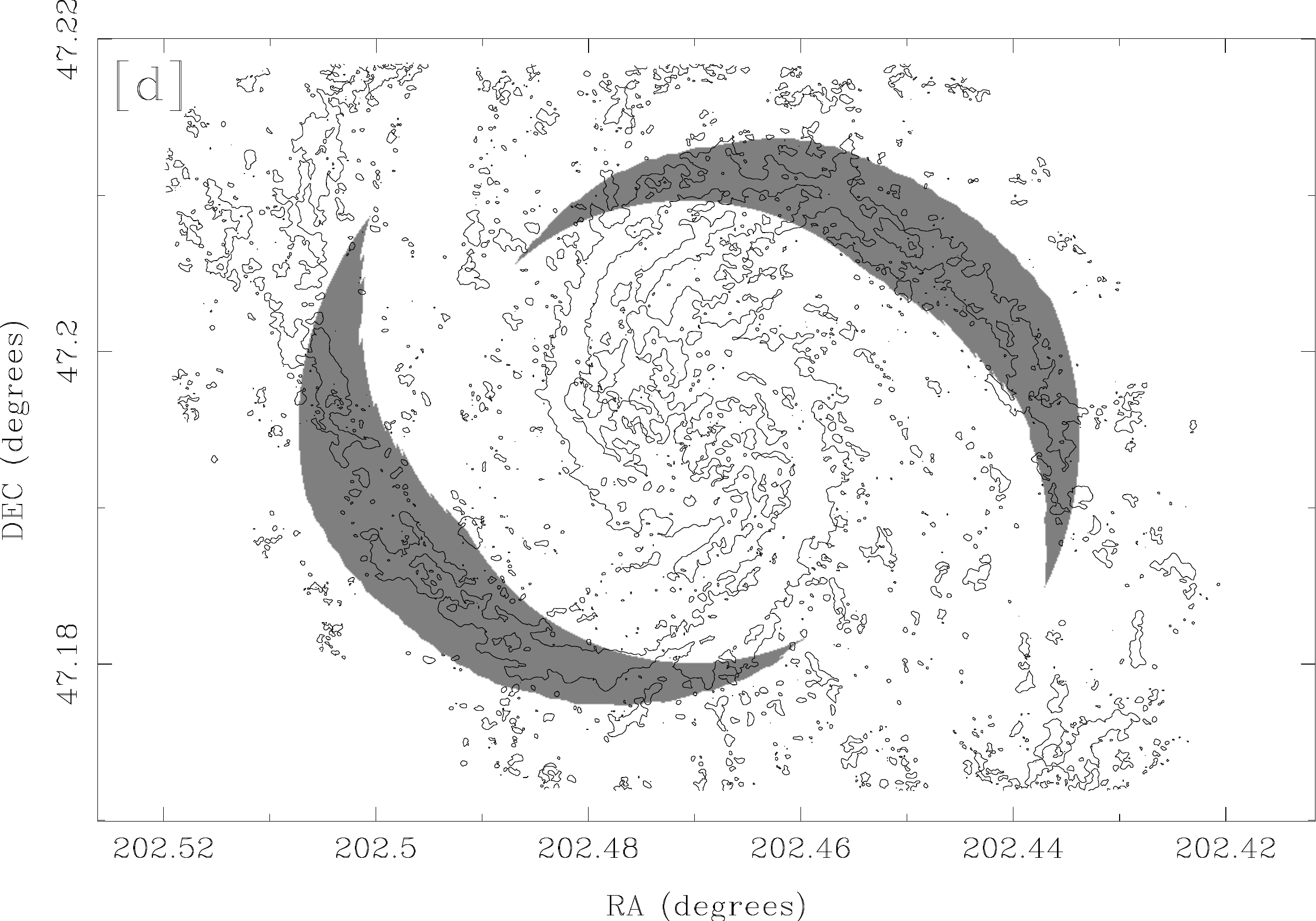}
\includegraphics[width=50mm,angle=270]{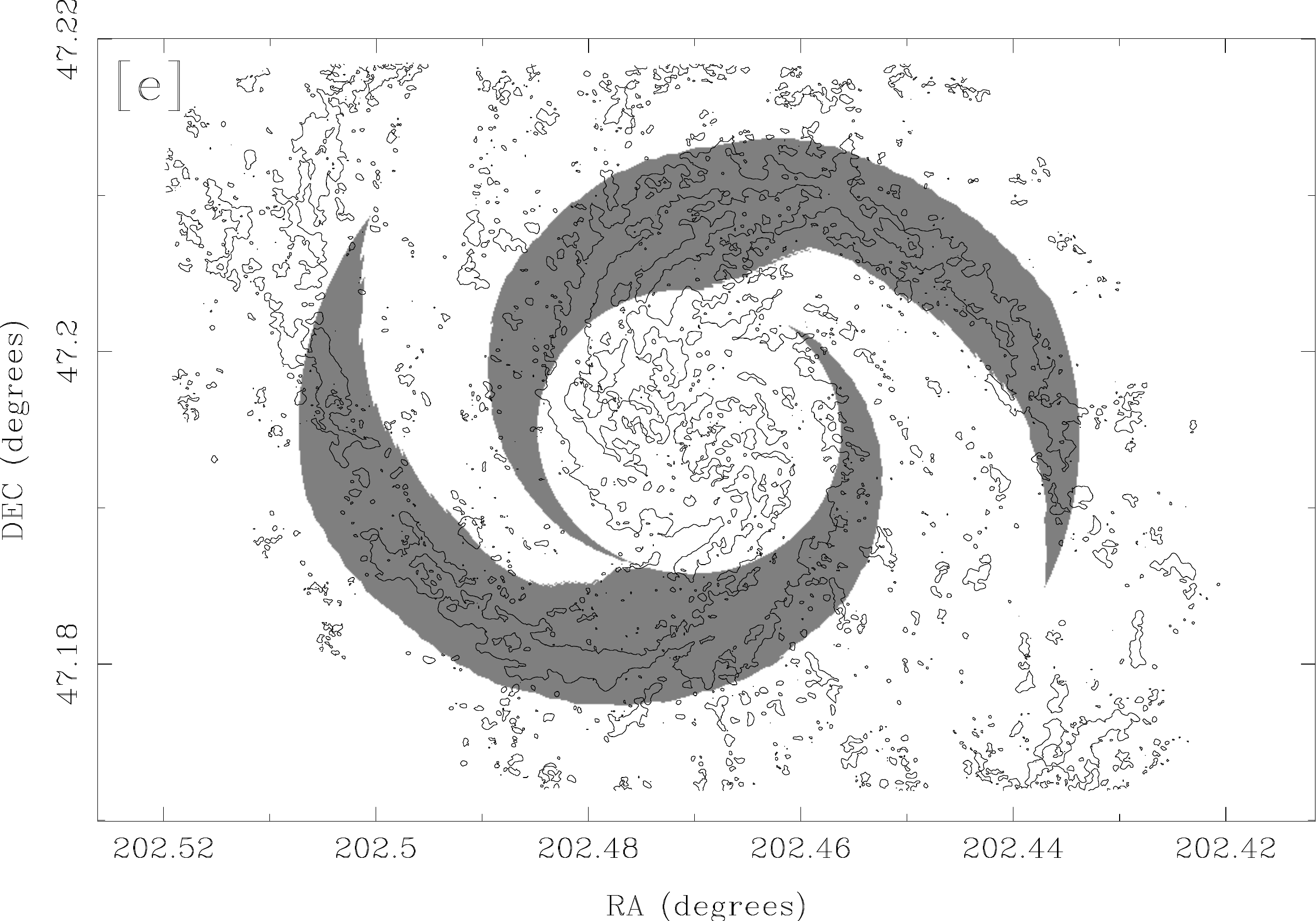}
\includegraphics[width=50mm,angle=270]{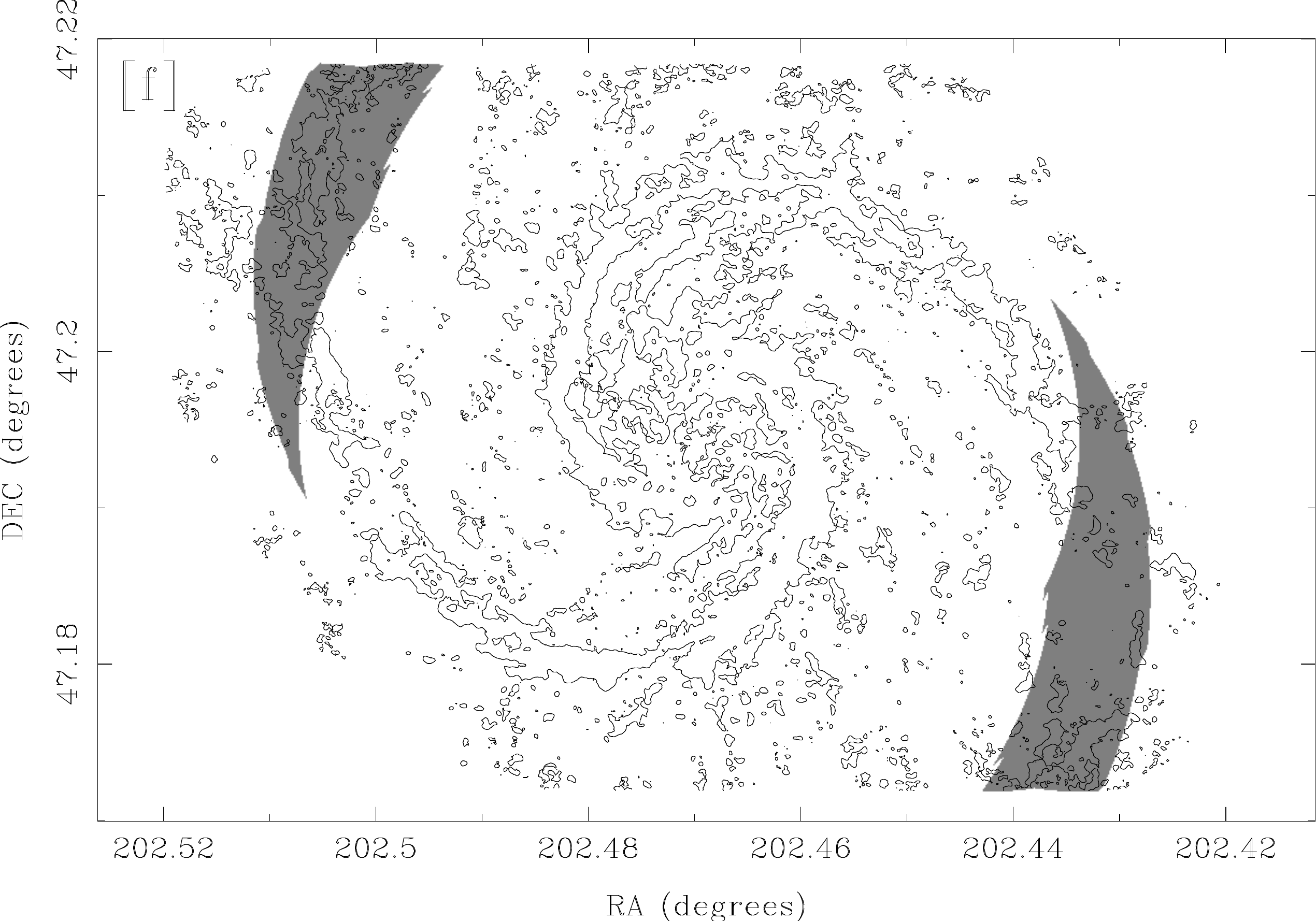}
\includegraphics[width=50mm,angle=270]{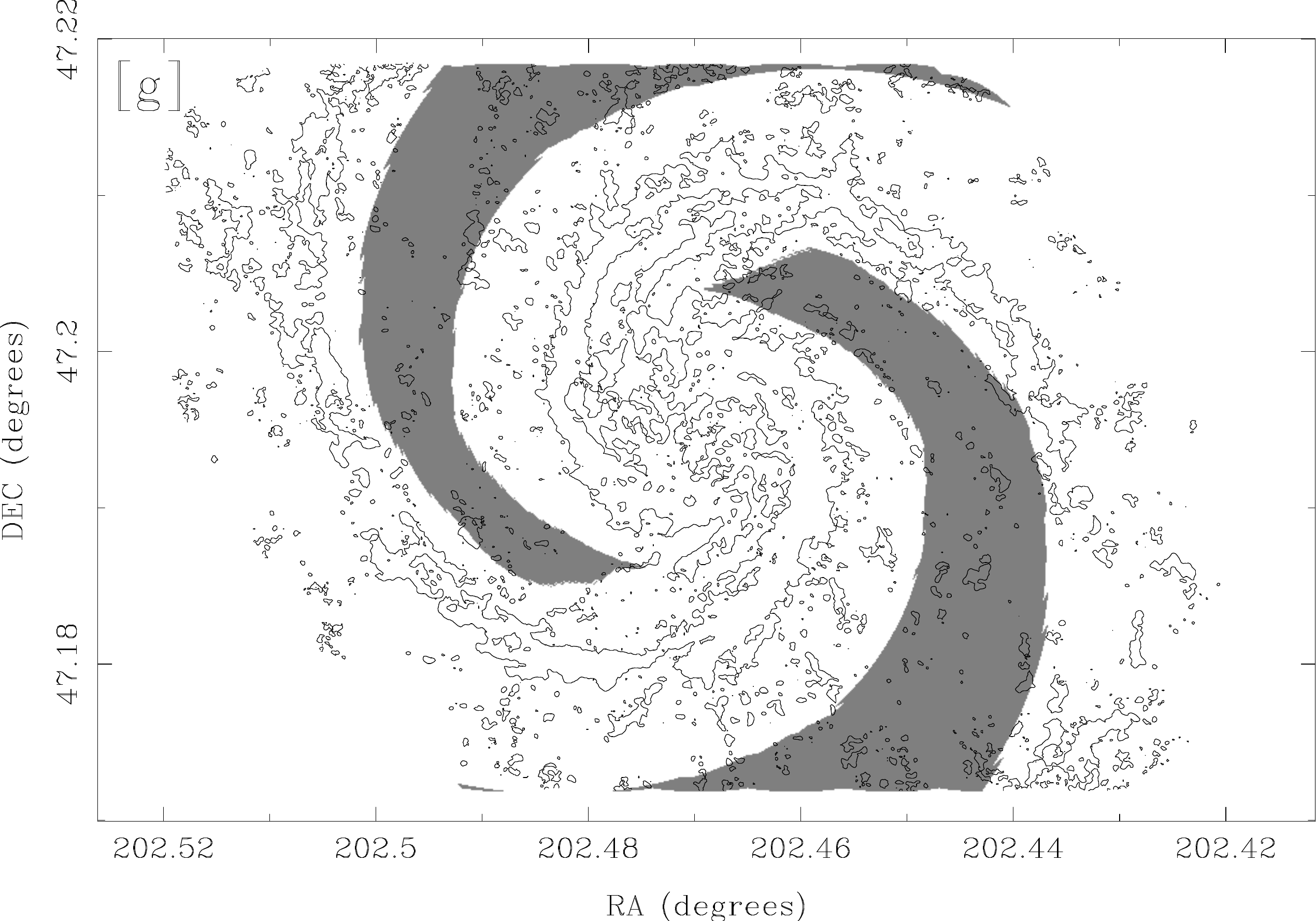}
\includegraphics[width=50mm,angle=270]{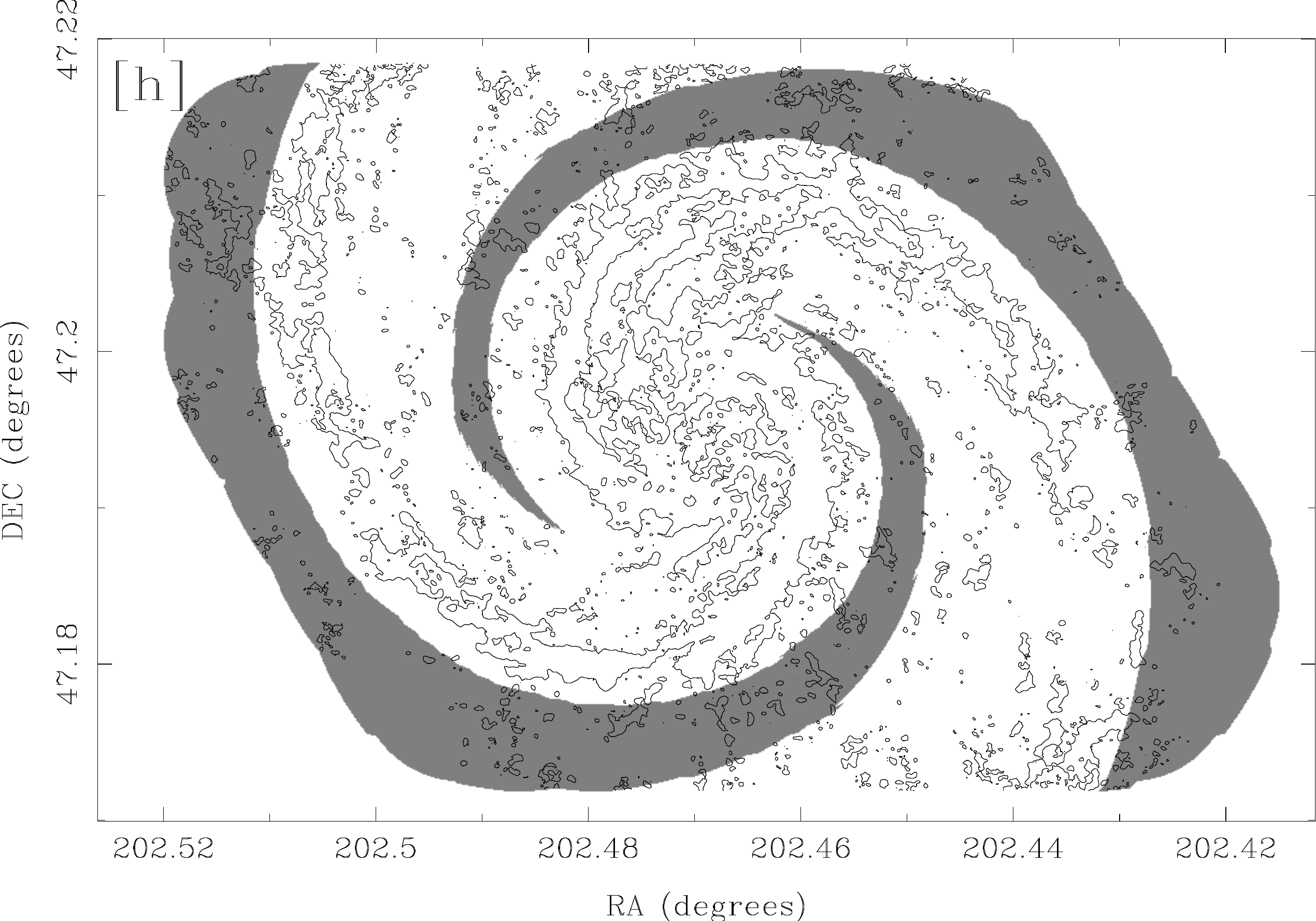}
 \caption{\small The different environments within the PAWS field that
   we analyse in this paper, indicated using grey shading in each
   panel. (a) Nuclear bar (bar); (b) Molecular ring (ring); (c) First
   spiral pattern inside corotation (A1I); (d) First spiral pattern
   outside corotation (A1O); (e) First spiral pattern (density wave
   arm, A1); (f) Second spiral pattern (material arm, A2); (g)
   Interarm region upstream of the spiral arms (up); (h) Interarm
   region downstream of the spiral arms (down). The black contours in
   all panels indicate $\ico = 25$\,\kkms, as measured by PAWS. }
\label{fig:enviropanels}
\end{center}
\end{figure*}

\noindent Here, we analyse seven regions within the PAWS field where
we expect the molecular gas to experience different dynamical effects
(see Figure~\ref{fig:enviropanels}). We define the different spiral
arm regions according to the direction of gas flows driven in response
to the underlying gravitational potential, which we derive from a map
of M51's stellar mass distribution \citep{meidtetal12}. The width of
the spiral arms are defined with respect to observed gas
kinematics. We determine the zone of enhanced spiral streaming
centered around the arm by measuring the (rotational) auto-correlation
of azimuthal streaming velocities in the PAWS field (Colombo et al.,
submitted). We construct azimuthal profiles of the auto correlation
signal in a series of radial bins and take the width of the signal at
95\% maximum as our measure of the kinematic arm width. The average
kinematic width from along the two arms is centered on the spiral arm
ridge line, defined by eye using the PAWS map of CO peak
brightness. Both the location of the ridge and the width are assumed
to be symmetric. The interarm region is divided into upstream and
downstream by the midpoint of the spiral arm ridge lines. The
definition of the spiral arm regions is based on the identification of
distinct spiral patterns within the galactic disk
\citep[cf.][]{vogeletal93,shettyetal07,meidtetal08,dobbsetal10} that
we refined and describe in detail elsewhere (Meidt et al., submitted;
Colombo et al., submitted).\\

\noindent The seven zones that we use to conduct our analysis are:
\begin{itemize}
\item{{\it Nuclear bar:} the region at galactocentric radii
  $R<23$\arcsec. The boundary is defined by the bar corotation
  resonance, inside of which the bar exerts negative torques and
  drives gas radially inwards.}
\item{{\it Molecular ring:} the region $23 < R < 35$\arcsec. Here the
  gas is influenced by both the bar and innermost spiral arms. Outside
  the bar corotation resonance, gas is driven radially outwards, while
  the spiral drives gas radially inwards inside its own
  corotation. These opposing torques accumulate gas in a ring-like
  structure. The region hosts some of the most active high-mass star
  formation in M51.}
\item{{\it Inner density-wave spiral arm:} the arm region $35 < R <
  55$\arcsec. The inner boundary is defined by the molecular ring, and
  the outer boundary is the corotation radius of the density-wave
  spiral arms. Within this zone, gas is driven radially inward by
  negative spiral arm torquing. Despite the high gas surface densities
  in this region, there is little star formation -- as traced by
  \ha\ and 24\,\m\ emission -- that is directly associated with the
  brightest CO emission (Schinnerer et al., in preparation).}
\item{{\it Outer density-wave spiral arm:} the arm region $55 < R <
  85$\arcsec. This region extends from the density-wave corotation
  resonance to the start of the material spiral. Within this zone, gas
  is driven radially outward by positive spiral arm torquing.}
\item{{\it Material spiral arm:} the arm region $R >
  85$\arcsec. This region extends from the boundary of positive arm
  torques associated with the density wave spiral to the edge of the
  PAWS field. There is some indication that gas flows radially inwards
  in this zone.}
\item{{\it Interarm region, downstream of the spiral arms.}}
\item{{\it Interarm region, upstream of the spiral arms.}}
\end{itemize}

\noindent Finally, we note that the projected area of the seven
regions is still quite large (between $\sim2$ and 17\,kpc$^{2}$) and
each contains a statistically significant number of GMCs
($\gtrsim100$). The PDFs of CO emission in these regions are therefore
more comparable to the PDFs of simulated galactic disks than the PDFs
of individual clouds. \\

\begin{figure*}
\begin{center}
\hspace{-0.5cm}
\includegraphics[width=160mm,angle=0]{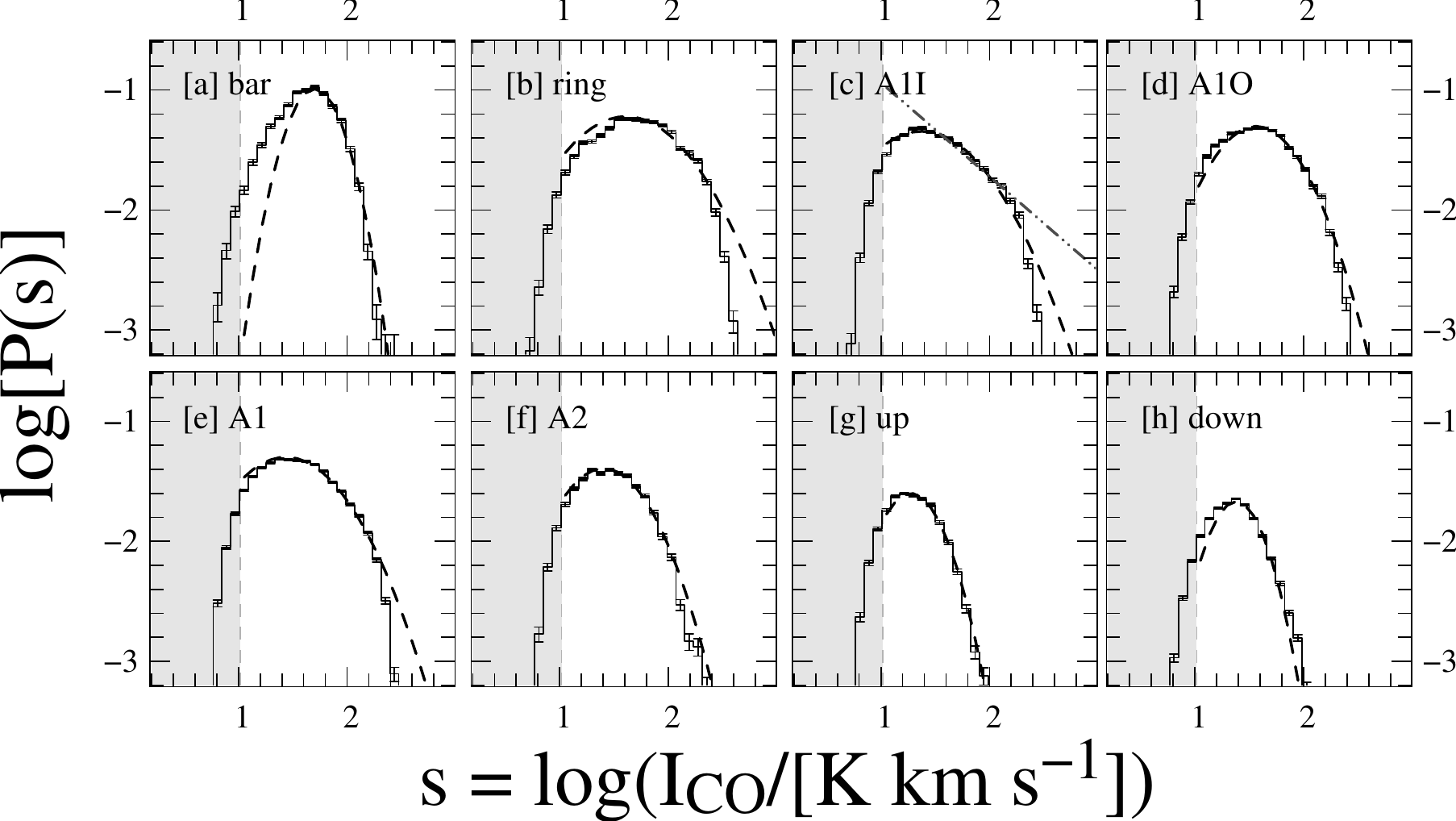}
\caption{\small The \ico\ PDFs for different regions within the PAWS
  field. The grey shaded region represents values beneath our nominal
  $3\sigma_{RMS}$ sensitivity limit of 10.5\,\kkms. Where a LN
  (power-law) function provides an good description of the PDF, it is
  indicated by a dashed (dot-dot-dashed) line. The vertical error bars
  represent the uncertainty associated with simple counting
  ($\sqrt{N}$) errors. }
\label{fig:pdf_ico_enviros}
\end{center}
\end{figure*}

\noindent Figure~\ref{fig:pdf_ico_enviros} shows that the \ico\ PDFs
for different M51 environments exhibit diverse shapes. The panels of
Figure~\ref{fig:pdf_ico_enviros} are ordered such that the PDF
amplitudes decrease from top left to bottom right, which reflects the
fact that CO emission is more prevalent in the arms and central region
of M51 than in the interarm region. The PDFs also tend to decrease in
width, indicating that the fraction of pixels with bright CO emission
declines with the overall frequency of CO detections. The PDFs of the
spiral arms are notably wider than for the interarm environments; it
is also evident that the PDF corresponding to the first spiral pattern
(A1, i.e. the density wave spiral arm) is wider than the PDF for the
second spiral (A2, the material arm). Since \ico\ is the integral of
the CO brightness over the line profile, this variation in the PDF
width would seem consistent with the results of previous studies
\citep[e.g.][]{garciaburilloetal93,kunonakai97,aaltoetal99,schusteretal07}
that find that the average CO integrated intensity and CO linewidth
decreases with increasing distance along the arms, and from the arm to
the interarm region. \\

\noindent The differences in the width of the PDF between M51
environments are reflected in the $IDI$ values that we derive, which
become more positive as the number of pixels with $\ico >
60$\,\kkms\ increases (see Table~\ref{tbl:bdi}). The development of
more high brightness emission appears to be accompanied by a change in
the PDF shape: a LN function is a better description of the PDFs in
the interarm region than in the spiral arm and ring regions. The
\ico\ PDFs for the first spiral pattern, especially inside corotation
(A1I), appear more like broken or truncated power-laws than LN
functions. The PDFs in the center of M51 also diverge from a LN shape:
the \ico\ distribution in the molecular ring is essentially flat
between $\sim20$ and 150\,\kkms, while the PDF of the bar is the only
region with an unambiguous decline at low intensities ($\ico \lesssim
50$\,\kkms). Several PDFs appear truncated near $\ico \sim
300$\,\kkms: this is seen most clearly in the molecular ring, but the
distributions in the first spiral arm regions also decline steeply for
$\ico \gtrsim 300$\,\kkms.\\

\begin{figure*}
\begin{center}
\hspace{-0.5cm}
\includegraphics[width=160mm,angle=0]{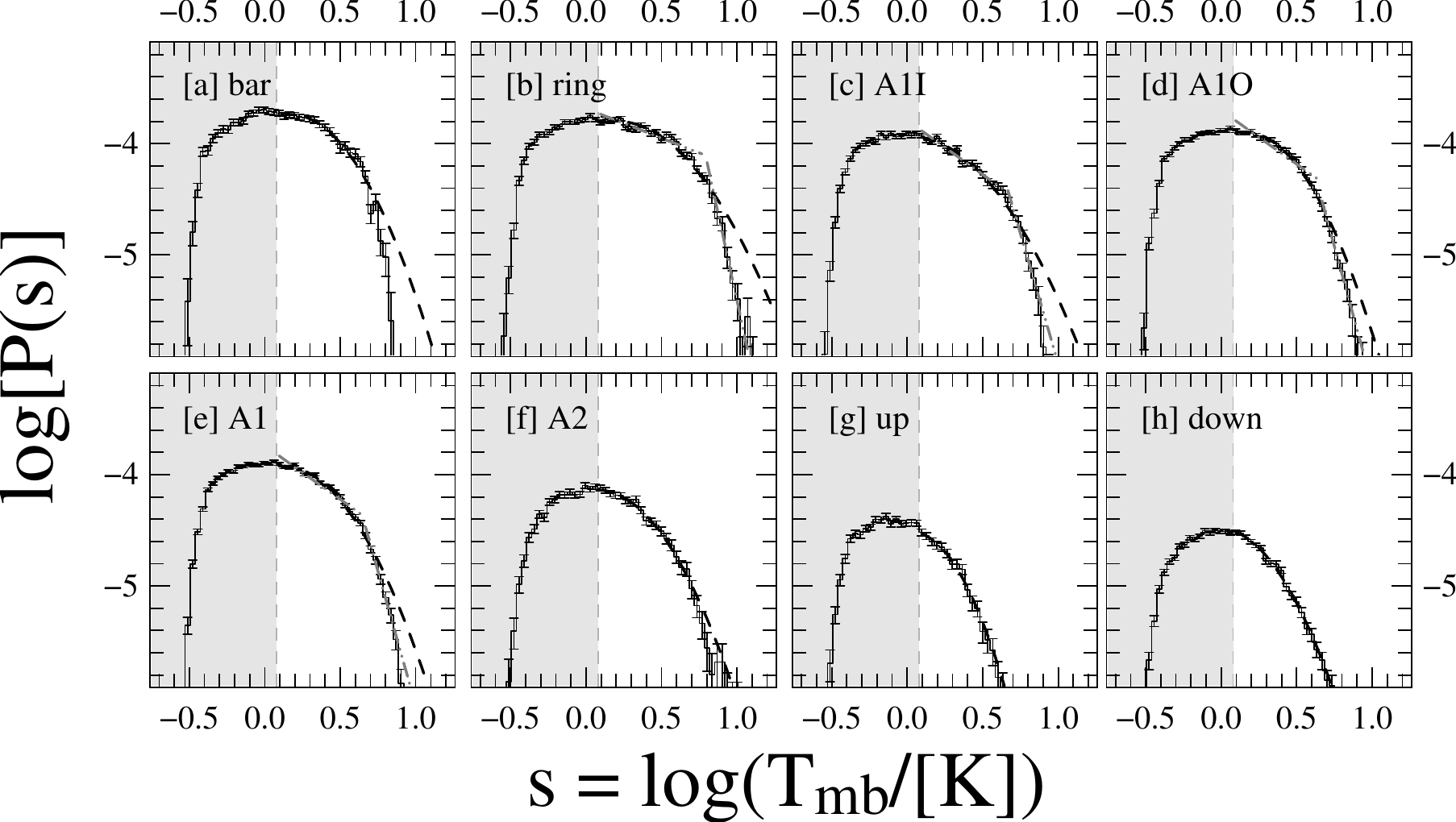}
\caption{\small The PDFs of CO brightness for different regions within
  the PAWS field. In each panel, the grey shaded region represents
  values beneath our nominal $3\sigma$ sensitivity limit of
  1.2\,K. When appropriate, the best-fitting LN (power-law) function
  is indicated by a dashed (dot-dot-dashed) line. The vertical error bars
  represent the uncertainty associated with simple counting
  ($\sqrt{N}$) errors.}
\label{fig:pdf_cube_enviros}
\end{center}
\end{figure*}

\noindent In Figure~\ref{fig:pdf_cube_enviros}, we present the PDFs of
CO brightness for the different M51 environments. The distributions
are more uniform than those of integrated intensity, but variations
similar to those identified for the \ico\ PDFs are still evident. The
PDF amplitude tends to decrease from panel [a] to [h], and only the
interarm regions and second spiral arm (A2) yield PDFs that are
approximately LN across the observed range of $T_{\rm mb}$ values (see
Table~\ref{tbl:pdffits}). As noted for the \ico\ PDFs, regions with a
relatively wide $T_{\rm mb}$ PDF (e.g. the ring, bar and spiral arms)
have more positive $BDI$s and also tend to diverge from a LN shape, in
this case developing a pronounced change of slope near $T_{\rm mb}
\gtrsim 6$\,K. This effect is most clearly seen for the PDFs of the
molecular ring (panel [b]) and the first spiral pattern inside
corotation (panel [c]), but generally it appears that an increase in
the fraction of high brightness CO emission is associated with a PDF
that more resembles a truncated power law than a LN function. We
discuss this result in relation to similar trends observed for PDFs of
CO emission in the Galaxy in Section~\ref{sect:prev_obs}.\\

\begin{table*}
\centering
\caption{\small Brightness and Integrated Intensity Distribution Index for M51 Environments}
\label{tbl:bdi}
\par \addvspace{0.2cm}
\begin{threeparttable}
{\small
\begin{tabular}{@{}lccc}
\hline 
Region  & $L_{\rm CO}$ & $BDI$ & $IDI$ \\
        & [$10^{7}$\,\lcou]    &       &       \\
\hline
Global                          & 70.4 &  -0.66 & 0.40  \\
Nuclear Bar                     & 6.6  &  -0.85 & 0.76  \\
Molecular Ring                  & 16.6 &  -0.13 & 1.08  \\
Arm 1 inside corotation (A1I)   & 11.9 &  -0.59 & 0.52  \\
Arm 1 outside corotation (A1O)  & 17.6 &  -0.83 & 0.55  \\
Arm 1 (A1)                      & 29.6 &  -0.72 & 0.54  \\
Arm 2 (A2)                      & 6.7  &  -0.76 & 0.12  \\
Upstream                        & 4.7  &  -1.73 & -0.95 \\
Downstream                      & 6.2  &  -1.50 & -0.65 \\
\hline
\end{tabular}
}
{\small 
\begin{tablenotes}
\item[]{The total CO luminosity (column 2), brightness distribution
  index ($BDI$, column 3) and integrated intensity distribution index
  ($IDI$, column 4) for the different M51 environments (see
  Figure~\ref{fig:enviropanels}). The $BDI$ and $IDI$ values are
  calculated according to Equations~\ref{eqn:bdi} and~\ref{eqn:idi}
  respectively. More positive $BDI$ and $IDI$ values indicate PDFs
  that have a larger fraction of pixels at high CO intensities.}
\end{tablenotes}}
\end{threeparttable}
\end{table*}

%%%%%%%%%%%%%%%%%%%%%%%%%%%%%%
\subsection{Comparison between M51, M33 and the LMC}
%%%%%%%%%%%%%%%%%%%%%%%%%%%%%%
\label{sect:pdfs_allgals}

\noindent Finally, we can compare the PDFs of CO integrated intensity
and CO brightness for the inner disk of M51 to the corresponding PDFs
for other nearby galaxies. As we discuss in
Appendix~\ref{app:pdfs_resn_sens}, the shape of the PDF is sensitive
to the resolution and sensitivity of the data. Prior to constructing
the PDFs, we therefore degraded the M51 and LMC data cubes to the same
spatial resolution as the M33 cube ($\sim53$\,pc), and folded the M33
and LMC data cubes along the velocity axis to the same channel width
as the M51 cube ($5\,\kms$). We interpolated all the cubes onto an
($x,y$) grid with the same pixel dimensions in physical space ($15
\times 15$\,pc). Significant emission was identified according to the
method outlined in Section~\ref{sect:analysis}. The resulting masks
were applied to the original data cubes, and the integrated intensity
images were constructed by summing unblanked pixels across the full
velocity bandwidth of each survey.\\

\begin{figure*}
\begin{center}
\hspace{-0.5cm}
\includegraphics[width=130mm,angle=0]{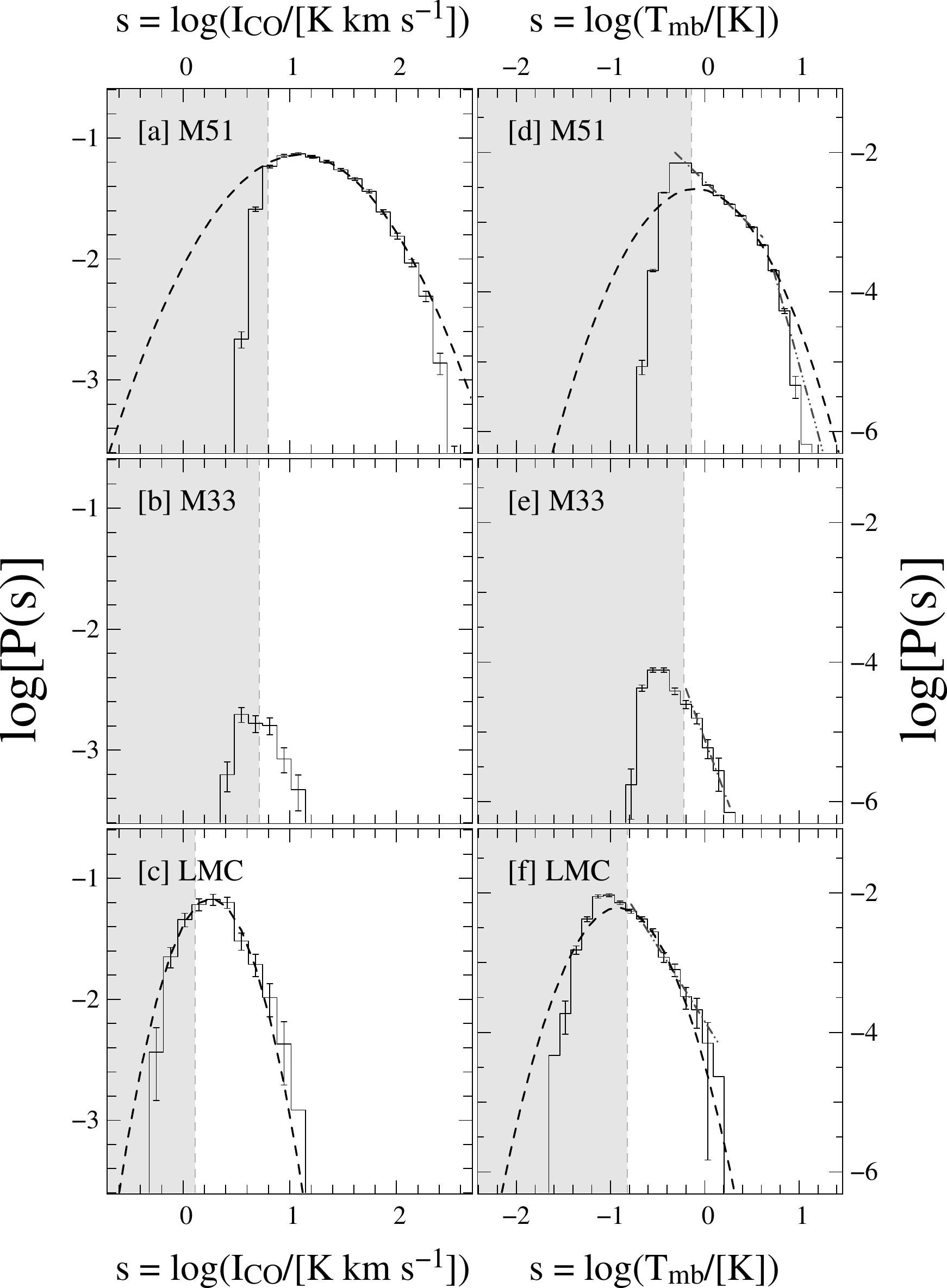}
\caption{\small PDFs of \ico\ (panels [a] to [c]) and $T_{\rm mb}$
  (panels [d] to [f]) for M51, the LMC and M33, constructed using a
  dilated mask technique. The data sets have been smoothed to the same
  spatial scale, and interpolated onto an $(x,y,v)$ grid with the same
  spatial and spectral dimensions. In all panels, the grey shaded
  region corresponds to pixels beneath the 3$\sigma_{RMS}$ sensitivity
  limits of the individual data cubes. The vertical error bars
  represent the uncertainty associated with simple counting
  ($\sqrt{N}$) errors. Where relevant, the best-fitting LN
  (power-law) function is indicated by a dashed (dot-dot-dashed) line.}
\label{fig:pdf_3gals}
\end{center}
\end{figure*}

\noindent The PDFs obtained from the \ico\ maps of M33, the LMC and
M51 are shown in the left panels of Figure~\ref{fig:pdf_3gals}. The
shape of the \ico\ PDF for M33 is highly uncertain due to the modest
sensitivity of the BIMA+FCRAO data cube. Never the less, it is obvious
that CO emission in the LMC and M33 are alike in the sense that the
maximum observed \ico\ intensities are $\sim10$\,\kkms\ at the
resolution of our analysis, and not a few times 100\,\kkms, as
observed for M51.  The \ico\ PDF for the LMC appears to be
well-represented by a narrow LN function with mean $\langle \ico
\rangle = 2$\,\kkms, and a logarithmic width $x = 0.3$. Since MAGMA is
a targeted rather than spatially complete survey of the LMC disk (see
Section~\ref{sect:data}), the amplitude of its PDF is biased high
compared to that of M51 and M33; normalizing by the full projected
area of the LMC's \hi\ disk, rather than the MAGMA field-of-view would
reduce the amplitude by more than an order of magnitude. The MAGMA
survey strategy of targeting the brightest clouds in the NANTEN
catalog also means that the shape of the PDF is biased towards high CO
intensities; extending MAGMA to fainter clouds would recover a greater
fraction of pixels with low CO brightness and narrow the PDFs in
Figure~\ref{fig:pdf_3gals}[c] and~[f]. Relative to observed \ico\ PDF
of the LMC, the M51 \ico\ distribution peaks at higher CO intensity,
$\langle \ico \rangle \sim 12$\,\kkms, and is also wider by a factor
of $\sim2$ in the logarithm (see Table~\ref{tbl:pdffits}). The
difference between the best-fitting LN function derived for the M51
distribution in Figure~\ref{fig:pdf_3gals}[a] and that in
Figure~\ref{fig:pdf_paws}[a] is consistent with the differences that
we observe for different masking techniques for identifying
significant emission within the data cube (see
Appendix~\ref{app:pdfs_methods}). \\

\noindent The PDFs of CO brightness for the three galaxies are
presented in the right panels of Figure~\ref{fig:pdf_3gals}. For M51,
we fit the distribution of CO brightness with a LN function with mean
$\langle T_{\rm mb} \rangle = 0.8$\,K, logarithmic width $x=0.4$. This
is consistent with the best-fitting LN function derived for the PAWS
data in Figure~\ref{fig:pdf_paws}[b], which used a more sophisticated
masking technique to identify significant emission. Alternatively, a
broken power-law with a shallow slope of $\sim-1.3$ between $T_{\rm
  mb} = 1$ and 4\,K, and a much steeper slope ($\sim-7.1$) above 4\,K
also fits the $T_{\rm mb}$ distribution for M51 reasonably well. For
the LMC, the best-fitting LN function has mean $\langle T_{\rm mb}
\rangle = 0.1$\,K and logarithmic width $x=0.3$; there is no sign of a
truncation. A simple power law with a slope of $\sim-2.2$ also
adequately represents the distribution. There are insufficient pixels
with significant emission in the M33 data cube to attempt to fit the
$T_{\rm mb}$ PDF with a LN function. A simple power-law with a slope
of $\sim-3.6$ provides a reasonable fit to the distribution for pixel
values $T_{\rm mb} \gtrsim 0.3$\,K. We note that the slopes of the
power-laws that fit the $T_{\rm mb}$ PDFs reproduce the trends
observed for the giant molecular cloud (GMC) mass distribution in the
three galaxies, i.e. a shallow slope for M51 (below $T_{\rm mb} \sim
4$\,K) and a much steeper brightness distribution for the low-mass
galaxies (Hughes et al., in preparation). We discuss the connection
between the shape of the CO PDFs and the GMC mass distribution further
in Section~\ref{sect:pdfs_vs_gmcs}.\\

%%%%%%%%%%%%%%%%%%%%%%%%%%%%%%
\section{Discussion}
%%%%%%%%%%%%%%%%%%%%%%%%%%%%%%
\label{sect:discussion}

%%%%%%%%%%%%%%%%%%%%%%%%%%%
\subsection{Comparison to Previous Observations}
\label{sect:prev_obs}
%%%%%%%%%%%%%%%%%%%%%%%%%%%

\noindent Both high resolution and wide-field coverage are necessary
to characterise the CO emission in galaxies on spatial scales that are
relevant for star formation, hence few extragalactic studies have
produced PDFs of CO brightness and/or integrated intensity that
represent a significant fraction of a galactic disk. One exception is
an analysis of the LMC by \citet{wongetal11}, which found that on
$\sim10$\,pc scales the \ico\ PDF was roughly consistent with a narrow
lognormal function ($\langle \Sigma_{H_{2}} \rangle = 16$\,\mpcsq, $x
\sim 0.3$\,dex) at high column densities. The authors noted some
evidence for a truncation around $\Sigma_{\rm H_{2}} = 200$\,\mpcsq,
which they tentatively attributed to opacity effects in the
\aco\ line. Notably, however, the MAGMA data show no evidence for a
power-law excess at high column densities, as has been observed on
parsec scales within star-forming Galactic clouds
\citep[e.g.][]{kainulainenetal09}, suggesting that the structure of
LMC molecular clouds is still dominated by turbulence on $\sim10$\,pc
scales.\\

\noindent The CO emission in M51 itself has been analysed by numerous
authors
\citep[e.g.][]{vogeletal88,rand93,aaltoetal99,helferetal03,shettyetal07}. With
the exception of the recent survey by \citet{kodaetal09}, however,
observations with high spatial resolution have mostly focussed on a
spiral arm segment \citep[e.g.][]{schinnereretal10,egusaetal11}, while
studies covering a significant fraction of the disk
\citep[e.g.][]{garciaburilloetal93,schusteretal07} have had a
resolution of a few hundred parsecs or greater, i.e. insufficient
resolution to resolve individual GMCs. Rather than PDFs, these
lower-resolution studies have typically examined radial trends in
properties such as the gas velocity dispersion, Toomre's Q parameter
and the molecular gas depletion time $\tau_{\rm H_{2}}$
\citep[e.g.][]{schusteretal07,hitschfeldetal09}. From the PDFs in
Section~\ref{sect:pdfs_enviros}, we would expect to observe a radial
decline in the average value of $\Sigma_{\rm H_{2}}$ constructed from
azimuthal averages, since the fraction of bright CO emission (as
parameterised by the $IDI$ values) decreases along the spiral arms and
also because the interarm region occupies an increasing fraction of
the disk area with increasing galactocentric radius. Since the mass
surface density is an important input for the determination of
Toomre's Q and $\tau_{\rm H_{2}}$, radial trends in these quantities
may likewise reflect a combination of differences between the arm and
interarm zones and variations along the spiral arms (an interpretation
that would seem to be supported by the map of Toomre's Q presented in
figure~15 of \citet{hitschfeldetal09} for example). \\

\noindent More generally, we note that differences in the basic
properties of the CO emission (i.e. peak brightness, velocity
dispersion) between the arm and interarm regions of M51 have been
reported by several previous studies, and often interpreted as
evidence for changes in the physical state of the molecular gas as it
passes through the spiral arms. While the precise identification of
M51's arm and interarm zones varies (usually because M51's gaseous
spiral arms appear wider at lower spatial resolution), CO emission in
the interarm has been shown to have lower velocity dispersion
\citep[e.g.][]{garciaburilloetal93,aaltoetal99,hitschfeldetal09},
lower peak brightness
\citep[e.g.][]{garciaburilloetal93,tosakietal02}, lower star formation
efficiency \citep[as inferred from the ratio of \ha\ to \aco\ emission,
  e.g.][]{rand93,tosakietal02}, higher \aco/\cco\ isotopic ratios
\citep[e.g.][]{tosakietal02} and lower \bco/\aco\ transitional ratios
\citep[e.g.][]{kodaetal12} than emission in the spiral arms. Most of
these results suggest that molecular gas in the interarm region has a
lower characteristic density than gas within the spiral arms. \\

\noindent Spatial resolution is rarely a limitation for studies of
molecular gas in the Milky Way, although previous analyses of Galactic
CO emission have tended to focus on the physical properties of GMCs
\citep[e.g.][]{solomonetal87,romanduvaletal10}, which were quickly
recognized to be the preferred -- perhaps only -- site of high-mass
star formation in the Galaxy. Recent work has emphasized, however,
that faint spatially extended CO emission contributes significantly to
a region's total CO flux
\citep[e.g.][]{goldsmithetal08,heyeretal09,lisztpetylucas10}. Very
recently, \citet{sawadaetal12} have presented PDFs of \aco\ and
\cco\ brightness for an $0\fdg8 \times 0\fdg8$ field towards the
Galactic plane at $l\approx 38\D$, observed using the Nobeyama Radio
Observatory (NRO) 45\,m telescope. These authors find clear
differences between the PDFs constructed from the emission at radial
velocities corresponding to the Sagittarius arm and those
corresponding to the interarm regions, showing that the structural
proprties of the molecular gas vary in response to Galactic
structure. They conclude that compact, high brightness CO structures
develop downstream of the molecular spiral arms, where they are
spatially coincident with signatures of active star formation
(e.g. \hii\ regions). \citet{sawadaetal12} point out that their result
is a rediscovery of a conclusion that had already been drawn by
earlier studies. \citet{sandersetal85}, for example, observed a
connection between the location of Galactic molecular clouds in
longitude-velocity space and their peak brightness temperature:
``hot'' (i.e. high brightness) clouds were preferentially located in
the spiral arms traced by \hii\ regions
\citep{georgelingeorgelin76}. \citet{egusaetal11} present a
qualitatively similar scenario for a $\sim2$\,kpc segment of M51's
inner spiral arm, showing that both high mass ($\sim10^{6}$\,\msol) CO
clumps and \hii\ regions are preferentially located downstream of the
spiral arm ridge line (note, however, that the high brightness CO
structures described by \citet{sawadaetal12} occur on much smaller
spatial scales than the structures observed by \citet{egusaetal11} in
M51). \\

\noindent Although the spatial resolution of the PAWS data is
considerably worse than that of Galactic surveys, the PDFs in
Figure~\ref{fig:pdf_cube_enviros} show similar trends as those
reported by \citet{sawadaetal12}. As we noted in
Section~\ref{sect:pdfs_enviros}, the PDFs of the interarm region
resemble narrow lognormal functions, while the PDFs in the central and
spiral arms regions reach higher maximum intensities and tend to be
better represented by broken power-laws. These variations in shape are
reflected by the $BDI$ and $IDI$ values: low brightness emission
dominates the total flux in both arm and interarm environments, but
the relative contribution from bright emission increases in the spiral
arms. Bright emission is most dominant in the center of M51, where
$\sim25$\% of the total CO flux arises from pixels with $T_{\rm mb} >
4$\,K (by comparison, less than 5\% of the emission in the interarm is
brighter than 4\,K). Similar to \citet{sawadaetal12}, we find that the
$BDI$ and $IDI$ values are higher on the downstream side of the spiral
arms than on the upstream side. Tracers of high-mass star formation,
e.g. \ha, 24\,$\mu$m and far ultra-violet (FUV) emission, also appear
to be preferentially located downstream of arms (Schinnerer et al., in
preparation), again consistent with the Galactic results. We discuss
the connection between star formation and the shape of the CO PDFs for
different M51 environments in more detail in
Section~\ref{sect:pdfs_vs_sf}. \\

\noindent The fact that a similar relationship between CO emission
properties and spiral arm structure is observed in both M51 and the
Milky Way would seem to support the argument by \citet{sawadaetal12}
and \citet{sawadaetal12b} that the arm-interarm variations they
observe reflect genuine changes in the density distribution of the
Galactic CO-emitting gas. One important caveat, however, is that the
PAWS data has much lower spatial resolution ($\sim40$\,pc) than the
Galactic NRO data ($\lesssim1$\,pc). In particular, the lower CO
brightness temperatures that we observe in M51 ($T_{\rm mb} \lesssim
16$\,K) indicate that our PAWS measurements reflect a combination of
the average kinetic temperature and the filling factor of the
CO-emitting gas within a resolution element. Variations in CO
brightness on $\lesssim1$\,pc scales, by contrast, should mostly track
variations in gas temperature and/or density since beam dilution
should be minimal on these scales. Some of the variation between high
and low $BDI$ values in M51 will reflect changes in the covering
fraction of the CO emission for the different M51 environments, as
well as differences in the intrinsic brightness temperature of the
CO-emitting structures that are more directly comparable to the
variations described by \citet{sawadaetal12}.\\

%%%%%%%%%%%%%%%%%%%%%%%%%%%
\subsection{Comparison with GMC properties}
\label{sect:pdfs_vs_gmcs}
%%%%%%%%%%%%%%%%%%%%%%%%%%%

\noindent In Sections~\ref{sect:pdfs_enviros}, we described variations
in the characteristic shape of the CO PDFs for different M51
environments. To what extent are these differences manifested in
variations of the properties of GMCs within each environment, or of
the ensemble properties of a GMC population (e.g. its mass
distribution)?  Intuitively we would expect some connection between
GMCs and the presence of high brightness CO emission, since most
methods for identifying GMCs from CO data cubes invoke either a
brightness threshold or local maximum in the CO brightness
distribution in order to define cloud structure. The connection may be
rather indirect, however, since the fraction of CO emission above the
PAWS sensitivity limit that is associated with the
observationally-defined GMCs varies between 40 and 65\%, depending on
galactic environment (Colombo et al., submitted). \\

\begin{figure*}
\begin{center}
\hspace{-0.5cm}
\includegraphics[width=140mm,angle=0]{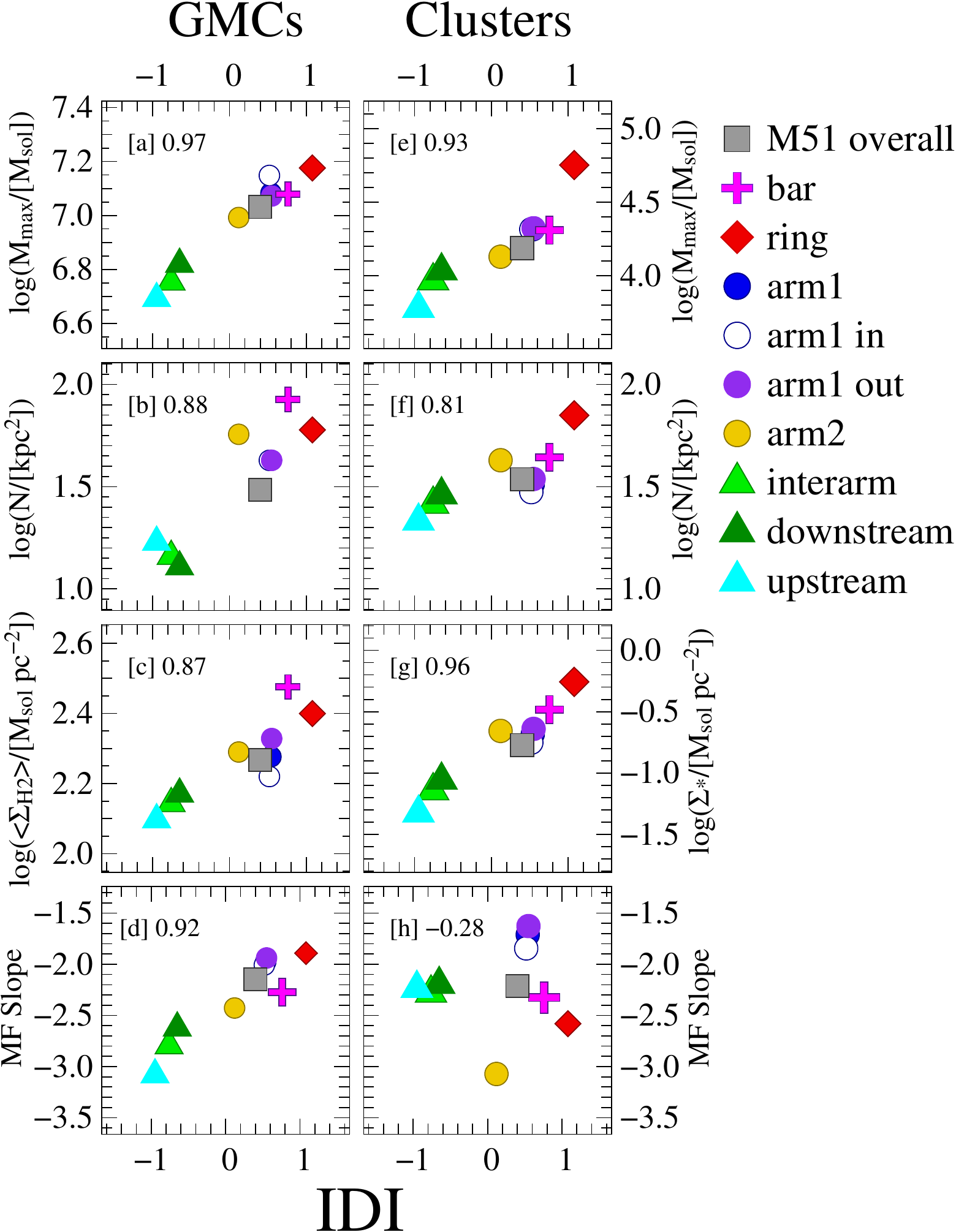}
\caption{\small Properties of the GMC (left column) and young
  ($<10$\,Myr) stellar cluster (right column) populations in different
  M51 environments, compared to the shape of the CO PDFs.  The
  Spearman rank correlation is indicated at the top left of each
  panel.}
\label{fig:idi_vs_gmcycprops}
\end{center}
\end{figure*}

\noindent We examined the relationship between GMC properties and the
shape of the CO PDFs using the cloud catalog presented by Colombo et
al. (submitted) and the $BDI$ and $IDI$ values calculated in
Section~\ref{sect:pdfs_enviros}. Since the $BDI$ and $IDI$ themselves
exhibit a tight one-to-one correlation, for simplicity we refer only
to the $IDI$ in the following sections. We illustrate some of these
correlations in the left column of
Figure~\ref{fig:idi_vs_gmcycprops}. Environments where bright CO
emission is more dominant (i.e. with more positive $IDI$ values) are
associated with a higher maximum GMC mass $M_{gmc,95}$, which we
estimate using the 95th percentile of the GMC virial mass distribution
(panel [a]), a greater number surface density of GMCs
$\mathcal{N}_{gmc}$ (panel [b]), and a higher average surface density
for individual GMCs $\langle \Sigma_{\rm H_{2}} \rangle$ (panel
[c]). The $IDI$ is also strongly correlated with the slope of the GMC
mass spectrum $\gamma_{gmc}$: in environments with more bright
emission, the mass spectrum is shallower (panel [d]). For observations
with low resolution (i.e. where a single resolution element is much
larger than the characteristic size of a GMC), a good correlation
between the prevalence of bright CO emission and the mass and mass
surface density of identified cloud structures might arise simply due
to higher filling factors of CO emission, i.e. increases in the
measured CO integrated intensity reflect a greater number of
CO-emitting clouds within the telescope beam, rather than changes in
the intrinsic properties of GMCs. We do not consider this to be the
cause of the good correlations in Figure~\ref{fig:idi_vs_gmcycprops},
however, since the PAWS resolution ($\sim40$\,pc) is well-matched to
the characteristic size of an individual Galactic GMC \citep[50\,pc,
  e.g.][]{blitz93}, and considerably less than the typical spacing
between the identified GMCs (a few times 100\,pc or greater, Colombo
et al. submitted). The peak CO brightness temperatures of the GMCs
range from $\sim2$ to 16\,K, which is comparable to the values
observed for Galactic GMCs \citep[5 to 10\,K,][]{solomonetal87}. Since
the molecular gas in M51 GMCs appears to have a similar kinetic
temperature as in Galactic GMCs
\citep[$\sim10$\,K,][]{schinnereretal10}, this again suggests that the
filling factor of the CO emission in M51 GMCs is close to unity. \\

\noindent It is remarkable that the GMC properties are often more
strongly correlated with the shape of the CO PDFs than other
quantities with which they might also be expected to correlate. In
particular, we note that $M_{gmc,95}$ and $\gamma_{gmc}$ are more
tightly correlated with the $IDI$ than with the total CO luminosity or
the total number of GMCs in each region (the latter plots are not
shown). This would seem to confirm that an increase in the maximum GMC
mass is not simply due to an increase in the available gas reservoir
and more adequate sampling of the top-end of the GMC mass function
(i.e. a size-of-sample effect), and that the good correlation between
the $IDI$, $M_{gmc,95}$ and $\gamma_{gmc}$ arises because the density
distribution of the molecular ISM plays a role in regulating the GMC
mass distribution. It is also noteworthy that the $IDI$ increases with
both the number density of GMCs $\mathcal{N}_{gmc}$ {\it and} the
average surface density of the individual clouds $\langle \Sigma_{\rm
  H_{2}} \rangle$.  This suggests that a distinction that is sometimes
drawn by empirical studies of extragalactic star formation between an
increase in the number of GMCs per resolution element and variations
in the \hh\ surface density on the scale of individual clouds
\citep[e.g.][]{bigieletal08} is somewhat artificial: at least in the
inner disk of M51, clouds in environments with more GMCs per unit area
also tend to have higher average surface densities.\\

\noindent The good correlation between the $IDI$ and $\gamma_{gmc}$ in
panel [d] of Figure~\ref{fig:idi_vs_gmcycprops} is especially
noteworthy. Considerable theoretical and observational effort has been
devoted to showing how the shape of the stellar initial mass function
might be inherited from the density structure of interstellar gas
\citep[e.g.][ and references therein]{hopkins12,chabrierhennebelle10},
with many studies adopting the shape of the GMC mass function as a
description of the latter. A major problem with using GMC mass spectra
for this purpose, however, is that the decomposition algorithm has a
major impact on the identification and parameterization of cloud
structures and hence the shape of the resulting mass distribution
\citep[e.g.][]{wongetal11,reidetal10}. Moreover, many widely-used
decomposition methods are not flux conservative, discarding a
considerable fraction of the CO emission that is unambiguously
detected within a spectral line data cube. As a description of how
dense gas is distributed within galaxies, PDFs avoid these ambiguities
even though, as we show in the Appendix~\ref{app:pdfs_resn_sens}, the
resolution of the observational data must be well-matched to the
physical scales of interest in order to capture the shape of the PDF
accurately. The PDF also conveys no information about the
characteristic size of dense gas structures, moreover, so a more
complete description of the organization of the dense ISM strictly
requires an analysis of the CO PDF in conjunction with a metric such
as the spatial power spectrum \citep[as suggested by
  e.g. ][]{bournaudetal10}. Nonetheless, the plots in the left column
of Figure~\ref{fig:idi_vs_gmcycprops} indicate that there is a strong
relationship between the shape of the PDF and the mass distribution
and properties of GMCs within M51 environments, suggesting that in
real galactic disks the presence of bright emission and the
development of massive molecular structures are physically
linked. Testing whether a similar connection between the GMC mass
function and shape of the PDF holds across a range of galaxy types is
a project that should become feasible once ALMA acquires cloud-scale
imaging of the CO emission across the full galactic disk for a large
sample of nearby galaxies.

%%%%%%%%%%%%%%%%%%%%%%%%%%%
\subsection{Comparison with properties of stellar clusters}
\label{sect:pdfs_vs_sf}
%%%%%%%%%%%%%%%%%%%%%%%%%%%

\noindent Another motivation for constructing the CO PDFs across a
range of M51 environments is to assess whether there are connections
between empirical tracers of star formation activity and the density
distribution of molecular gas. Several empirical calibrations for the
star formation (SF) rate exist in the literature \citep[for a detailed
  comparison of the limitations and assumptions of different methods,
  see][]{leroyetal12}, but here we restrict our analysis to comparing
the shape of the CO PDFs to the properties of young stellar clusters
identified by \citet{chandaretal11} using multi-colour images of M51
obtained by the Advanced Camera for Surveys on board the {\it Hubble
  Space Telescope} \citep{mutchleretal05}. The interested reader is
referred to \citet{chandaretal11} for a description of the methods
used to select clusters and to derive physical quantities such as
their age and mass. The relationship between CO emission and other SF
tracers within the PAWS field is discussed in several companion papers
(Meidt et al., submitted, Schinnerer et al., in preparation). \\

\noindent As for GMCs, we find evidence for a strong connection
between the prevalence of bright CO emission and M51's young ($\tau
\leq 10^{7}$\,Myr) cluster population. In particular, more positive
$IDI$ values are associated with a higher maximum young cluster mass
$M_{yc,95}$ (defined analogusly to $M_{gmc,95}$), and with a higher
number surface density $\mathcal{N}_{yc}$ and combined mass surface
density $\mathcal{M}_{yc}$ of young clusters
(Figure~\ref{fig:idi_vs_gmcycprops}[e] to~[g]). The origin of these
trends would seem to lie in a physical -- as opposed to statistical --
connection between young clusters and GMCs: $M_{yc,95}$,
$\mathcal{N}_{yc}$ and $\mathcal{M}_{yc}$ are better correlated with
the average GMC mass surface density ($\langle \Sigma_{\rm H_{2}}
\rangle$) and maximum GMC mass ($M_{gmc,95}$) than with the total
number of young clusters or GMCs within each M51 environment.\\

\noindent An exception to the good correspondence between the shape of
the CO PDFs and the properties of GMCs and young stellar clusters is
the slope of the cluster mass distribution: while there is a clear
trend for the GMC mass spectrum to become shallower in regions where
bright CO emission is more prevalent, a connection between the slope
of the young cluster mass function and the shape of the CO PDFs is
less obvious (cf panels~[d] and~[h] of
Figure~\ref{fig:idi_vs_gmcycprops}. There is some indication that the
mass distribution of the young cluster populations in the arm and
interarm regions follow the same trend with $IDI$ as GMCs, but within
the central kiloparsec of M51 (i.e. the molecular ring and nuclear bar
regions), the young cluster mass distributions are steep
($\lesssim-2.5$) even though bright CO emission is relatively dominant
there. \\

\begin{figure*}
\begin{center}
\hspace{-0.5cm}
\includegraphics[height=70mm,angle=0]{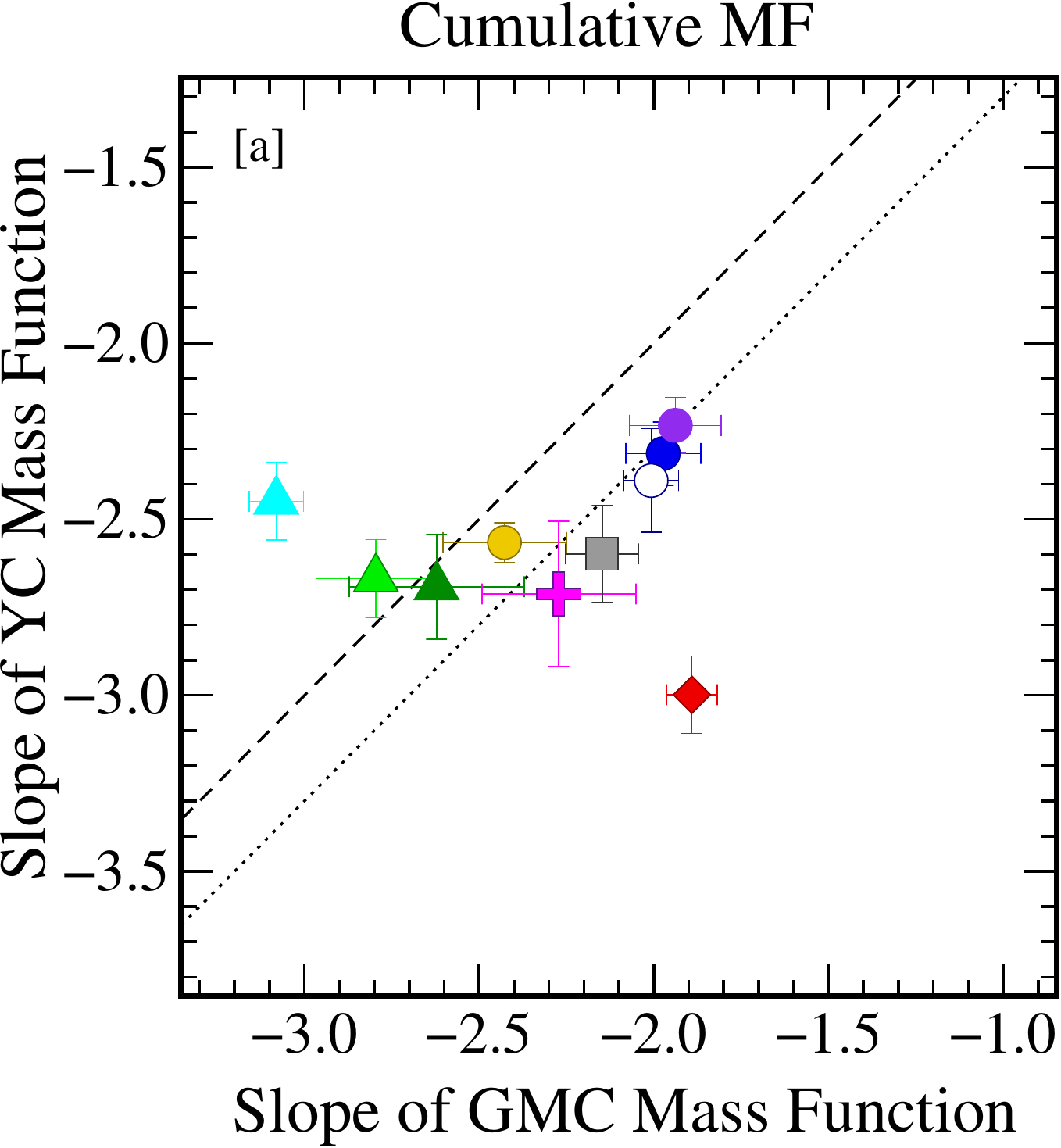}
\includegraphics[height=70mm,angle=0]{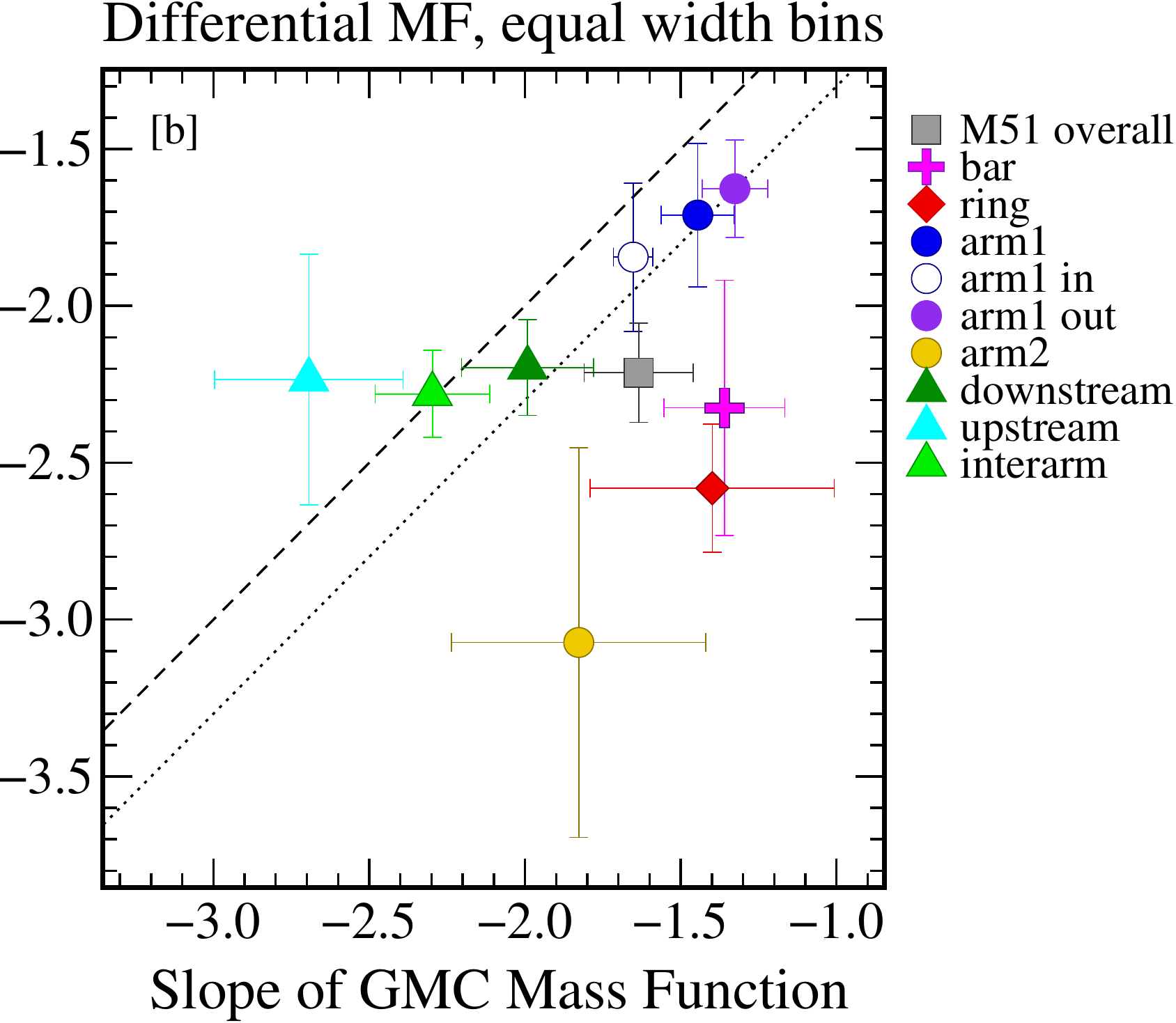}
\caption{\small Slope of the young cluster mass function versus the
  GMC mass function for different dynamical environments within
  M51. Panel [a] represents the results using a cumulative formulation
  for the mass distribution; panel [b] shows the results for a
  differential representation. In both panels, the error bars reflect
  the dispersion in the estimated slopes (see
  Appendix~\ref{app:mspectests}). The dashed diagonal line indicates
  equality, and the dotted diagonal line represents $\gamma_{gmc} =
  \gamma_{yc} + 0.3$ The plot annotations are the same as in
  Figure~\ref{fig:idi_vs_gmcycprops}.}
\label{fig:gmc_yc_mfs}
\end{center}
\end{figure*}

\noindent In Figure~\ref{fig:gmc_yc_mfs}, we plot the slope of the
young cluster mass spectrum directly against the slope of the GMC mass
spectrum for the different M51 environments. Various techniques for
estimating the slope of the mass spectrum have been used by empirical
studies of young clusters and GMCs (e.g. differential versus
cumulative mass distributions, bins of equal width versus bins
containing an equal number of objects), and the derived slope is known
to be sensitive to factors such as cloud decomposition algorithm, the
adopted low-mass completeness limit, and undersampling and/or the
existence of a physical truncation to the distribution at high
masses. We constructed both differential and cumulative mass
distributions for the GMC and young cluster populations, and for each
variant, we estimated the slope multiple times using different binning
strategies (in the case of the differential mass distributions) and
mass ranges to calculate the fit. We describe these tests more fully
in Appendix~\ref{app:mspectests}. The two panels of
Figure~\ref{fig:gmc_yc_mfs} represent the results using a cumulative
(panel [a]) and differential representation (panel [b]) for the mass
distributions. In both panels, the error bars reflect the dispersion
in the estimated slopes of the GMC and young cluster mass spectra in
each environment. Despite the systematic uncertainties that limit the
accuracy of any individual measurement of the mass distribution slope,
we can therefore confidently draw two conclusions from
Figure~\ref{fig:gmc_yc_mfs}. The first is that while there is
reasonable agreement between $\gamma_{gmc}$ and $\gamma_{yc}$ for the
arm and interarm environments of M51, the slope of the mass
distribution is not universal, i.e. the same values of $\gamma_{gmc}$
and $\gamma_{yc}$ do not hold everywhere within M51. Second, agreement
between the slopes of the GMC and young cluster mass distributions is
not ubiquitous. Averaged across the entire PAWS field and for some of
the arm and interarm regions, $\gamma_{gmc}$ and $\gamma_{yc}$ agree
to within $\sim0.3$\,dex, but for the molecular ring and upstream
environments $\gamma_{gmc} \not\approx \gamma_{yc}$, regardless of the
method used to represent the mass functions.\\

\noindent The trends in Figure~\ref{fig:gmc_yc_mfs} are remarkable
since the observed similarity between $\gamma_{gmc} \approx -1.7$ and
$\gamma_{yc} \approx -2.0$ is frequently cited as evidence for the
weak mass dependence of both the efficiency of star formation in GMCs
and probability of cluster disruption. In their investigation of
stellar feedback and disruption of GMCs, for example,
\citet{falletal10} derive relations between $\gamma_{gmc}$ and
$\gamma_{yc}$, which for isolated, bound systems in the absence of
magnetic support are linked via the slope of the mass versus size
relationship of GMCs. \citet{falletal10} examine the limiting regimes
of energy- and momentum-driven feedback, arguing that the constant
surface density of GMCs ($M \propto R^{2}$) ensures that $\gamma_{gmc}
\sim \gamma_{yc}$ regardless of the type of feedback that dominates
GMC disruption. \\

\noindent In a globally-averaged sense, the young cluster and GMC
populations of M51 would seem to conform to the model outlined by
\citet{falletal10}. In this case, $\gamma_{gmc} = -1.7$, $\gamma_{yc}
= -2.1$ and for momentum-driven feedback, the predicted slope of the
GMC mass-size relation is $\alpha = 1.8$, in acceptable agreement with
the observed value ($\alpha_{obs} = 2.1$, Colombo et al.,
submitted). However, our results for individual environments within
M51 suggest a more nuanced interplay between molecular gas, young
clusters and galactic structure. Downstream of the spiral arms, for
example, $\gamma_{gmc} \sim \gamma_{yc} \approx -2.0$ and
$\alpha_{obs} = 1.8$, in good agreement with the model prediction for
energy-driven feedback. Yet upstream of the spiral arms (where
$\gamma_{gmc} \sim \gamma_{yc} \approx -3.0$), in the molecular ring
($\gamma_{gmc} \approx -1.4$, $\gamma_{yc} \approx -2.7$) and in the
first spiral arm pattern of M51 ($\gamma_{gmc} \approx -1.5$,
$\gamma_{yc} \approx -2.1$), the exponent of the mass-size
relationship predicted by \citet{falletal10} is too shallow compared
to the observed value by 0.4 to 1.0\,dex, regardless of the assumed
feedback mechanism. This suggests that there may be regions within
galactic disks where physical processes that would seem to be excluded
by \citet{falletal10}, e.g. cluster coalescence, mass-dependent
cluster/GMC disruption or a dominant role for energy-driven feedback,
are in fact important. More generally, however, it highlights how
valuable information can be lost in the calculation of galaxy-wide
averages. With datasets that yield statistically significant samples
of GMCs and other star-forming phenomena within $\sim$kiloparsec-scale
regions, it is timely that physical quantities (for example, based on
a consideration of galaxy dynamics and/or ISM properties) determine
the environments where GMC properties and extragalactic star formation
are investigated, rather than relying solely on radial profiles and/or
apertures that are `blind' to their location with respect to galactic
structure.\\

%%%%%%%%%%%%%%%%%%%%%%%%%%%
\subsection{Comparison to Numerical Simulations of Galactic Disks}
\label{sect:models}
%%%%%%%%%%%%%%%%%%%%%%%%%%%

\noindent Our analysis in this paper was prompted, in part, by
numerical simulations showing that the gas density distribution in
galactic disks is well-represented by a single lognormal function
spanning several orders of magnitude \citep[e.g.][]{wadanorman07}. If
this is an accurate description of real galactic disks, then the
result offers insight into the origin of the KS law, which can be
reproduced from a lognormal density PDF with a limited number of
plausible assumptions, such as a critical density threshold for star
formation \citep{elmegreen02}, or a direct proportionality between the
local gas and star formation rate densities
\citep[e.g.][]{kravtsov03}. While the overall \ico\ and $T_{\rm mb}$
PDFs for the PAWS field are roughly lognormal, our results in
Section~\ref{sect:pdfs_enviros} show that this average PDF shape
obscures considerable diversity among the PDFs observed for different
$\sim$kiloparsec-sized regions within M51. Since the regions that we
use to investigate the PDFs are defined according to dynamical
criteria, our basic result is that large-scale dynamical processes in
M51's inner disk have an observable effect on the density (and column
density) distribution of M51's molecular ISM.\\

\noindent There are several further characteristics of our observed CO
PDFs that are noteworthy in relation to the simulation results. For
example, \citet{wadanorman07} find an increase in the logarithmic
width of the PDF of $\sim$0.3\,dex for an order of magnitude increase
in the mean gas density. This is roughly consistent with the variation
in the width of the \ico\ PDFs of M51 and the LMC, between which the
average \hh\ column density also varies by a factor of $\sim10$ (see
Section~\ref{sect:pdfs_allgals}). However, we caution that several
effects limit the extent to which we can compare our observed CO PDFs
to the density PDFs from simulations. The first is that the
simulations describe gas density across six orders of magnitude,
corresponding not only to the molecular ISM but also to the atomic and
warm ionized phases. In practical terms, this makes the numerical
result difficult to verify, since different observational tracers must
be used to probe different phases of the interstellar gas, and each of
them are sensitive to a much narrower range of densities (and column
densities) than the full dynamic range of the simulated PDFs.\\

\noindent The few simulations that include explicit treatment of
molecular chemistry tend to find a range of densities between $\sim1$
and $10^{3}$\,\ccc\ for the \hh\ gas in galactic disks, and that the
distribution exhibits a sharp cut-off below the \hh\ self-shielding
limit \citep[$n \lesssim 5\,\ccc$, see e.g. figure~11
  of][]{dobbsetal08}. The use of \aco\ emission to trace the
\hh\ column density should narrow the observed PDF even further, since
at moderately low extinction \citep[$A_{\rm V} \sim 1$\,mag,
  e.g.][]{wolfireetal10} \hh\ can self-shield while CO molecules are
photodissociated. At high \hh\ column densities ($N(H_{2}) \gtrsim
10^{22}$\,\cc), on the other hand, the CO-emitting structures within a
GMC will start to overlap and shadow each other, leading to a
saturation of \ico\ intensities \citep[e.g.][]{shettyetal11}.
\citet{feldmannetal12} show that this saturation should occur at
\ico\ intensities near a few 100\,\kkms. In general, however, the
\ico\ PDFs in Figure~\ref{fig:pdf_ico_enviros} show no evidence for a
peak at these intensities caused by a `pile-up' of saturated pixels,
but instead more closely resemble their pure $N(H_{2})$ PDFs after
they exclude pixels with low CO integrated intensities
\citep[$<0.2$\,\kkms, see figure~8 of][]{feldmannetal12}. We suggest
that this is because the typical CO linewidths in M51 are much greater
than the two possibilities considered by \citet{feldmannetal12}
(i.e. a constant linewidth of 3\,\kms, or virial scaling of the
linewidth with mass surface density). As a consequence, the majority
of the CO-emitting molecular structures in M51 do not shadow each
other in velocity space and hence \ico\ remains a relatively good
tracer of the \hh\ column density. This is consistent with recent
studies of gas and dust in M51, which suggest that the \xco\ factor
has a roughly Galactic value throughout the disk
\citep[e.g.][]{schinnereretal10,tanetal11,mentuchcooperetal12}. \\

\noindent What insights can models provide regarding the diversity
of distribution shapes that we observe? As noted in the Introduction,
the PDF for supersonically turbulent isothermal gas is lognormal when
the influence of gravity is negligible. In this case, the logarithmic
width of the PDF $x$ varies with the Mach number $\mathcal{M}$
according to $x^{2} \approx \ln (1 + 0.25 \mathcal{M}^{2})$
\citep{padoanetal97}. Across a galactic disk, however, the temperature
and average density of the molecular gas will vary with
location. Thus, a more realistic expectation for the observed density
PDF on global to kiloparsec scales may be the convolution of the local
lognormal PDF (reflecting the distribution of densities within a
region over which the average gas density $\rho_{\rm ave}$ and Mach
number are relatively constant) with the PDFs of $\rho_{\rm ave}$ and
$\mathcal{M}$ within the galactic disk. \citet{elmegreen11} has
recently considered such a model, presenting convolution PDFs for
several idealized cases of gas clouds with different radial density
profiles, and variable Mach numbers (see his figure~1). The
resulting PDFs clearly diverge from a pure lognormal shape. As clouds
become more centrally condensed (and hence more dominated by
self-gravity), the convolution PDFs develop a power-law tail at high
densities, with a slope that varies inversely with the slope of the
density profile. If the Mach number decreases at
higher average densities, on the other hand, the convolution PDF
appears truncated relative to a pure lognormal since the local PDFs get
narrower with increasing $\rho_{\rm ave}$.\\

\noindent In broad terms, the analysis by \citet{elmegreen11} suggests
that the diverse shapes of the \ico\ PDFs in
Figure~\ref{fig:pdf_ico_enviros} reflect large-scale variations in the
average density, temperature and/or velocity fluctuations for the
molecular gas within different M51 environments. In reality, these
properties are likely to vary simultaneously, so attributing a
specific PDF morphology to a variation in one physical quantity and/or
process is not straightforward. Nevertheless, it is suggestive that
the \ico\ PDFs in the interarm region -- where we might expect the
temperature, density and velocity structure of the molecular gas to be
determined by cloud-scale processes -- resemble the pure lognormals
expected for isothermal supersonically turbulent molecular gas,
whereas the \ico\ PDFs in M51's spiral arms -- where the molecular gas
not only reaches higher densities, but its velocity structure can be
influenced by large-scale dynamical effects such as streaming motions
and the spiral shock -- more obviously diverge from a lognormal
shape. The molecular ring region, where the \ico\ PDF is very broad
and almost flat-topped, is arguably the extreme case both in terms of
dynamical effects and star formation activity. Although shear and
large-scale non-circular motions should be low in the ring, the
molecular gas accumulates here due to opposing bar and spiral torques
and the average gas velocity dispersion is relatively high (Colombo et
al., submitted). The level of star formation activity in the ring
is also high (Schinnerer et al., in preparation), so feedback from
nascent stars may also have a strong effect on the distribution of gas
densities in this region. \\

\noindent Finally, we note that \citet{hopkinsetal12} have recently
shown that the dominant mode of stellar feedback (and not just total
amount of star formation) has an observable effect on the shape of the
density PDF for the cold gas component in their simulated
galaxies. Like \citet{wadanorman07}, they find that the width of the
density PDF decreases for systems with lower average gas
densities. The density PDFs of their simulated gas disks show striking
departures from lognormality (see their figures~10 and~11), however,
which they attribute to the inclusion of cooling, self-gravity and a
physically-motivated implementation of different feedback mechanisms
in their simulations. In particular, they find that radiation pressure
is crucial for suppressing a pile-up of gas with high densities ($n
\gtrsim 10^{4}$\,\ccc), since pure gas heating (e.g. by supernovae,
stellar winds and \hii\ photoionization) is ineffective in disrupting
dense gas clumps where the cooling time is much shorter than the
dynamical time. While the overall shapes of the \ico\ and $T_{\rm mb}$
PDFs for different M51 environments almost certainly reflect the
combined action of several distinct physical processes, the absence of
a secondary peak in the PDFs at high CO intensities would seem to
confirm that the dominant feedback mechanism in M51 must be effective
at preventing the build-up of high-density material.

%%%%%%%%%%%%%%%%%%%%%%%%%%%%%%
\section{Conclusions}
%%%%%%%%%%%%%%%%%%%%%%%%%%%%%%
\label{sect:conclusions}

\noindent In this paper, we have presented the probability
distribution functions (PDFs) of CO integrated intensity
and CO brightness within the inner disk of M51, using new
high resolution ($\sim40$\,pc) data from the PdBI Arcsecond Whirlpool
Survey (PAWS, Schinnerer et al., in preparation). We have compared the PDFs
of these properties for different environments within the PAWS field,
and to PDFs constructed using high resolution CO datasets for two
nearby dwarf galaxies. We report the following results and
conclusions:\\

\noindent 1. On $\sim40$\,pc scales, the distribution of
\ico\ integrated intensities within the inner $\sim11 \times 7$\,kpc
of M51 spans $\sim1.5$ orders of magnitude above the $3\sigma$
sensitivity limit of the PAWS data. The shape of the \ico\ PDF is
consistent with a lognormal function with a mean of 20\,\kkms\ and a
logarithmic width of 0.4. Relative to this LN function, there is some
evidence that the observed PDF is truncated for \ico\ values greater
than $\sim200$\,\kkms. \\

\noindent 2. The CO brightness temperatures that we measure for the
inner disk of M51 span $\sim1$ to 10\,K, where the lower limit
corresponds to our survey's $3\sigma$ sensitivity limit. The shape of
the $T_{\rm mb}$ PDF can be represented by a lognormal function with a
mean of 0.9\,K and a logarithmic width of 0.3, but a broken power law
with a slope of $\sim-1.4$ for $1 < T_{\rm mb} < 5$\,K and a much
steeper slope of $\sim-4.9$ for $T_{\rm mb} > 5$\,K is an equally
adequate description of the distribution.\\

\noindent 3. The CO PDFs that describe the emission in the inner disk
of M51 are clearly different to the PDFs obtained for M33 and the
Large Magellanic Cloud (LMC). The maximum \ico\ and $T_{\rm mb}$
values observed in M51 are $1$ to $1.5$ dex higher than in the two
low-mass galaxies. The CO PDFs in M51 are also wider, consistent with
numerical results indicating that the width of the density and column
density PDF increases with the average gas density of a galactic disk
\citep[e.g.][]{wadanorman07}. \\

\noindent 4. The CO PDFs for different dynamical environments within
M51's inner disk exhibit diverse shapes. The CO PDFs in the interarm
regions are narrower than in the spiral arms, nuclear bar and
molecular ring regions. The distributions of \ico\ and $T_{\rm mb}$
are approximately lognormal in the interarm, while the PDFs in the
arms, ring and bar exhibit strong departures from lognormality such as
power-law slopes and/or truncations at high CO intensities. While a
lognormal function may provide an adequate description for the overall
gas distribution within a galaxy, phenomena such as streaming motions,
spiral arm shocks and star formation feedback produce observable
changes to the gas density distribution for $\sim$kiloparsec-sized
within galaxies.\\

\noindent 5. To avoid assuming a particular functional form for the CO
PDFs in M51, we characterised their shape using the brightness (or
integrated intensity) distribution index, originally devised by
\citet{sawadaetal12}, a simple parameter that specifies the ratio
between bright and faint emission. With this, we showed that the shape
of the CO PDFs for dynamically-defined, kiloparsec-scale environments
within M51 are strongly correlated with physical properties of the
GMC and young stellar cluster populations of those environments and,
we infer, their star formation activity. The implications of this
result for interpreting the observational scatter in extragalactic
star formation laws are explored in several companion papers (Meidt et
al. submitted, Schinnerer et al., submitted, Leroy et al., in preparation). \\

\noindent 6. Consistent with the predictions from numerical
simulations \citep[e.g.][]{wadanorman07}, we find a shallow increase
in the width of the PDF with increasing average gas surface
density. The dynamic range of the observed \ico\ PDFs is also in
approximate agreement with the distributions of \hh\ column density
obtained by simulations that include explicit treatment of molecular
chemistry \citep[e.g.][]{dobbsetal08,feldmannetal12}, but we do not
observe a secondary peak in the PDFs at high CO intensities
corresponding to CO saturation. We suggest that this is because the CO
linewidths in M51 are typically larger than the linewidths adopted by
the simulations, so the `mist-model' explanation \citep{dickmanetal86}
of the Galactic \xco\ factor remains valid even at high \hh\ column
densities. \\

\noindent 7. We show that the diverse shapes of the CO PDFs in M51 are
qualitatively similar to the deviations from lognormality expected
from the combined action of star formation feedback and large-scale
variations in density, temperature and velocity structure throughout
M51's inner disk. Our results suggest that star formation feedback on
small scales and dynamical effects on large scales (e.g. the influence
of the stellar bar and spiral density way) together regulate the
velocity structure of the molecular gas, and that these processes in
combination with gas self-gravity determine the shape of the CO PDFs. Isolating
the dominant physical process responsible for the morphology of each
PDF will require a more detailed comparative analysis with theoretical
models however.\\

\noindent 8. The precise shape of the \ico\ and $T_{\rm mb}$ PDFs is
sensitive to several non-physical effects including resolution,
sensitivity, and the method used to identify significant emission
within a spectral line cube. These caveats should be kept in mind by
future studies that compare the PDFs derived from CO observations of
different galaxies, or aim to validate numerical models using
observational results. In particular, we note that the estimated
logarithmic width of the PDFs tends to decrease for datasets with
poorer sensitivity, and that degrading the observational resolution to
a spatial scale greater than the characteristic spacing between high
brightness structures can produce the appearance of a threshold in the
PDFs. This result is not unique to PDFs of CO emission and suggests
that observations of thresholds on kiloparsec scales should be
interpreted with care.\\

\acknowledgments

\noindent We thank R. Chandar for providing the catalog of M51 stellar
clusters to us. We thank the IRAM staff for their support during the
observations with the Plateau de Bure interferometer and the 30m
telescope.  DC and AH acknowledge funding from the Deutsche
Forschungsgemeinschaft (DFG) via grant SCHI 536/5-1 and SCHI 536/7-1
as part of the priority program SPP 1573 'ISM-SPP: Physics of the
Interstellar Medium'.  CLD acknowledges funding from the European
Research Council for the FP7 ERC starting grant project LOCALSTAR.
TAT acknowledges support from NASA grant \#NNX10AD01G.  During this
work, J.~Pety was partially funded by the grant ANR-09-BLAN-0231-01
from the French {\it Agence Nationale de la Recherche} as part of the
SCHISM project (\url{http://schism.ens.fr/}).  ES, AH and DC thank
NRAO for their support and hospitality during their visits in
Charlottesville.  ES thanks the Aspen Center for Physics and the NSF
Grant No. 1066293 for hospitality during the development and writing
of this paper.

\clearpage

\begin{appendix}

%%%%%%%%%%%%%%%%%%%%%%%%%%%%%%
\section{Masking Methods}
%%%%%%%%%%%%%%%%%%%%%%%%%%%%%%
\label{app:pdfs_methods}

\noindent As discussed by Pety et al. (submitted), a number of
different techniques for constructing CO integrated intensity images
have been presented in the literature
\citep[e.g.][]{helferetal03,dame11,wongetal11}. These include: 
\begin{enumerate}
\item{A {\it sigma-clipping} method, M1, whereby pixels containing
  emission below $n\sigma_{RMS}$ are blanked. $\sigma_{RMS}$ is the
  RMS of the noise variations, which we calculate for each independent
  line-of-sight. For the comparison in this Appendix, we adopted
  $n=3$.}
\item{A {\it dilated mask} method, M2, which identifies islands of
  significant emission by selecting peaks above a threshold of
  $t\sigma_{RMS}$ across two contiguous velocity channels. The
  preliminary mask is then expanded to include all contiguous pixels
  with emission above $e\sigma_{RMS}$. We adopted $(t,e) = (5,1.2)$.}
\item{A {\it smooth-and-mask} method, M3, which generates a version of
  the cube that has been spatially smoothed to an angular resolution of
  $\theta$, and identifies emission in the smoothed cube above a
  significance threshold $m\sigma_{RMS}$. The blanking mask is then
  transferred back to the original (i.e. full resolution) data
  cube. We adopted $(\theta,m) = (3\farcs 6,5)$.}
\item{An {\it \hi\ velocity prior} method, M4, which assumes all of a
  galaxy's CO emission arises in velocity channels within a restricted
  interval, $\Delta V$, around the radial velocity corresponding to
  the peak of the \hi\ line profile for each line-of-sight. We used
  $\Delta V = 50$\,\kms.}
\end{enumerate}
\noindent In addition to these, we defined a final mask for the PAWS
cube that optimized flux recovery while eliminating anomalous features
in the map of M51's velocity field (M5). The construction of this mask is
described in detail by Colombo et al. (in preparation). Example
\ico\ maps for the PAWS field constructed using each of the five
techniques are shown in figure~23 of Pety et al. (submitted). \\

\begin{figure*}
\centering
\hspace{-0.5cm}
\includegraphics[width=80mm,angle=0]{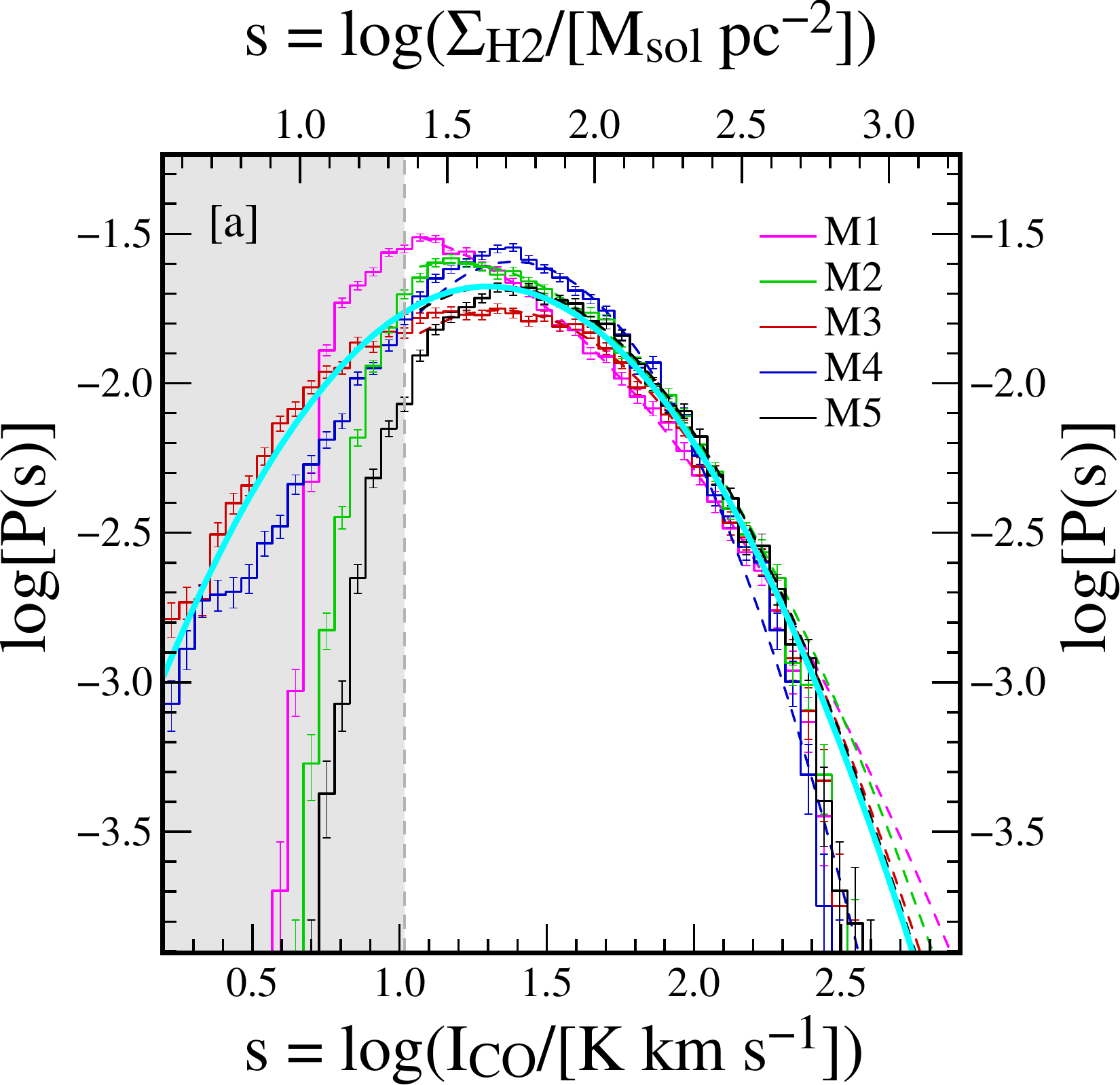}
\hspace{0.5cm}
\includegraphics[width=80mm,angle=0]{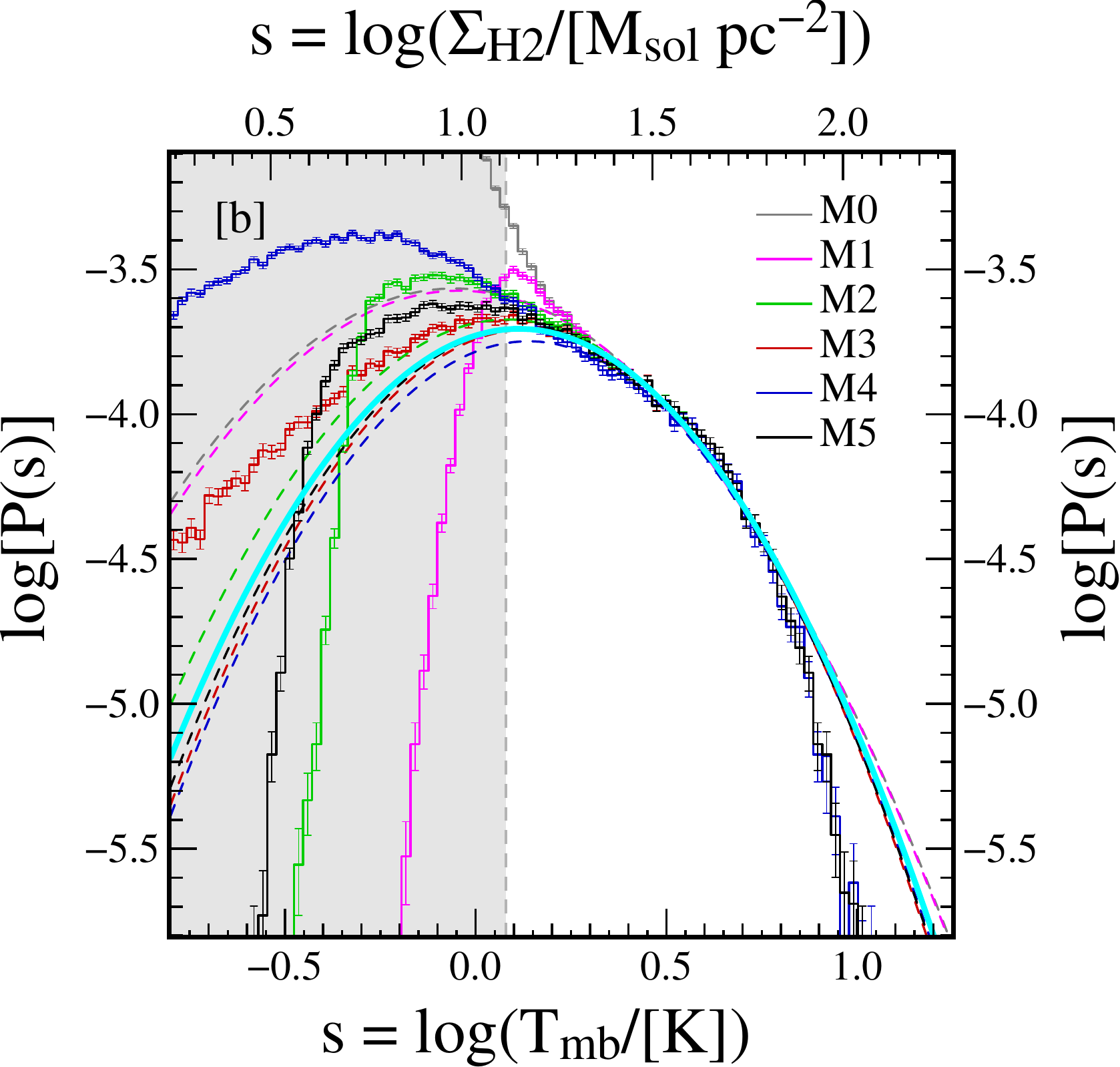}
\caption{\small PDFs of [a] \ico\ and [b] $T_{\rm mb}$ for the PAWS
  field. The different colours represent PDFs obtained from different
  masking techniques to identify significant emission within the data
  cube (see text). A dashed parabola indicates the LN function that provides the
  best-fit to the corresponding PDF. The thick cyan line in panel [a]
  represents a LN function with mean $\langle \ico \rangle =
  20$\,\kkms\ and logarithmic width $x=0.45$; in panel [b], the cyan
  line represents a LN function with mean $\langle T_{\rm mb} \rangle
  = 1.3$\,K and logarithmic width $x=0.35$. In both panels, the
  grey shaded region represents values beneath the nominal $3\sigma_{RMS}$
  sensitivity limit. The top horizontal axis indicates the equivalent
  \mh\ surface density for the \ico\ or $T_{\rm mb}$ value on the
  lower axis, assuming $\xco = 2 \times 10^{20}$\,\xcou\ and a helium
  contribution of 1.36 by mass. The error bars represent the
  uncertainty associated with simple counting ($\sqrt{N}$) errors. }
\label{fig:pdf_methods}
\end{figure*}

\noindent In Figure~\ref{fig:pdf_methods}[a], we show the PDFs
obtained from the different \ico\ maps. It is clear that the map
construction method affects the shape of the \ico\ PDF. Differences
between the curves are apparent up to $\sim60$\,\kkms, which is
considerably greater than our nominal $5\sigma_{RMS}$ sensitivity
limit ($\sim18$\,\kkms). The mean ($s_{0}$) and logarithmic width
($x$) of the best-fitting LN function to each PDF in
Figure~\ref{fig:pdf_methods}[a] are listed in
Table~\ref{tbl:pdffits_appendix}. The PDF from the \ico\ map
constructed using the dilated mask (M2) is the most similar to the PDF
from our preferred mask (M5), although it recovers fewer pixels than
M5 with $10 < \ico < 60$\,\kkms. The same applies to the PDF
corresponding to the \hi\ velocity prior method (M4), although this
PDF also slightly underestimates the number of high intensity pixels
($\ico \gtrsim 160$\,\kkms). Inspection of the CO velocity dispersion
map for the PAWS field shows that there is a small fraction of pixels
with FWHM linewidths greater than 17\,\kms\ (mostly in the nuclear bar
and southern spiral arm regions), so our chosen velocity interval of
50\,\kms\ excludes some genuine emission in the wings of these line
profiles. The smooth-and-mask technique (M3) recovers the least
emission at intermediate \ico\ values. This is because we used a
relatively large smoothing kernel and high significance threshold, so
compact regions with moderate significance in the original data cube
are excluded from the final map. The large smoothing kernel is also
the reason why the M3 map does not show a sharp cut-off at a low
\ico\ value, as it incorporates many pixels with low significance that
are adjacent to high brightness regions. Unlike the other methods,
which tend to peak around $\ico \sim 20$\,\kms, the sigma-clipping
method (M1) peaks at our nominal $3\sigma_{RMS}$ sensitivity
limit. Even though we define $\sigma_{RMS}$ locally, i.e. we estimate
$\sigma_{RMS}$ for each line-of-sight, inspection of the \ico\ map
constructed using M1 reveals that many of these pixels come from the
edge of the field where the brightness sensitivity of PAWS
declines. Although this masking method is relatively common, we regard
the resulting PDF to be the least reliable measure of the
\ico\ distribution in M51's inner disk. In summary, while the
different methods for constructing the \ico\ map yield PDFs that are
different in detail, all the \ico\ PDFs except that obtained using M1
are roughly consistent with a LN function with mean $s_{0} \sim
20$\,\kkms\ and logarithmic width $x = 0.4$, which we have indicated
by a thick cyan line in Figure~\ref{fig:pdf_methods}[a] (see also
Table~\ref{tbl:pdffits_appendix}). Future studies should keep in mind
that estimates for the shape of the \ico\ PDF are more likely to be
dominated by systematic uncertainties due to different techniques for
identifying significant emission than by simple counting and/or
measurement errors. \\

\begin{table*}
\centering
\caption{\small Fit Parameters for CO PDFs in
  Figures~\ref{fig:pdf_methods} and~\ref{fig:pdf_sensitivity}.}
\label{tbl:pdffits_appendix}
\par \addvspace{0.2cm}
\begin{threeparttable}
{\footnotesize
\begin{tabular}{@{}cllccc}
\hline 
\hline 
Figure & CO Property      &  Description  & Mean   & Logarithmic Width & Goodness-of-fit \\
       &          &               & $s_{0}$ & $x$               & $\epsilon$ \\
\hline
\ref{fig:pdf_methods}[a] & \ico\             & M1       & 6.3\,\kkms           & 0.62 & 0.10                    \\
\ref{fig:pdf_methods}[a] & \ico\             & M2       & 15.3\,\kkms          & 0.50 & 0.08                    \\
\ref{fig:pdf_methods}[a] & \ico\             & M3       & 21.8\,\kkms          & 0.45 & 0.07                   \\
\ref{fig:pdf_methods}[a] & \ico\             & M4       & 24.4\,\kkms          & 0.36 & 0.05                   \\

\ref{fig:pdf_methods}[b] & $T_{\rm mb}$          & M0       & 0.9\,K           & 0.40  & 0.19                  \\
\ref{fig:pdf_methods}[b] & $T_{\rm mb}$          & M1       & 0.9\,K           & 0.40  & 0.18                  \\
\ref{fig:pdf_methods}[b] & $T_{\rm mb}$          & M2       & 1.2\,K           & 0.36  & 0.20                 \\
\ref{fig:pdf_methods}[b] & $T_{\rm mb}$          & M3       & 1.4\,K           & 0.34  & 0.19                 \\
\ref{fig:pdf_methods}[b] & $T_{\rm mb}$          & M4       & 1.4\,K           & 0.34  & 0.20                  \\

\ref{fig:pdf_sensitivity}[a] & \ico\             & $\sigma_{RMS} = 0.4$\,K, $(t,e)=(5,1.2)$        & 15.4\,\kkms          & 0.50 & 0.18                    \\
\ref{fig:pdf_sensitivity}[a] & \ico\             & $\sigma_{RMS} = 0.6$\,K, $(t,e)=(5,1.2)$        & 21.0\,\kkms          & 0.44 & 0.09                    \\
\ref{fig:pdf_sensitivity}[a] & \ico\             & $\sigma_{RMS} = 1.0$\,K, $(t,e)=(5,1.2)$        & 22.1\,\kkms          & 0.43 & 0.14                   \\
\ref{fig:pdf_sensitivity}[a] & \ico\             & $\sigma_{RMS} = 2.0$\,K, $(t,e)=(5,1.2)$        & 62.0\,\kkms          & 0.31 & 0.14                   \\
\ref{fig:pdf_sensitivity}[c] & \ico\             & $\sigma_{RMS} = 0.4$\,K, $(t,e)=(3.5,2)$        & 10.1\,\kkms          & 0.66 & 0.13                    \\
\ref{fig:pdf_sensitivity}[c] & \ico\             & $\sigma_{RMS} = 0.6$\,K, $(t,e)=(3.5,2)$        & 21.1\,\kkms          & 0.39 & 0.12                    \\
\ref{fig:pdf_sensitivity}[c] & \ico\             & $\sigma_{RMS} = 1.0$\,K, $(t,e)=(3.5,2)$        & 28.9\,\kkms          & 0.33 & 0.11                   \\
\hline
\hline
\end{tabular}
}
{\footnotesize 
\begin{tablenotes}
\item[]{Parameters of best-fitting functions to PDFs in
  Figure~\ref{fig:pdf_methods}. The parameters of the LN functions are
  determined from a Levenberg-Marquardt fit to
  Equation~\ref{eqn:lmfit}. We use the logarithmic dispersion of the
  fit residuals to estimate the goodness-of-fit.}
\end{tablenotes}}
\end{threeparttable}
\end{table*}

\noindent As well as the \ico\ PDF, we tested the effect of different
masking techniques on the shape of the $T_{\rm mb}$ PDF. The results
for the central 140\arcsec $\times 90$\arcsec\ of the PAWS field are
presented in Figure~\ref{fig:pdf_methods}[b]; we do not construct PDFs
using the entire field since the noise increases significantly towards
the edge of the map (see Figure~\ref{fig:icomap}) which makes the
resulting PDFs harder to interpret. For completeness, we also show the
$T_{\rm mb}$ PDF of the PAWS cube without applying any mask (M0, grey
histogram). Even though the total CO flux of the different masked and
unmasked cubes agrees to within $\sim30$\% (see table~8 of Pety et
al., submitted), the grey and magenta histograms (M0 and M1) only
converge with the other PDFs for $T_{\rm mb} \gtrsim 2$\,K. Below
2\,K, the M0 and M1 histograms begin to curve upwards, departing from
the roughly LN shape of the distribution at higher intensities. The
remaining PDFs are more similar, suggesting that the masking
techniques that use additional criteria (e.g. proximity to a bright
peak) are successful at retaining genuine emission at
$\sim3\sigma_{RMS}$, whereas a simple $3\sigma$ clip (i.e. masking
method M1) may retain a significant number of isolated noise
peaks. Never the less, some genuine low brightness emission could be
masked by methods M2 to M5, although its spatial distribution is
difficult to determine. In our analysis (Sections~\ref{sect:pdfs_ico}
to~\ref{sect:pdfs_allgals}), we focus on the shape of the \ico\ and
$T_{\rm mb}$ PDFs at relatively high brightness (i.e. brighter than
$4\sigma_{RMS}$), so our results and interpretation should not be
affected by the presence of such a faint emission component.\\

\noindent As for the \ico\ PDF, the smooth-and-mask method (M3)
recovers the least emission at intermediate $T_{\rm mb}$ values ($1 <
T_{\rm mb} < 3$\,K) because compact, high brightness regions are
diluted beneath our $5\sigma_{RMS}$ threshold in the smoothed
cube. The dilated mask technique (M2) yields a PDF that appears more
like a broken power-law with a turnover at $\sim3$\,K. This variation
is reasonable, since M2 should exclude some isolated regions of
genuine emission that fall beneath our $5\sigma_{RMS}$ threshold, but
include some noise at the edges of the mask with $T_{\rm mb} \sim
2\sigma_{RMS}$. For values above $\sim 3\sigma_{RMS}$, the $T_{\rm
  mb}$ PDFs for M2, M4 and M5 are practically identical, exhibiting a
mean $T_{\rm mb} \sim 1.3$\,K and logarithmic width $x \sim 0.35$.\\

\noindent We use our tailored mask, M5, for our analysis of the CO
emission in the PAWS field (Section~\ref{sect:pdfs_ico}) and within
different M51 environments (Section~\ref{sect:pdfs_enviros}). As it
closely reproduces the M5 results for M51, but is much simpler to
implement across multiple data sets, we use the dilated mask technique
(M2) for our comparative analysis of M51, the LMC and M33
(Section~\ref{sect:pdfs_allgals}).

%%%%%%%%%%%%%%%%%%%%%%%%%%%%%%
\section{Resolution and Sensitivity Effects}
%%%%%%%%%%%%%%%%%%%%%%%%%%%%%%
\label{app:pdfs_resn_sens}

\begin{figure*}
\begin{center}
\hspace{-0.5cm}
\includegraphics[width=75mm,angle=0]{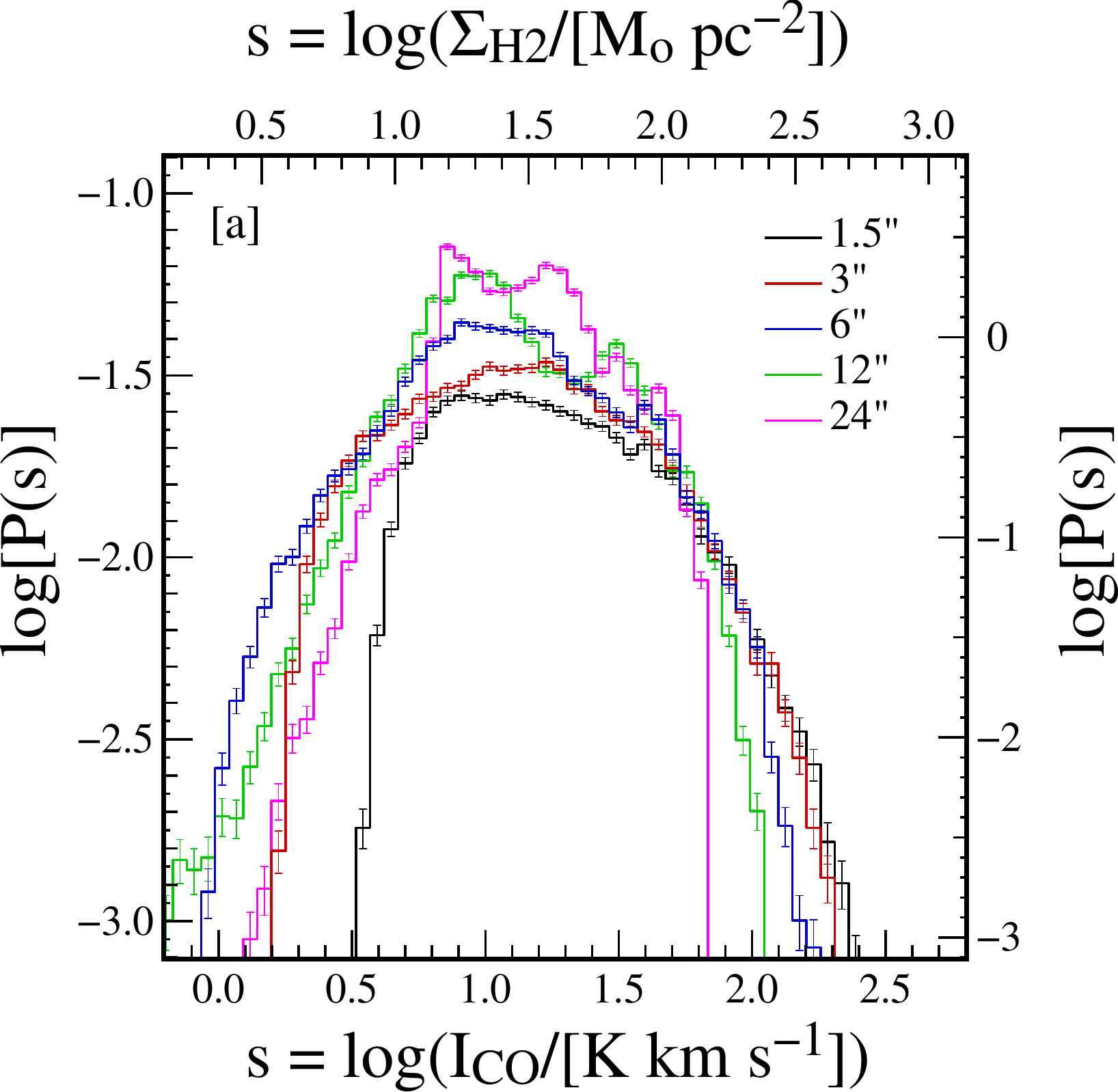}
\hspace{0.5cm}
\includegraphics[width=75mm,angle=0]{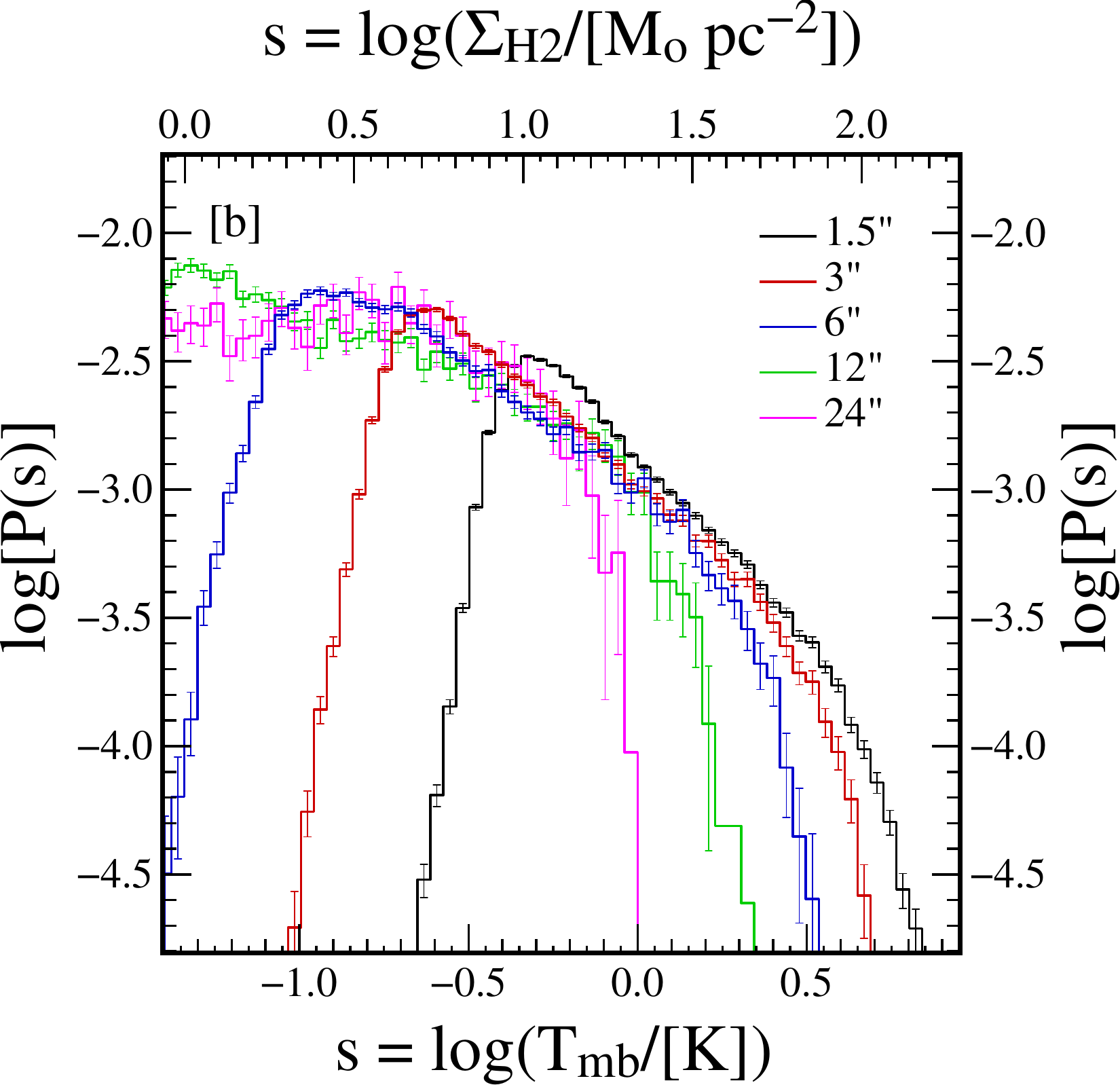}
\caption{\small The PDFs of [a] \ico\ and [b] $T_{\rm mb}$ within the
  PAWS field, obtained after convolving the original PAWS
  data cube with Gaussian smoothing kernels of varying width (see
  text). The PDFs are obtained from cubes where significant emission
  is identified using a dilated mask method. The error bars represent
  the uncertainty associated with simple counting ($\sqrt{N}$)
  errors.}
\label{fig:pdf_resolution}
\end{center}
\end{figure*}

\noindent Previous studies of the column density PDF for individual
molecular clouds have shown that the shape of the PDF depends on the
spatial resolution of the data \citep[e.g.][]{froebrichrowles10}. To
assess the importance of this effect, we smoothed the original M51
data cube to angular resolutions of 1.5, 3, 6, 12 and 24\arcsec,
corresponding to linear scales of 60, 110, 230, 450 and 910\,pc
respectively for our assumed distance to M51. We constructed
\ico\ maps from all the cubes after applying the dilated mask method
to identify significant emission. The resulting \ico\ PDFs are
presented in Figure~\ref{fig:pdf_resolution}[a], while the PDFs of CO
brightness are shown in Figure~\ref{fig:pdf_resolution}[b]. For the
\ico\ PDFs, moderate variations in the resolution (i.e. up to
$\sim200$\,pc) produce PDFs with a similar shape over a restricted
range of \ico\ values (20 to 100\,\kkms). It is striking, however,
that the shape of the PDF at high intensities steepens as the angular
resolution of the datacube is degraded.  At $\sim1$\,kpc resolution,
the \ico\ PDF appears to show a threshold at $\ico \sim 60$\,\kkms;
smoothing the data over even larger scales has no further effect. A
kiloparsec corresponds roughly to the radius of the central region,
and also to half the distance between the spiral arms. The appearance
of an upper threshold would therefore seem to occur because the
emission from these high brightness regions is averaged together once
the resolution is coarser than this spatial scale.\\

\noindent We obtain similar, though not identical, results for the
$T_{\rm mb}$ PDFs. The slope of the PDFs at intermediate intensities
(i.e. from $\sim5\sigma_{RMS}$ to the $T_{\rm mb}$ value where the PDF
begins to steepen) appears relatively constant, flattening slightly as
the resolution degrades from a slope of $\sim-1.4$ at 60\,pc
resolution, to $\sim-0.7$ at 450\,pc resolution. We note that this is
the opposite trend to what is observed for the \ico\ PDFs, which
become steeper at lower resolution.  Like the \ico\ PDFs, however, the
$T_{\rm mb}$ PDFs show a truncation that shifts to lower CO
intensities as the smoothing scale increases. By $\sim1$\,kpc
resolution, the PDF has a sharp cut-off at $T_{\rm mb} \sim 1$\,K for
the same reason that the \ico\ PDF shows a threshold at this
scale. The results of our resolution tests suggest that the appearance
of thresholds and/or truncations in PDFs of gas emission tracers
should therefore be interpreted with some caution, although these
effects would appear to be most severe once the resolution of a
dataset becomes comparable to the characteristic spacing between
regions of high brightness (e.g. the spacing between spiral arms
and/or between star formation complexes). Since the PAWS observations
resolve these spatial scales, we do not regard resolution effects to
be the main driver of the deviations from lognormality that we observe
for the CO PDFs in Figures~\ref{fig:pdf_ico_enviros}
and~\ref{fig:pdf_cube_enviros}. The fact that the CO PDFs of M51's
interarm regions do not show truncations at high intensities would
further tend to support our interpretation that the truncations
observed for the PDFs of the spiral arm and central environments are
not solely due to limited spatial resolution. Since Galactic GMCs
exhibit variations in their CO surface brightness, we would expect
observations that spatially resolve internal structure of extrgalactic
clouds to reveal more features in the shape of the \ico\ and $T_{\rm
  mb}$ PDFs at high CO intensities \citep[such as power-law tails due
  to the formation of strongly self-gravitating clumps, see
  e.g.][]{kainulainenetal09}. While resolution effects do not seem to
be the primary explanation for the variation in PDF shapes with
galactic environment that we describe in Section~\ref{sect:results},
we therefore echo the recommendation by \citet[][see their
  figure~7]{wadanorman07} that well-matched resolution is critical for
comparative studies between observational datasets, or between models
and observations. \\

\begin{figure*}
\begin{center}
\hspace{-0.5cm}
\includegraphics[width=75mm,angle=0]{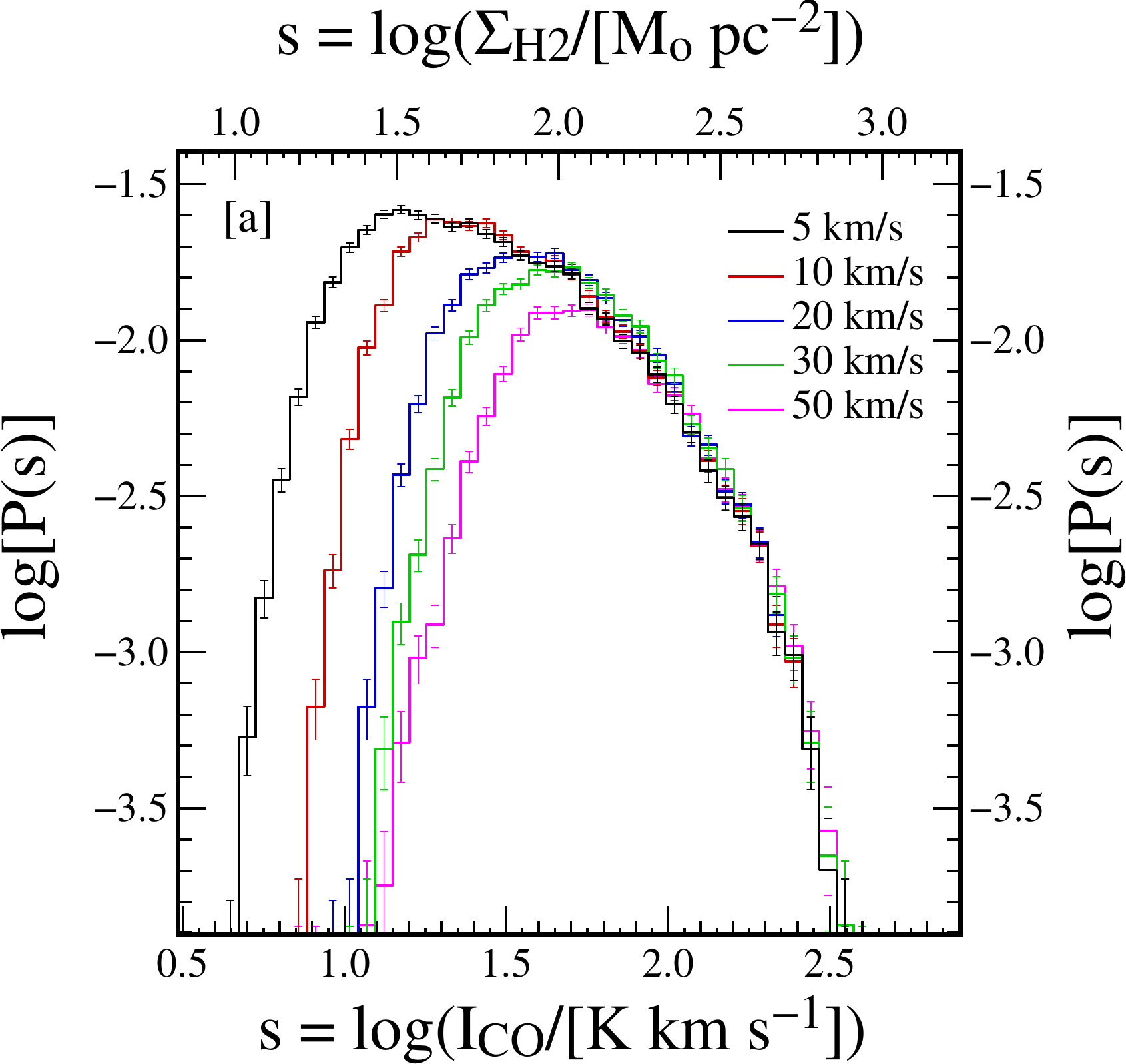}
\hspace{0.5cm}
\includegraphics[width=75mm,angle=0]{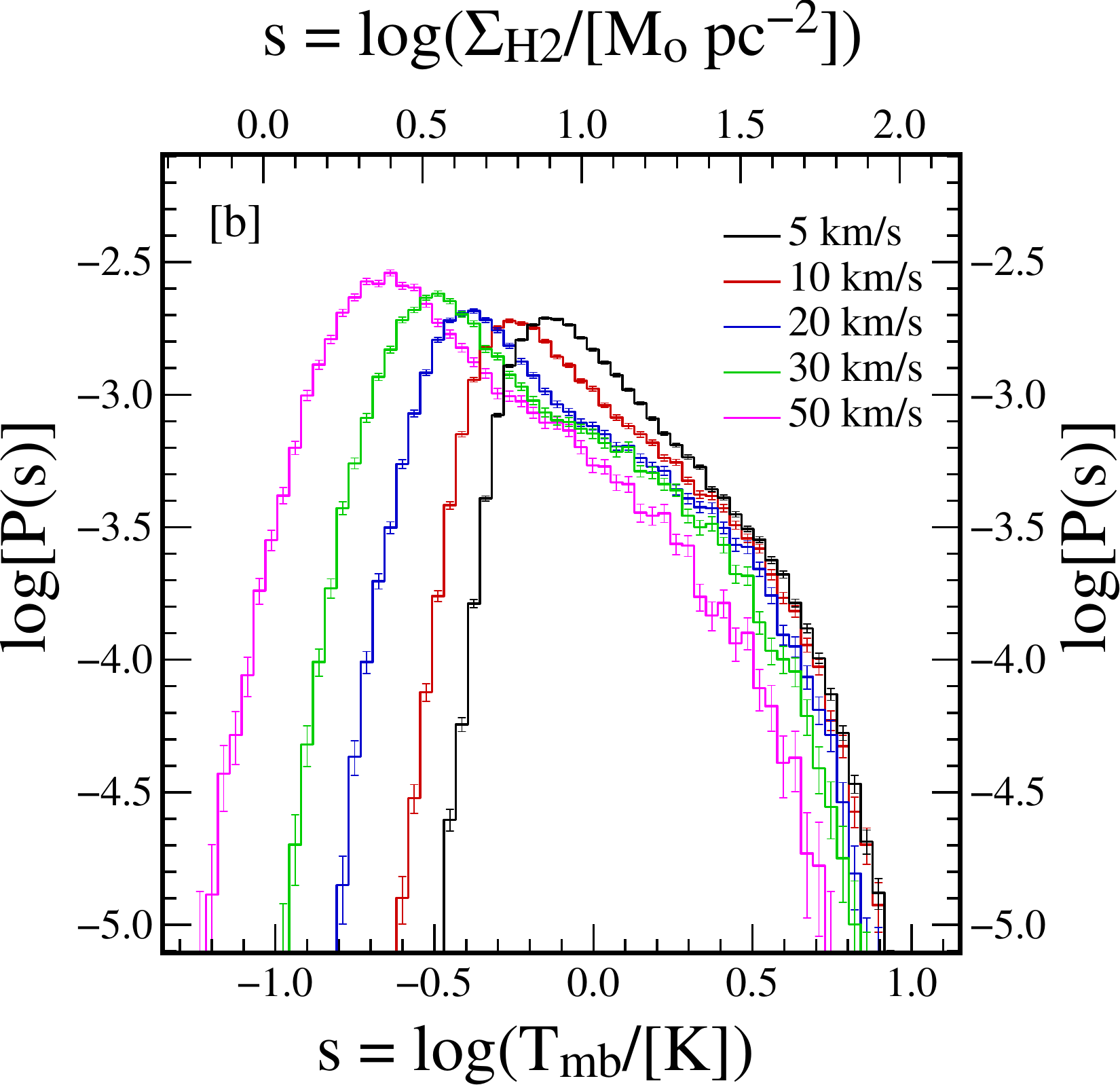}
\caption{\small The PDFs of [a] \ico\ and [b] $T_{\rm mb}$ within the
  PAWS field, obtained after folding the original PAWS data
  cube along the spectral axis to have velocity channels of varying
  width. The PDFs are obtained from the resulting cubes after
  identifying significant emission using a dilated mask method. The
  error bars represent the uncertainty associated with simple counting
  ($\sqrt{N}$) errors.}
\label{fig:pdf_sresolution}
\end{center}
\end{figure*}

\noindent In addition to the effect of spatial resolution, we checked
whether variations in the width of the velocity channels, i.e. the
cube's spectral resolution, produced systematic changes in the shape
of the \ico\ and $T_{\rm mb}$ PDFs. We constructed PDFs from cubes
that had been folded along the velocity axis to 10, 20, 30 and
50\,\kms, and show the results in Figure~\ref{fig:pdf_sresolution}. As
the width of a velocity element increases, the \ico\ PDF narrows but
the shape of distribution at high intensities ($\ico \gtrsim
50$\,\kkms) remains unchanged. The $T_{\rm mb}$ PDF broadens with
increasing channel width, but its shape is also relatively robust. The
maximum observed CO brightness declines by $\sim3$\,K as the channel
width increases from 5 to 50\,\kms; this is due to the spectral
equivalent of beam dilution. \\

\begin{figure*}
\begin{center}
\hspace{-0.5cm}
\includegraphics[width=80mm,angle=0]{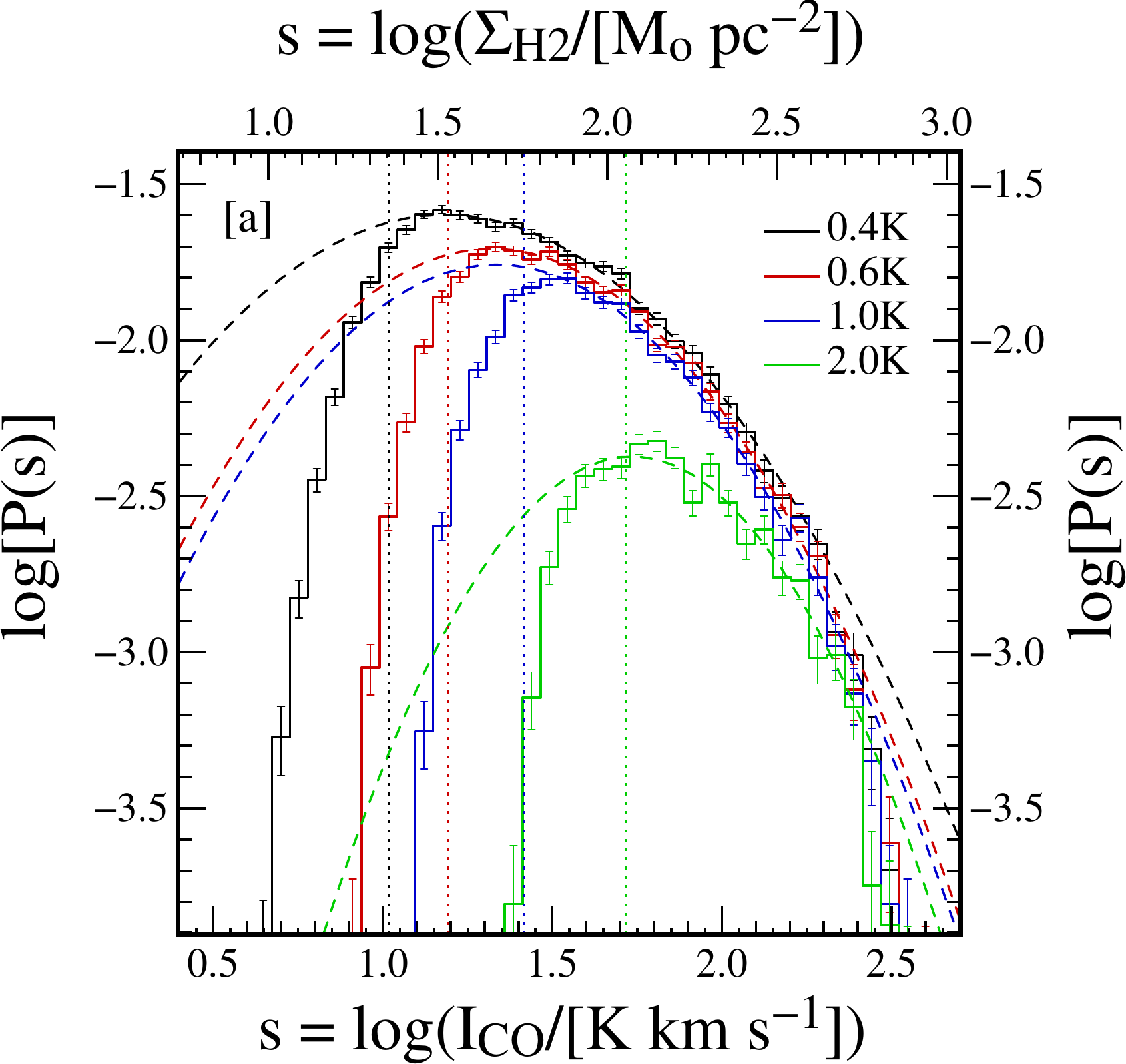}
\hspace{0.5cm}
\includegraphics[width=80mm,angle=0]{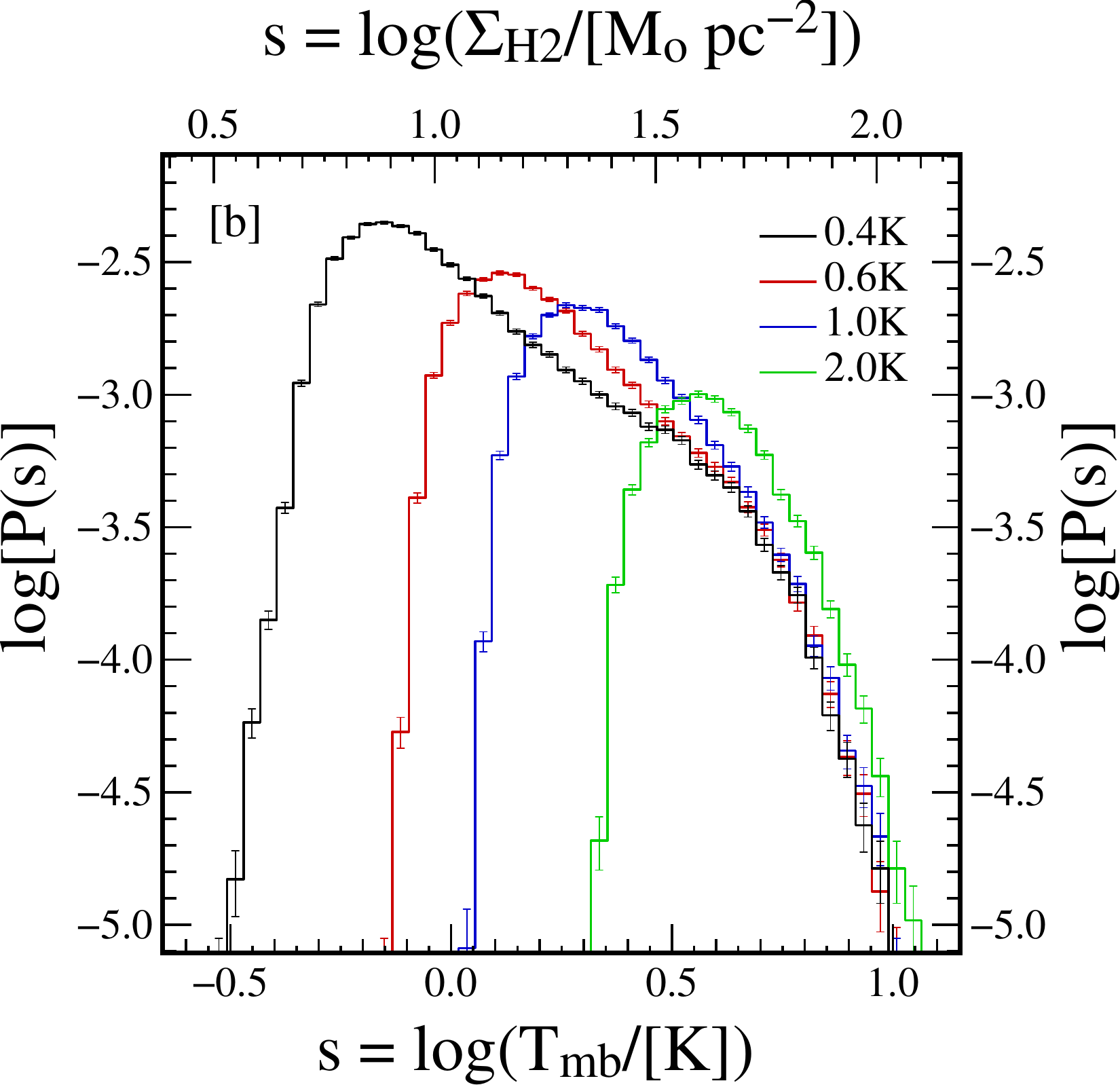}
\par \addvspace{0.3cm}
\hspace{-0.5cm}
\includegraphics[width=80mm,angle=0]{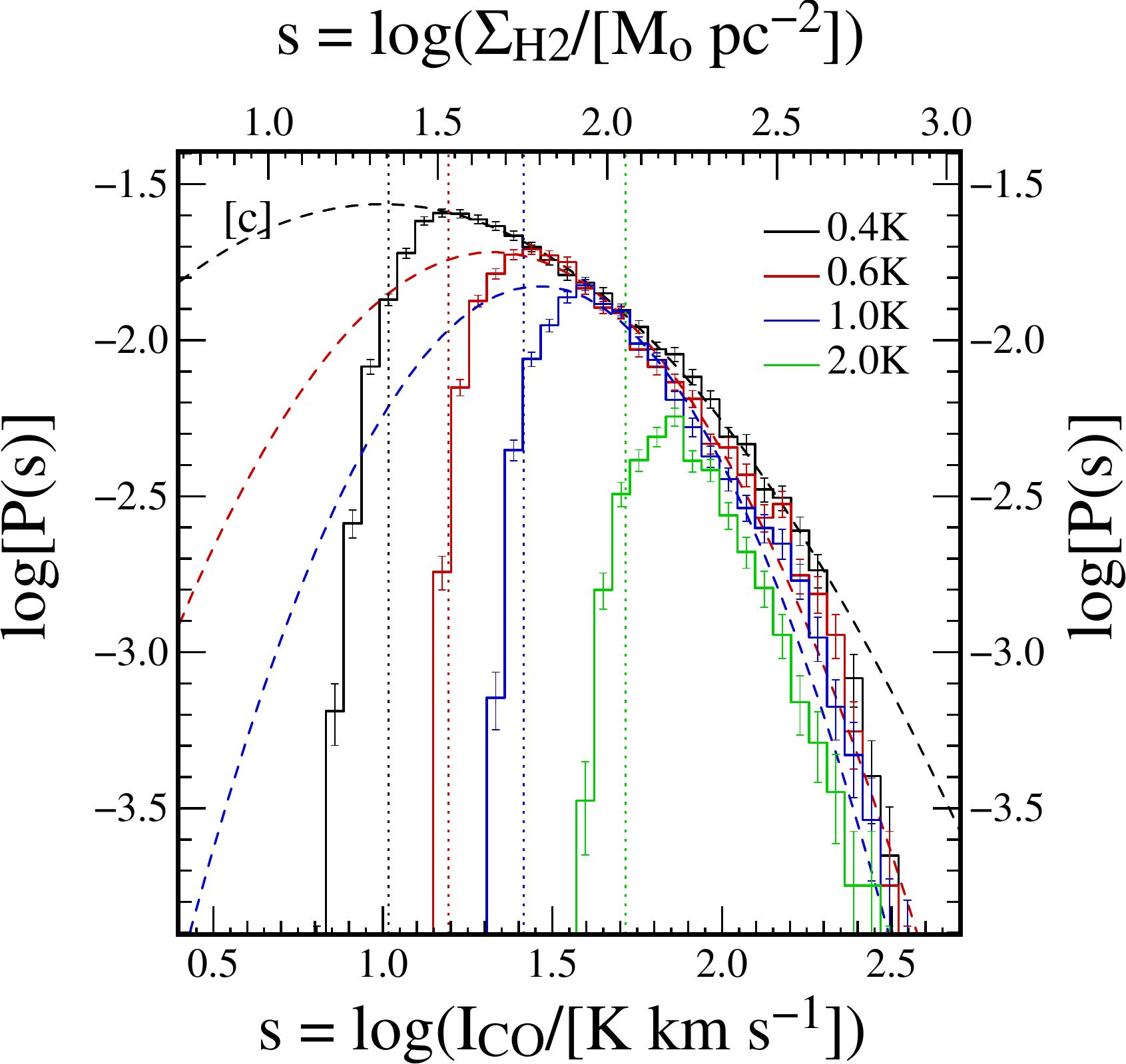}
\hspace{0.5cm}
\includegraphics[width=80mm,angle=0]{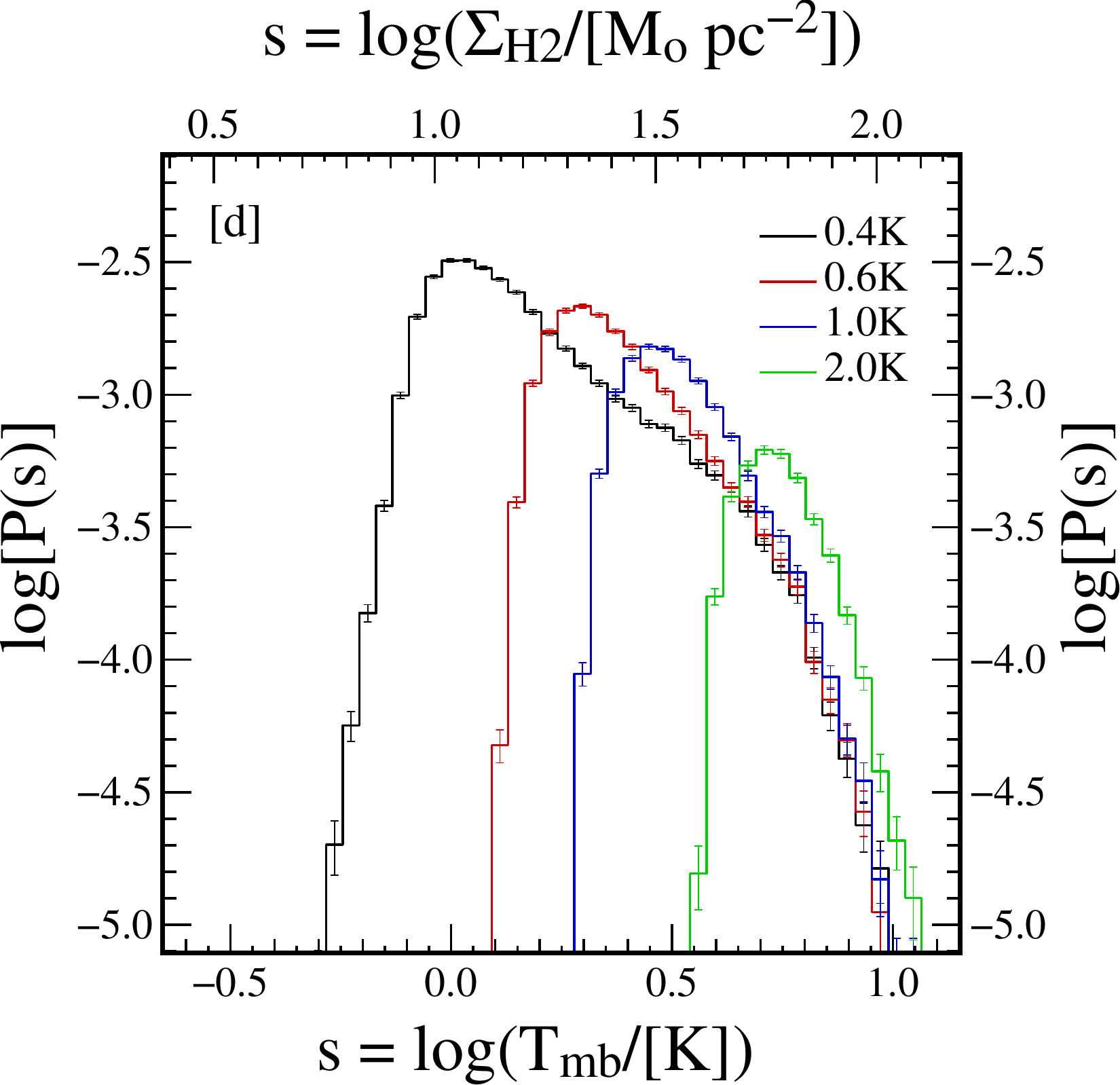}
\caption{\small PDFs of \aco\ integrated intensity (panels [a] \& [c])
  and CO brightness (panels [b] \& [d]) within the PAWS field,
  obtained after adding Gaussian noise to the datacube at the beam
  scale. The PDFs were obtained from cubes where significant emission
  was identified using the dilated mask method with $(t,e)=(5,1.2)$
  (panels [a] \& [b]) and $(t,e)=(3.5,2)$ (panels [c] \& [d]). In
  panels [a] and [c], a dashed parabola indicates the LN function that
  provides the best-fit to the corresponding PDF. The vertical dotted
  lines indicate our nominal $3\sigma_{RMS}$ sensitivity limits. The
  PDFs in the panels [b] and [d] were obtained using the central
  quarter of the PAWS field only, to suppress effects that are caused
  by the lower signal-to-noise at the edge of the PAWS field. The
  error bars represent the uncertainty associated with simple counting
  ($\sqrt{N}$) errors.}
\label{fig:pdf_sensitivity}
\end{center}
\end{figure*}

\noindent In principle, the shape of the \ico\ and $T_{\rm mb}$ PDF at
high intensities should be relatively robust to noise, provided that
the signal-to-noise ratio is sufficiently high. In practice, however,
the sensitivity of most extragalactic \aco\ mapping surveys is
limited. In Figure~\ref{fig:pdf_sensitivity}[a] and~[b] we show the
PDFs of integrated intensity and CO brightness, after adding
increasing levels of Gaussian noise at the beam scale and using a
dilated mask with $(t,e)=(5,1.2)$ to identify regions of significant
emission. For a moderate decrease in sensitivity ($\sigma_{RMS} \in
[0.4,1.0]$\,K), the \ico\ PDFs retain their shape at high intensities
(i.e. above $\sim50$\,\kkms). Once the noise level is increased to
$\sigma_{RMS} = 2$\,K, however, the PDF begins to diverge
significantly from the distribution obtained for the original cube,
even at high intensities. This is because several regions in the
nuclear bar and molecular ring regions are excluded from the mask once
this noise level is added to the cube. High \ico\ measurements in
these regions are often due to line profiles that are unusually wide
(FWHM greater than $\sim30$\,\kms) but have moderate brightness in
individual channels, so they do not possess a bright core that lies
above $5\sigma_{RMS} = 10$\,K. Even for moderate levels of added noise
($\sigma_{RMS} \leq 1$\,K), however, the best-fitting LN functions to
the \ico\ PDFs in Figure~\ref{fig:pdf_sensitivity}[a] and~[c] appear
to narrow as the sensitivity decreases (see
Table~\ref{tbl:pdffits_appendix}). This occurs even though we restrict
the range of CO intensities used to estimate the fit to pixels where
the emission is brighter than $4\sigma_{RMS}$. The dependence of the
\ico\ PDF width on brightness sensitivity should be kept in mind when
comparing the M51 PDFs to results from other galaxies or numerical
simulations.\\

\noindent The $T_{\rm mb}$ PDFs from cubes with differing noise levels
show more variation. Not surprisingly, the PDFs become narrower as the
noise increases (since less emission satisfies our criteria for
significance) but they also become steeper at low CO intensities,
losing the appearance of a truncation at $T_{\rm mb} \sim 5$\,K that
is observed for the original cube. Since the construction of the
dilated mask depends on the signal-to-noise, we examined whether
similar trends were observed when we used different $(t,e)$
combinations to identify significant emission. As an example,
\ico\ and $T_{\rm mb}$ PDFs for varying levels of noise using a
dilated mask with $(t,e) = (3.5,2)$ are shown in
Figure~\ref{fig:pdf_sensitivity}[c] and~[d] respectively. While these
PDFs are not identical to the PDFs in panels [a] and [b], the effects
of increasing noise on the shape of PDFs that we have described are
not sensitive to the particular combination of $(t,e)$ that we adopt
to mask the input cubes. \\

%%%%%%%%%%%%%%%%%%%%%%%%%%%%%%
\section{Estimating the Slope of the GMC and Young Cluster Mass Functions}
%%%%%%%%%%%%%%%%%%%%%%%%%%%%%%
\label{app:mspectests}

\noindent For both GMCs and stellar clusters, the shape of the mass
distribution is an important empirical signature of the physical
processes that regulate their formation and disruption. Defined as the
number of objects per unit mass, $f(M) = dN/dM$, numerous studies have
found that the mass distribution can be well-described by a power law,
$f(M) \propto M^{\beta}$, with a typical exponent of $\beta \approx
-1.7$ \citep[e.g.][]{rosolowsky05,fukuikawamura10} for extragalactic
GMCs identified in \aco\ surveys, and $\beta \approx -2.0$ for young
star clusters \citep[e.g.][]{chandaretal10}. An accurate determination
of the shape of the mass distribution is crucial if it is to be used
as a metric to quantify differences between and/or among populations
of GMCs and clusters, and hence to argue for (or against) the
universality of the processes that determine their evolution. As noted
by several authors, however, it can be difficult to measure the shape
of the mass distribution robustly, especially when the sample of
observed objects is small. Well-recognized sources of uncertainty
include the method used to identify structures of interest (which
usually depend on both resolution and sensitivity), unambiguous
determination of the low-mass completeness limit, inadequate sampling
of the high-mass end of the distribution, uncertainty in mass
measurements of individual objects, and (for mass distributions that
model a differential formulation with a histogram) the choice of
binning parameters. An added complication is that there are several
`standard' methods for representing mass distributions employed by the
GMC and stellar cluster research communities. In broad terms, studies
of stellar cluster populations tend to adopt a differential
formulation of the mass distribution, either separating the cluster
mass measurements into bins of variable width with equal numbers of
clusters in each bin, or into bins of uniform logarithmic width but
with a variable number of clusters in each bin, while it has become
increasingly common for studies of GMC populations to represent the
GMC mass distribution in cumulative form.\\

\noindent In light of these uncertainties and variations in technique,
it is evident that any literature comparison between the parameters of
mass distributions must be made with caution. In this Appendix, we
estimate the uncertainty in the slope of the GMC and stellar cluster
mass functions for different M51 environments by constructing the
distributions using three common methods, and by using different mass
ranges in the linear regression that we use to fit the mass
distribution. Our analysis falls short of being a general study of
bias in mass distributions in at least two important ways. Firstly, we
model all the mass distributions as pure power-laws, and do not
investigate other functional forms (such as a truncated power-law or
Schechter function) that have been used to describe GMC and cluster
populations. Secondly, we only examine a simple estimator --
i.e. ordinary least squares linear regression -- to determine the
best-fitting power-law to the mass functions. Since our main goal is
to assess whether the trends discussed in
Section~\ref{sect:pdfs_vs_sf} are robust, we consider it sufficient
that we have used the same power-law model and statistical estimator
for all the GMC and cluster mass distributions in our analysis. We
refer the reader to other studies
\citep[e.g.][]{rosolowsky05,reidetal10} for more comprehensive
investigations of the general problem of estimating the true mass
distribution of objects from observational data.  \\

\begin{figure*}
\begin{center}
\hspace{-0.5cm}
\includegraphics[width=140mm,angle=0]{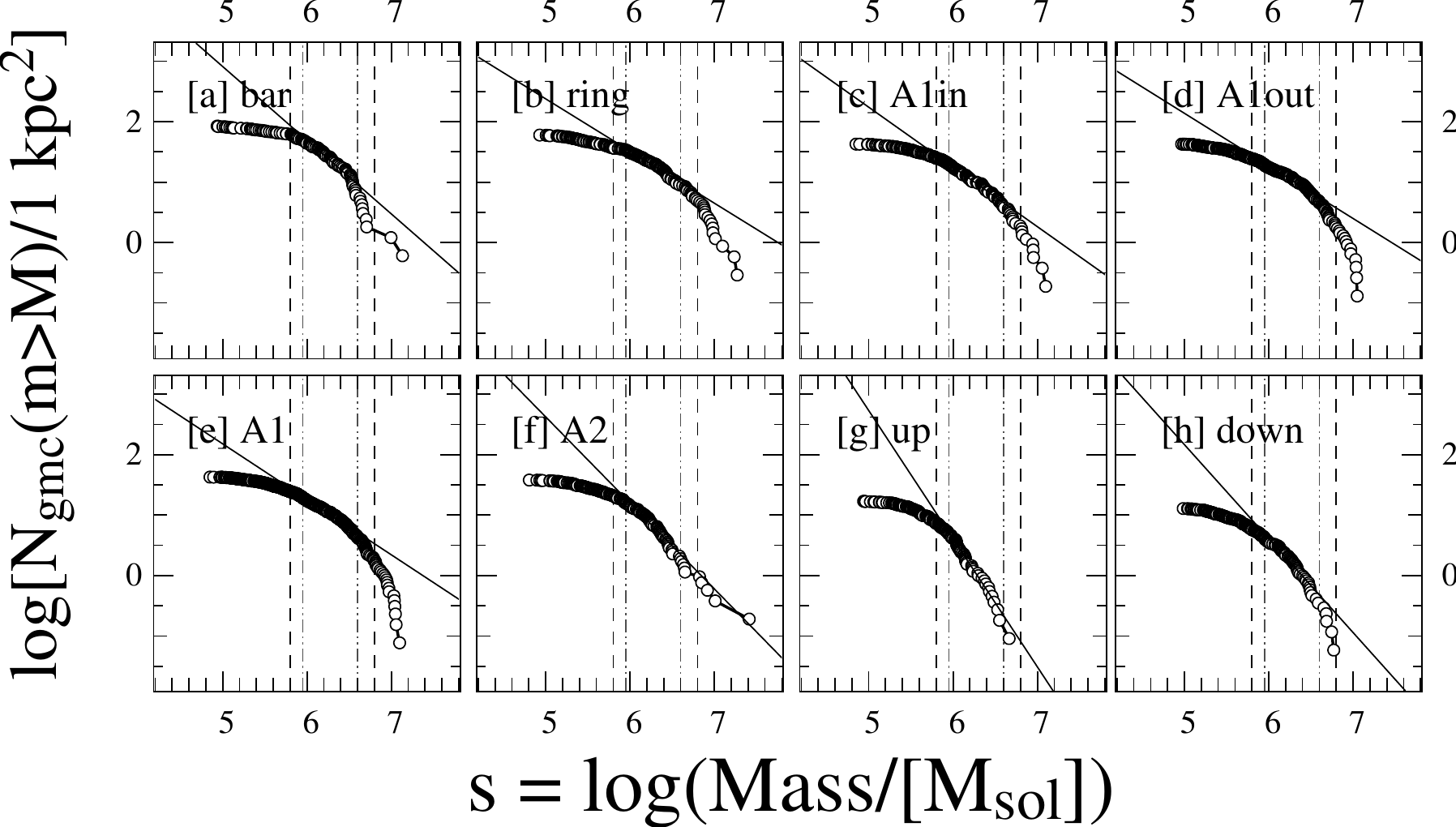}
\caption{\small Cumulative mass distributions for GMCs in the
  different M51 environments. In each panel, the vertical dot-dot-dashed
  lines indicate the mass range that was used to obtain the fit in the
  example shown. The black dashed lines indicate the limiting mass
  range over which we conduct the trials.}
\label{fig:gmc_cmf_eg}
\end{center}
\end{figure*}

\begin{figure*}
\begin{center}
\hspace{-0.5cm}
\includegraphics[width=140mm,angle=0]{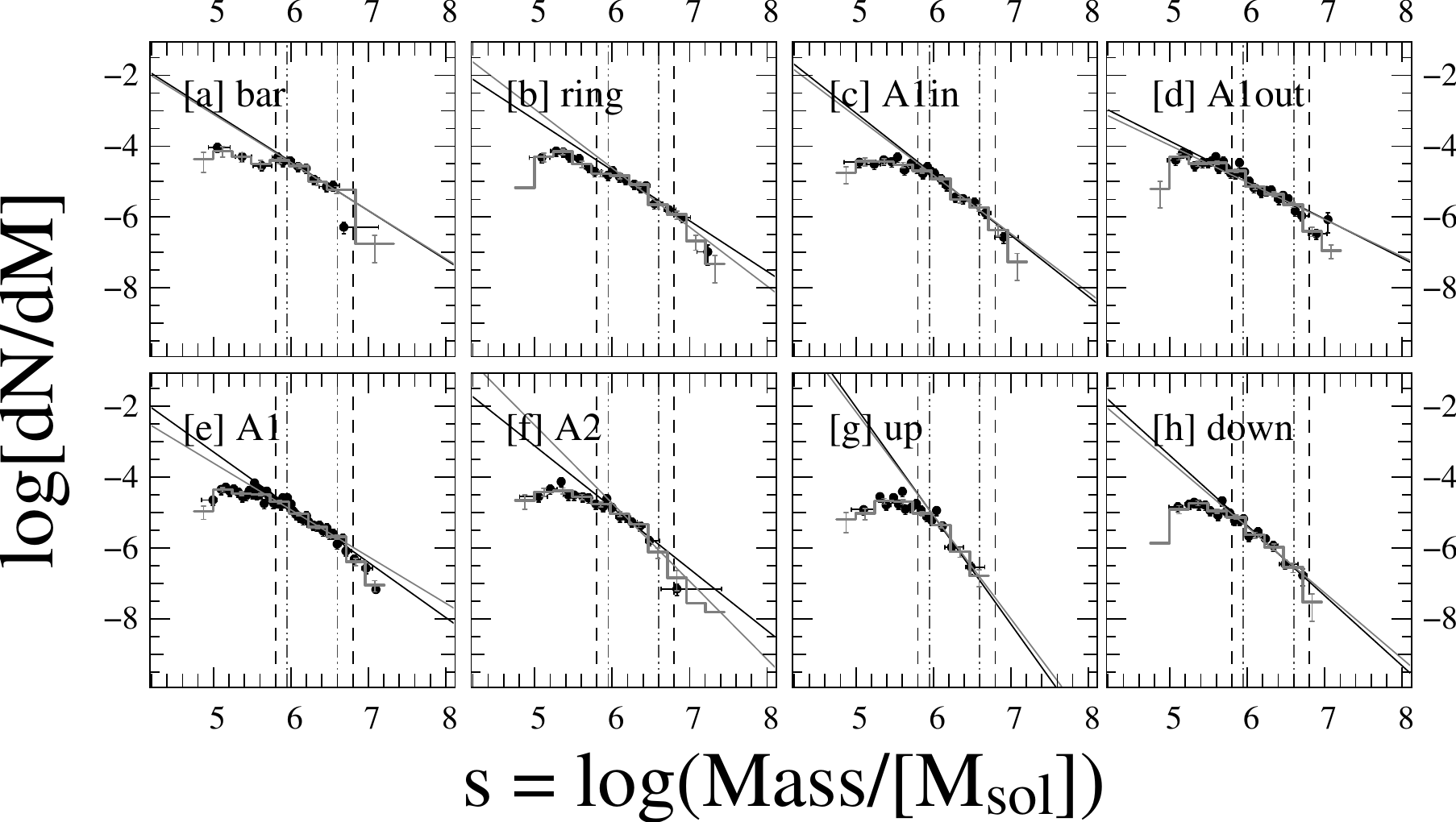}
\caption{\small Differential mass distributions for GMCs in the
  different M51 environments. In each panel, the grey lines indicate a
  histogram with bins of equal logarithmic width, while the black
  points represent a histogram with equal number of GMCs per
  bin. Other plot annotations are the same as in
  Figure~\ref{fig:gmc_cmf_eg}.}
\label{fig:gmc_dmspec_eg}
\end{center}
\end{figure*}

\begin{figure*}
\begin{center}
\hspace{-0.5cm}
\includegraphics[width=140mm,angle=0]{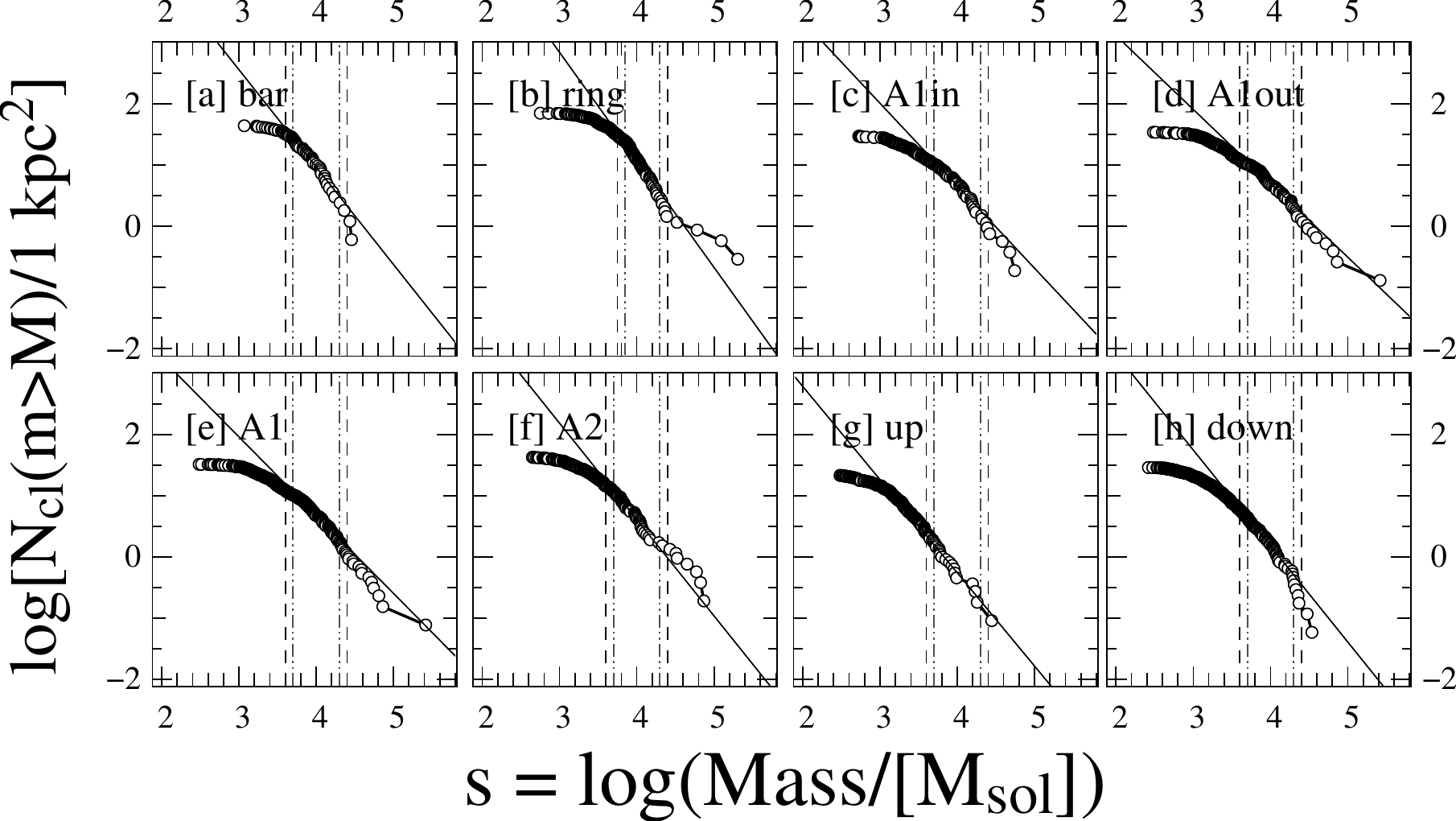}
\caption{\small Cumulative mass distributions for young ($<10$\,Myr)
  stellar clusters in the different M51 environments. Plot annotations
  are the same as in Figure~\ref{fig:gmc_cmf_eg}.} 
\label{fig:yc_cmf_eg}
\end{center}
\end{figure*}

\begin{figure*}
\begin{center}
\hspace{-0.5cm}
\includegraphics[width=140mm,angle=0]{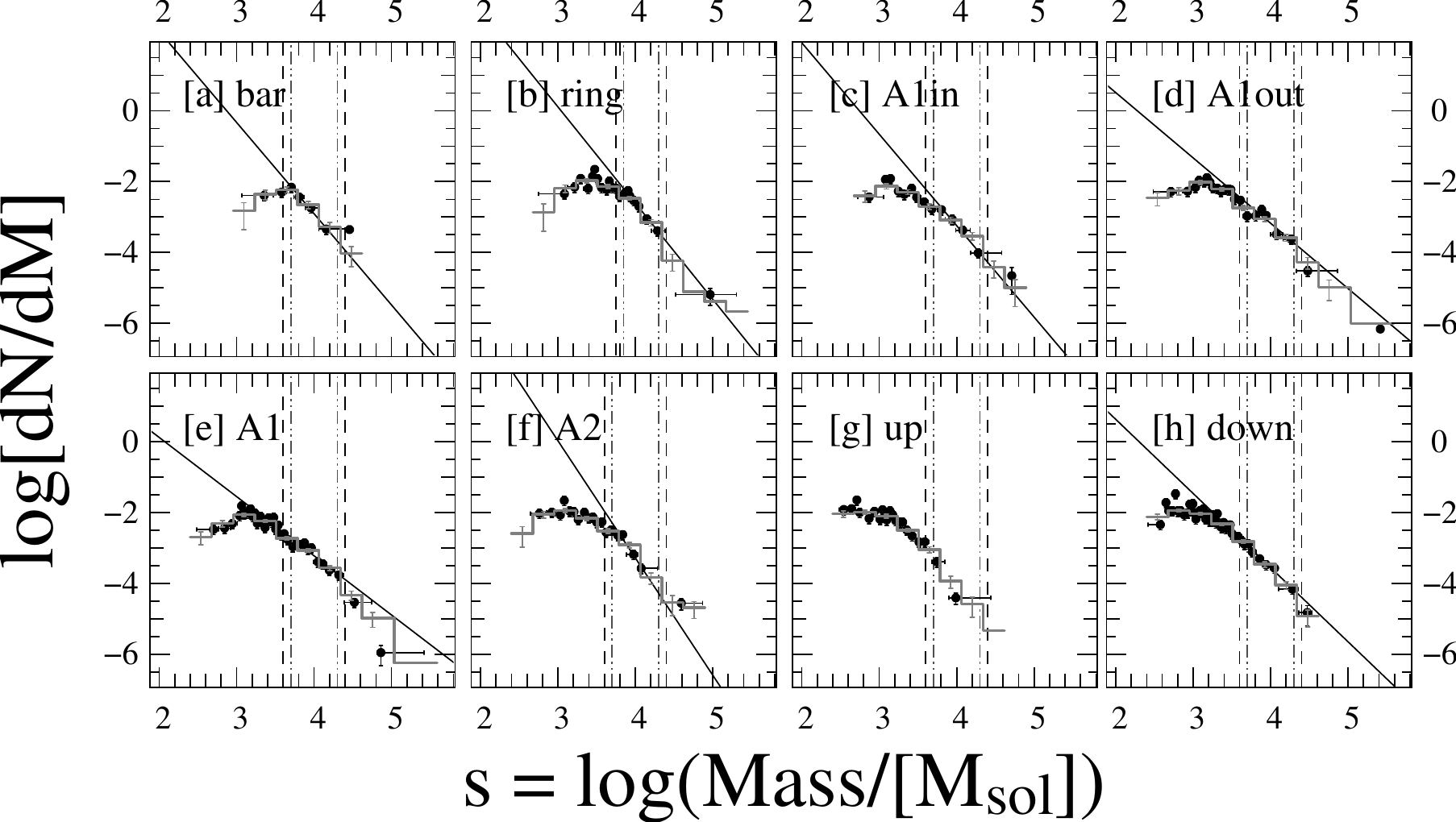}
\caption{\small Differential mass distributions for young 
  stellar clusters in the different M51 environments. Plot annotations
  are the same as in Figure~\ref{fig:gmc_dmspec_eg}.} 
\label{fig:yc_dmspec_eg}
\end{center}
\end{figure*}

\noindent In Figures~\ref{fig:gmc_cmf_eg} to~\ref{fig:yc_dmspec_eg},
we plot example mass distributions for the GMCs and young ($<10$\,Myr)
clusters in the eight M51 environments that we examine in this
paper. The distributions in Figures~\ref{fig:gmc_cmf_eg}
and~\ref{fig:yc_cmf_eg} are constructed using a cumulative
representation, while those in Figures~\ref{fig:gmc_dmspec_eg}
and~\ref{fig:yc_dmspec_eg} are differential mass distributions for the
same GMC/cluster populations, constructed using one example set of
binning parameters. In each panel of Figures~\ref{fig:gmc_dmspec_eg}
and~\ref{fig:yc_dmspec_eg}, we show both common forms of the
differential mass distribution, i.e. a histogram with bins of equal
logarithmic width (grey lines) and a histogram with equal number of
objects per bin (black points). The black vertical dashed lines
indicate the limiting mass range over which we estimate the fit for
each distribution; the limits of this range were chosen so that the
fit was calculated over an appreciable range of objects masses but
avoiding regions of the mass distributions that show clear evidence
for incompleteness effects (i.e. flattening) at low masses or
truncation and/or sampling effects at high masses. The grey dot-dot-dashed
lines indicate the actual mass range that was used to obtain the fit
in the examples shown. \\

\noindent The binning parameters and mass limits that were used in the
linear regression for our trial fits to the mass distributions are
listed in Table~\ref{tbl:mspectests}. For most environments, we tested
mass ranges with lower (upper) limits $\log M_{min,gmc} \in [5.8,6.1]$
($\log M_{max,gmc} \in [6.5,6.8]$) for GMCs and $\log M_{min,cl} \in [3.6,3.9]$
($\log M_{max,cl} \in [4.2,4.4]$) for young clusters. Additionally, for the
stellar clusters, we required that the mass range used to estimate the
fit was larger than 0.4\,dex, i.e. $\log M_{max} > \log M_{min} +
0.4$. For the fits to the differential mass spectra, bins containing
less than two objects were excluded from the fit, and the fit was only
estimated when three or more bins occupied the specified mass
range. For the ring environment, the range of lower mass limits for
stellar clusters was modified to $\log M_{min,cl} \in [3.75,3.9]$ because the
mass distribution (see panel [b] of Figure~\ref{fig:yc_cmf_eg})
flattens sharply below $M \sim 6000$\,\msol. In principle, this
flattening could have an observational (e.g. a higher completeness
limit due to crowding/extinction in this region) or a physical origin
(e.g. a higher probability of disruption for low-mass clusters
objects). These possibilities will be explored in a future paper that
investigates the shapes of the GMC and cluster mass distributions in
detail; here, we simply reduce the mass range that we fit to the part
of the distribution that conforms to a pure power-law. In total, we
obtained 35 (49) estimates of the cluster (GMC) mass distribution
slope for each of the cumulative mass distributions, and between 36
(362) and 385 (539) estimates for each of the differential mass
distributions. \\

\begin{table*}
\centering
\caption{\small Parameter Space Explored By GMC/Cluster Mass Distribution Trials}
\label{tbl:mspectests}
\par \addvspace{0.2cm}
\begin{threeparttable}
{\small
\begin{tabular}{@{}lcl}
\hline 
Parameter  & Range & Comment\\
           &       & \\
\hline
$n_{bin}$   & $[8,9,10,11,12,13,14,15,16,17,18]$ & Number of histogram bins \\
$n_{obj}$   & $[8,9,10,11,12,13,14,15,16,17,18]$ & Number of GMCs/cluster in each bin \\
$\log M_{min,cl}$\tnote{a} & $[3.6,3.65,3.7,3.75,3.8,3.85,3.9]$ & Lower mass limit for fit to cluster mass distribution\\
$\log M_{max,cl}$ & $[4.2,4.25,4.3,4.35,4.4]$          & Upper mass limit for fit to cluster mass distribution\\
$\log M_{min,gmc}$ & $[5.8,5.85,5.9,5.95,6.0,6.05,6.10]$ & Lower mass limit for fit to GMC mass distribution\\
$\log M_{max,gmc}$ & $[6.5,6.55,6.6,6.65,6.7,6.75,6.8]$  & Upper mass limit for fit to GMC mass distribution\\
\hline
\end{tabular}
}
{\small 
\begin{tablenotes}
\item[a]{Restricted to $\log M_{min,cl} \in [3.75,3.8,3.85,3.9]$ for
  fits to the cluster mass distribution in the ring region.}
\end{tablenotes}}
\end{threeparttable}
\end{table*}

\begin{figure*}
\begin{center}
\hspace{-0.5cm}
\includegraphics[width=140mm,angle=0]{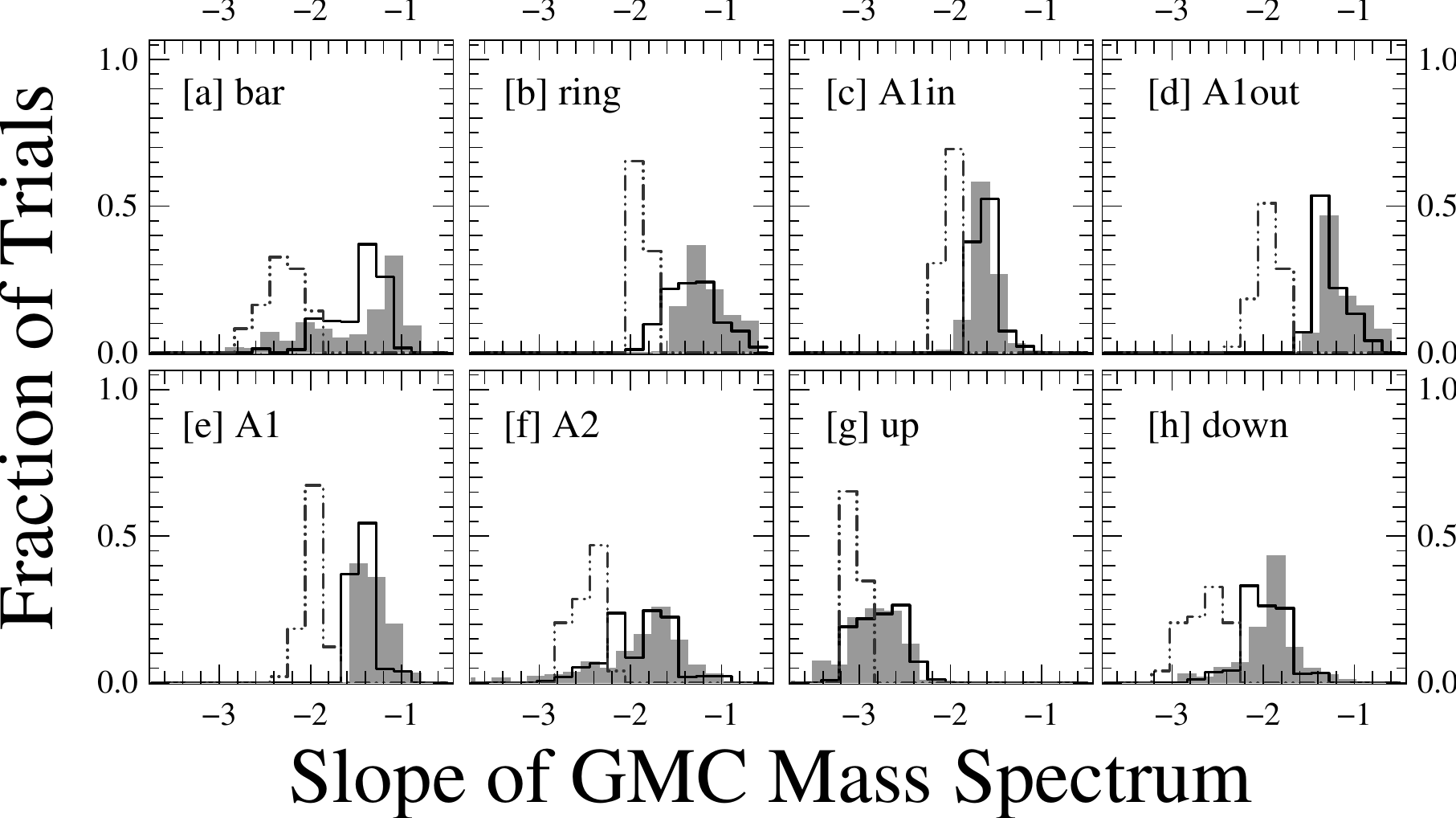}
\caption{\small Histograms showing the distribution of slopes that we
  obtain from the trial fits to the cumulative (dot-dot-dashed lines) and
  differential mass distributions of GMCs in different M51
  environments. The black histograms indicate the slopes obtained from
  distributions constructed using bins of equal logarithmic width. The
  filled grey histograms represent the distribution of slopes for a mass
  function constructed using an equal number of GMCs per bin.}
\label{fig:gmc_fithistos}
\end{center}
\end{figure*}

\begin{figure*}
\begin{center}
\hspace{-0.5cm}
\includegraphics[width=140mm,angle=0]{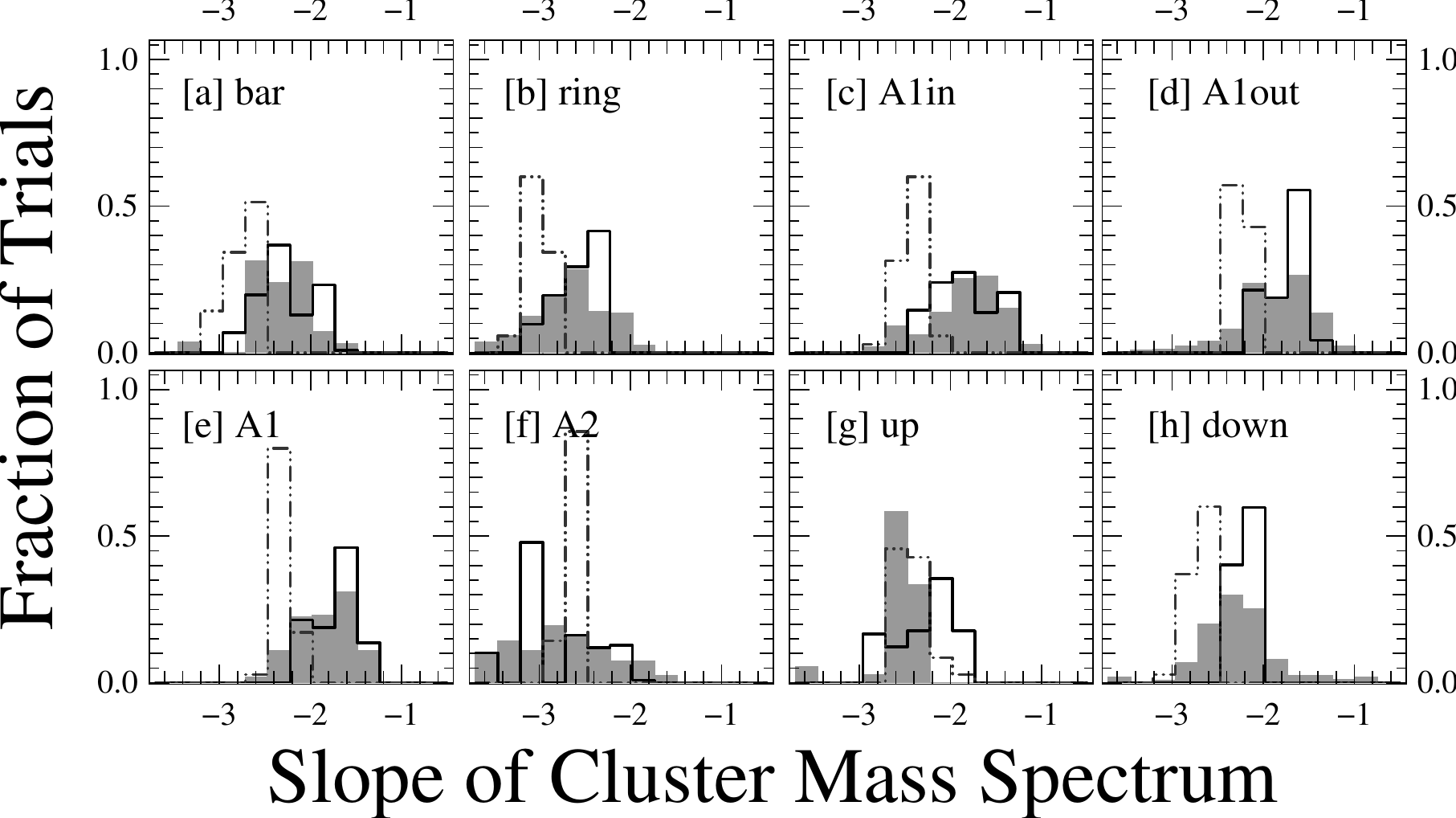}
\caption{\small The same as Figure~\ref{fig:gmc_fithistos}, but for
  young stellar clusters. }
\label{fig:yc_fithistos}
\end{center}
\end{figure*}

\noindent Histograms showing the distribution of slopes that we obtain
from the trial fits are shown in Figures~\ref{fig:gmc_fithistos}
and~\ref{fig:yc_fithistos}. For both GMCs and young clusters, it is
evident that the slopes obtained using a cumulative representation of
the mass distributions are systematically steeper (more negative) than
for the differential formulation, although the offset between the peak
of the histograms varies with environment. The discrepancy likely
reflects the fact that the observed mass distributions are not pure
power-laws, but tend to steepen continuously across the observed range of
masses. This steepening is not well-captured by the differential
distributions, since the values in each bin are weighted towards the
lower-mass objects (which are more common) and hence a systematically
shallower slope. The agreement between the two techniques is better
for environments (e.g. upstream of the spiral arms) where the mass
distribution more closely follows pure power-law behaviour, consistent
with this interpretation. \\

\noindent The slopes obtained for the differential mass distributions
constructed using different binning strategies (i.e. the black and filled grey
histograms in Figures~\ref{fig:gmc_fithistos}
and~\ref{fig:yc_fithistos}) are generally in good agreement for the
range of $n_{obj}$ and $n_{bin}$ values that we consider, but tend to
show a larger dispersion than the fits obtained from the cumulative
mass distributions. For some environments, the distribution of slopes
is especially broad, or shows evidence for bimodality. This is most
evident for environments where there is a relatively sharp bend in the
mass distribution within the mass range that we use to estimate the
slopes, e.g. GMCs in the bar region (panel [a] of
Figure~\ref{fig:gmc_fithistos}), and clusters in the material arm (A2,
panel [f] of Figure~\ref{fig:yc_fithistos}). Though we simply use the
median of each environment for our analysis in Section~\ref{sect:pdfs_vs_sf} and
capture the broader distribution in the uncertainties, it is worth
noting that the relevant value of the slope in these environments may
depend on the physical process under investigation, and range of
masses over which that process is likely to be acting.

\end{appendix}

\end{document}